\documentclass[a4paper,12pt,twoside]{report}
\usepackage{amsmath}
\usepackage{amssymb}
\usepackage{graphicx}
\usepackage[numbers]{natbib}

\newcommand{\be}{\begin{equation}}
\newcommand{\ee}{\end{equation}}
\newcommand{\bsplit}{\begin{split}}
\newcommand{\esplit}{\end{split}}
\newcommand{\chill}{$\chi$LL}
\newcommand{\dr}{\delta\rho}
\newcommand{\dra}{\delta\rho_\alpha}
\newcommand{\drb}{\delta\rho_\beta}
\newcommand{\rhoxx}{\rho_{xx}}
\newcommand{\rhoxy}{\rho_{xy}}
\newcommand{\rhor}{\rho_{R}}
\newcommand{\rhol}{\rho_{L}}
\newcommand{\rhoa}{\rho_{\alpha}}
\newcommand{\rhob}{\rho_{\beta}}
\newcommand{\sigmaz}{\sigma_{\alpha,\beta}^z}
\newcommand{\thetaa}{\theta_{\alpha}}
\newcommand{\thetab}{\theta_{\beta}}
\newcommand{\thetar}{\theta_{R}}
\newcommand{\thetal}{\theta_{L}}
\newcommand{\thetaj}{\theta_{J}}
\newcommand{\thetan}{\theta_{N}}
\newcommand{\chilll}{$\chi$LL~}
\newcommand{\phia}{\phi_{\alpha}}
\newcommand{\phib}{\phi_{\beta}}
\newcommand{\sign}{\mathrm{sign}}
\newcommand{\xa}{x_\alpha}
\newcommand{\xb}{{x_\beta}}
\newcommand{\sgn}[1]{\mathrm{sign}(#1)} 

\begin{document}
\begin{titlepage}
\begin{figure}[!h]
\includegraphics[width=2cm,clip]{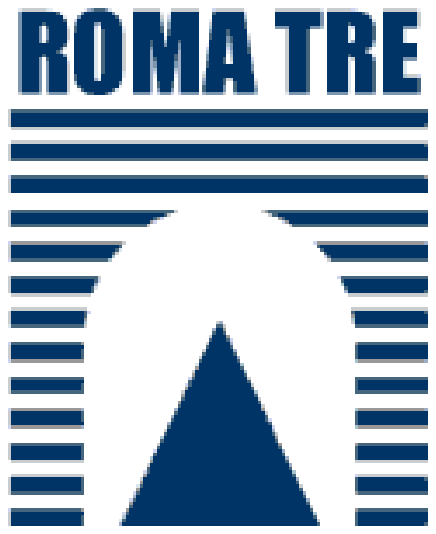}
\hspace{.3 cm}\parbox[b]{17cm}
{\Large{UNIVERSIT\`A DEGLI STUDI ROMA TRE} \\ \medskip
\medskip
\Large{Dipartimento di Fisica ``Edoardo Amaldi''} \\
\Large{Dottorato di Ricerca in Fisica - XV Ciclo}}
\end{figure}
\begin{center}
\vspace{1.0cm}
\end{center}
\begin{center}
\vspace{2.cm}
{\bf \LARGE 
Electrical transport properties of }\\
\medskip 
{\bf \LARGE bidimensional electron liquids in  }\\ 
\medskip
{\bf \LARGE the presence of a high magnetic field}\\
\vspace{1.5cm}
{\LARGE Roberto D'Agosta}
\end{center}
\vspace{4.cm}
\begin{table}[!h]
\begin{center}
\begin{tabular}{ccc}
\large \it Coordinatore & \qquad \qquad \qquad \qquad
& \large \it Tutore \\
& &  \\
\large prof. Filippo Ceradini & & \large dr. Roberto Raimondi\\
\end{tabular}
\end{center}
\end{table}
\vspace{1.cm}
\begin{center}
{\large Aprile 2003}
\end{center}
\end{titlepage}

\chapter*{}
\thispagestyle{empty}
\newpage

\setcounter{page}{1}
\tableofcontents
%\listoffigures
\par\vfill
\eject

\chapter{Classical and Quantum Hall Effect}
The research in condensed matter physics is very often closely related
to advances in material science.  The improvement of the experimental
techniques to fabricate novel devices and the achievement of very low
temperatures have opened new extraordinarily opportunities to
investigate the physics of the electron systems.  One of the most
impressive result in the last twenty years is the discovery of high
nonlinearities in the conductance of a two-dimensional electron liquid
in the presence of a high magnetic field \cite{Klitzing1980, Tsui1982, Laughlin1983}.
The name of Quantum Hall
Effect (QHE) was assigned to this phenomenon because it recalls, as we
will briefly discuss later, the classical Hall effect in metals.  This
discovery has started a very productive research field where new
concepts as Landau Levels, Incompressible Electron Liquid, Composite
Fermions and Edge States were introduced and proved to be very useful \cite{de-Picciotto1997, Griffiths2000, Comforti2002, Maasilta1997, Jain1989, Heinonen1998}.
Also new experimental ideas were developed and nowdays the Quantum
Hall Effect is used as a standard to define some physical constants.

\begin{figure}[!ht]
\begin{center}
\includegraphics[clip,width=10cm]{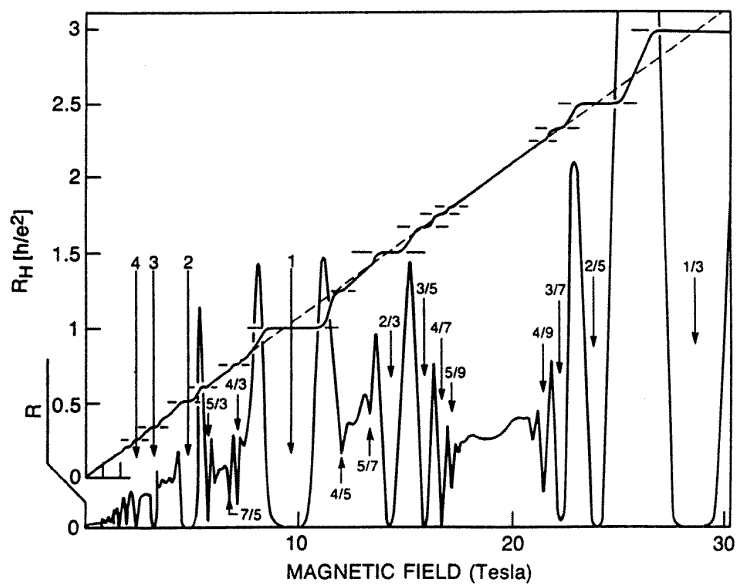}
\caption{The longitudinal and transversal resistance in Hall bar as 
functions of the magnetic field. In the transversal resistance the
 plateaus of quantized values are evident.  From \cite{Stormer1992}.}
\label{qhe_transport_data.eps}
\end{center}
\end{figure}

The experimental results for the transversal $\rhoxx$ and longitudinal
$\rho_{xy}$ resistance are summarized in
Fig. \ref{qhe_transport_data.eps} (data from \cite{Stormer1992}).
The transversal resistance $\rhoxy$ of this system shows an exact
quantization in its dependence on the magnetic field, $\rhoxy=h/e^2
\nu$ with $\nu$ an integer or rational.  The longitudinal resistance
$\rhoxx$, on the other hand, shows a very non trivial behavior,
vanishing or having a delta-like behavior when the transversal
conductance resides on a plateau or passes through two distinct
plateaus, respectively, (see for reference
Fig. \ref{qhe_transport_data.eps}).  The QHE can be expressed hence by
the relations\footnote{We will use the boldface to define vectors in
two dimensions and an arrow to define three dimensional vectors. In
the following we will also use a relativistic notation where the greek
letters will indicate the coordinates in $2+1$ dimensions and the
latin letter the spatial coordinates.}
\be
\begin{split}
{\bf j}=&\hat \sigma {\bf E},\\
\hat\sigma=&\frac{e^2 \nu}{h}\left(
\begin{array}{cc}
0 & 1\\ -1 & 0
\end{array}
\right).
\end{split}
\ee
The first equation is the Ohm's law in matrix form and relates the
current density to the electric field in the material.  In the
plateau regions the conductivity matrix becomes purely off-diagonal
and is expressed in terms of a universal conductance quantum ($e^2/h$)
and of a number $\nu$.  From the experiments we know that $\nu$, which
takes the name of {\it filling fraction}, assumes only integer and
fractional values.  We separate the case when $\nu$ is an integer,
and call this phase Integer Quantum Hall Effect (IQHE), or a
fractional number, namely the Fractional Quantum Hall Effect (FQHE).

It is widely believed that the IQHE and FQHE have a very different
physical origin. IQHE is thought to come from the interplay between
the Landau Levels quantization of the motion of a charged particle in
a magnetic field, and the disorder.  FQHE cannot be explained without
taking into account the electron-electron interaction. This leads to a
new state of matter in which one may have a new kind of
quasi-particles carring a fractional charge \cite{Laughlin1983}.  
A hierarchy of
fractional filling fractions was justified theoretically by assuming
that in the FQHE the quasi-particles which consistute the excitation
of the system can undergo a IQHE. The idea is that every
quasi-particle carries a number of magnetic flux quanta which
partially compensates for the external magnetic field. Hence the
quasi-particles may give rise to a IQHE phase when the effective
magnetic field, obtained as the difference of the external applied
field and the sum of the magnetic fields attached to the
quasi-particles, assumes certain values \cite{Jain1989}.

The quasi-particles defined in these theoretical 
approaches belong to the bulk 
of the system. 
However it was showed that the current
in the QHE is concentrated on the edges of the system
\cite{Halperin1982}.  This can be justified by a semiclassical approach. 
The finite size of the sample can be taken into account by an external
potential which bends up near the physical edges.  The electrons that
are well inside the system bulk move on circular orbits whose radius
is determined by the magnetic field.  Hence they are fully localized
and cannot contribute to the transport.  However the orbit of the
electrons which reside near one edge is not fully contained in the
system. From a classical point of view these electrons are reflected
by the barrier generated by the confining potential and they propagate
all in the same direction because the magnetic field fixes the sign of
the circulation.  Then there is the possibility of the formation of
`skipping orbits' which can travel through the system. From a quantal
point of view the electrons localized near the edge of the system
reside in the `edge states' whose properties we will discuss in the
following.

In this thesis we will be concerned with transport properties and edge
dynamics in the FQHE regime.  We will show that the charged
excitations of a two-dimensional electron liquid are concentrated in
the edges of the systems thereby forming a one-dimensional system.
Hence we can study the transport properties of the whole system by
studying the interaction beetween two one-dimensional systems.  It is
believed that the edges of a system in the FQHE regime are a nearly
perfect realization of the (Chiral) Luttinger Liquid (\chill)\cite{Haldane1981, Wen1990}.  The
Luttinger Liquid model describes one-dimensional electron liquid with
linear energy dispersion.  It is an exactly solvable model and it will
be a fundamental tool in this thesis.  We will describe its basic
properties in the following chapter.

Starting from some physical reasonable assumptions we will show that
the localization of the density excitations in the region of the edges
is independent of the presence of a developed QH phase. We will derive
an equation of motion for such excitations and then use the solutions
of this equation to calculate the electrical transport properties.

This research was stimulated by a collaboration with an experimental
group.  The main experimental aim was the realization of new
electronic devices able to detect the fractionally charged
quasi-particles. To this end, it was planned to study the transport
through a constriction in the two-dimensional electron liquid. This
kind of problems have been treated theoretically in the past by
considering the constriction as a quantum point contact in the sense
that in a very narrow region the edges of the Hall liquid come almost
in contact so enhancing the probability of the tunneling of
excitations between the edges
\cite{Wen1991b, Chamon1997, Kane1995}. 
In this thesis we will drop the assumption that the constriction has
zero extension, and we will show that, by taking into account a finite
size of the constriction, one obtains different results for the
tunneling current.  In particular, we may obtain a better qualitative
agreement with the experimental results in certain ranges of
temperature.

\section{The Classical Hall effect}

We will start our discussion of the QHE by recalling the Classical
effect. The Classical Hall effect occurs when a current flows in a
metal in the presence of a uniform magnetic field.  The motion of a
charged particle in a magnetic field can be separated in the motion
parallel to the direction of the magnetic field and in the motion
perpendicular to such direction.  We assume that the magnetic
field is parallel to the direction of the $\hat z$ axis
\be
{\vec B}=-B\hat z
\ee
and we choose a right handed set of orthogonal axis $(x,y,z)$. 
%When an
%electric field is also present the charged particle is subject to the
%Lorentz force. 
The equation of motion projected in the $\hat z$
direction is given by
\be
\ddot z=-e E_z
\ee
hence the particle is uniformly accelerated in this direction.

We neglect in the following such motion and consider the particle
confined in a two-dimensional plane $(x,y)$. The equation of motion
can be written as\footnote{The electron charge is negative then the
physical constant $e$ is positive.}
\be
m\ddot{\bf r}=-e{\bf E}-e{\bf v}\times {\bf B}.
\ee
When we consider the equilibrium situation we have that the
acceleration must be zero and we obtain
\be
{\bf J}=\hat \sigma {\bf E}
\label{classicalhall}
\ee 
where we have defined the current density
\be
{\bf J}=-ne{\bf v},
\ee 
and the conductivity matrix
\be
\hat\sigma=\left(
\begin{array}{cc}
0 & -\sigma_H\\
\sigma_H & 0 
\end{array}
\right).
\label{classicalsigma}
\ee  
The classical Hall resistance ($R_H$), defined by
\be
\sigma_H=R_H^{-1}=\frac{ne}{B},
\label{classicalsigmah}
\ee
where $n$ is the electron number density, is then linearly dependent
on the magnetic field.  

The measure of the Hall resistance as a
function of the magnetic field gives information on the electron
density in metals. Notice that the Hall resistance is measured in the
transversal direction with respect to the direction of the
current. The conductivity matrix however is not complete.  Indeed the
Eqs. (\ref{classicalhall}) and (\ref{classicalsigma}) imply that the
longitudinal conductivity is zero. Hence the current in that direction
will flow without dissipation. In fact the classical Hall effect and
the usual metal dissipation must be considered together then giving the
conductivity matrix for a metal in a magnetic field
\be
\hat\sigma=
\left(
\begin{array}{cc}
\sigma_D & -\sigma_H\\
\sigma_H & \sigma_D
\end{array}
\right)
\ee
where $\sigma_D$ is the Drude conductivity.

We can understand the classical Hall effect in the following way. When
the magnetic field is zero the electron current flows following the
gradient of the electric potential. If one turns on the magnetic field
the motion of the particle becomes circular as one can easily verify
be solving the equations of motion.  The important point to stress is
that the circulation is determined by the sign of the magnetic field
and hence all the particles, which run in the same direction, turn in the
same way.  This effect creates a mean transversal current and a charge
accumulation at one edge of the metal. To maintain the electrical
neutrality an equal amount of opposite charges must accumulate at the
other edge. This creates a transversal electrical field which will
equilibrate the transversal current and establishes a dynamical
equilibrium.  The measure of the transversal electric field
constitutes a way to measure the magnetic field.

We
introduce now a relativistic notation which will become useful
in the following.  The electromagnetic field can be defined by using
the vectorial and electric potentials $\vec{A}$ and $V$. In terms of
these potentials the electric and magnetic fields are defined as
\be
\begin{split}
\vec{E}&=-\partial_t \vec{A}-\vec{\nabla} V,\\
\vec{B}&=\vec{\nabla}\times \vec{A}.
\end{split}
\ee
We deal with only two spatial dimensions because, as we have seen,
the problem can be separated and in the direction of the magnetic
field the equation of motion is easily solved. On the other hand when
we will address the quantum problem we will see that the
particles are strongly confinated in the direction of the magnetic
field and the problem can again be separated.  We introduce the
covariant ``tri-vector" potential\footnote{Notice that even if the
magnetic field points out the $(x,y)$ plane the vector potential can
be defined as a function only of the in-plane variables.}
\be
A_\mu=(V,{\bf A})
\ee
where the index $\mu$ will run over $0,1,2$ with the convention that
$0$ will coincide with the temporal part of the tri-vectors, $1$ coincides
with the $x$ variable and $2$ with $y$. As an example the
tri-derivative is
\be
\partial_\mu=(\partial_{ct},-\partial_x,-\partial_y)=(\partial_{ct},-{\bf \nabla}).
\ee
We need to introduce also the metric
\be
g_{\mu\nu}=\left(
\begin{array}{ccc}
1 & 0 & 0\\ 0 & -1 & 0\\ 0 & 0 & -1
\end{array}
\right),
\ee
and the totally antisymmetric Levi-Civita tensor $\epsilon_{\mu\nu}$.
With these definitions, we may rewrite the magnetic and electric field
in the form
\be
\begin{split}
E_i&=-\partial_0 A_i+\partial_i A_0\\ B&=\epsilon_{jk}\partial^jA^k
\end{split}
\ee
where, as usual in the relativistic notation, repeated indices are
summed over.  We can combine the equation (\ref{classicalhall}) and
the definition (\ref{classicalsigmah}) in the compact form
\be
J^\mu=\sigma_H \epsilon^{\mu}_{\phantom{\nu}\nu\lambda} \partial^\nu
A^\lambda
\label{phe-qhe}
\ee
where $J^0=c \rho=-e c n$. This equation is the phenonemological
representantion of the Hall effect and it can be extended to the QHE
because it comes directly from a Lorentz transformation which is valid
if the system is translationally invariant. Its quantum counterpart
constitutes the starting point for the seminal papers of Wen
\cite{Wen1991b, Wen1990, Wen1991a, Wen1995} for the derivation of the
properties of the FQHE. The main task of a microscopic theory of the
QHE must be to derive this equation starting from a microscopic 
electron Hamiltonian.

%%%%%%%%%%%%%%%%%%%%%%%%%%%%%%%%%%%%%%%%%%%%%%%%%%%%%%%%%%%%%%%%%%%%%%%%%%%%%%%%%%%%%%%%%%%%%
\section{A first glance to the Quantum Hall Effect}
The classical theory of the Hall effect was well understood and proved very
useful in the experimental characterization of the properties of
non-magnetic metals. Hence the experimental result of von Klitzing
{\it et. al.}
\cite{Klitzing1980}, reported in Fig. \ref{qhe_transport_data.eps},
 was totally unexpected.

\begin{figure}[!ht]
\begin{center}
\includegraphics[clip,width=7cm,angle=-90]{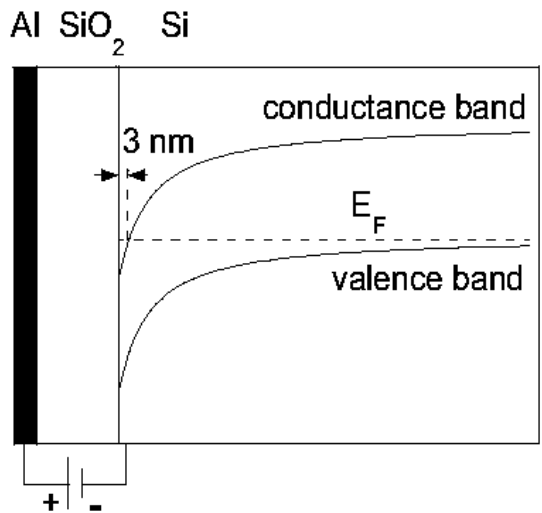}
\caption{A simple scheme of the electron confinement around an interface
 between two different semiconductors.}
\label{semiconductor-inter.eps}
\end{center}
\end{figure}

Before discussing the features of this experimental result, let us
consider briefly the experimental setup. The devices used in this type 
of experiments are semiconductor heterostructures or heterojunctions
where the electron gas resides at the interface between two different
semiconductor species. The polarization of the different
semiconductors generates a uniform electric field which confines the
electrons in a small region localized around the interface as it is
indicated in Fig.  \ref{semiconductor-inter.eps}.
Hence the electron gas can be considered as two-dimensional. This
 introduces a great simplification. Indeed it is well known that the
 resistivity $\rho$, and the resistance $R$ are related
 by
\be
R=\rho L^{(2-d)}
\ee
where $d$ is the dimension of the system and in two dimensions these
two quantities coincide.

\begin{figure}[!ht]
\begin{center}
\includegraphics[clip,width=10cm]{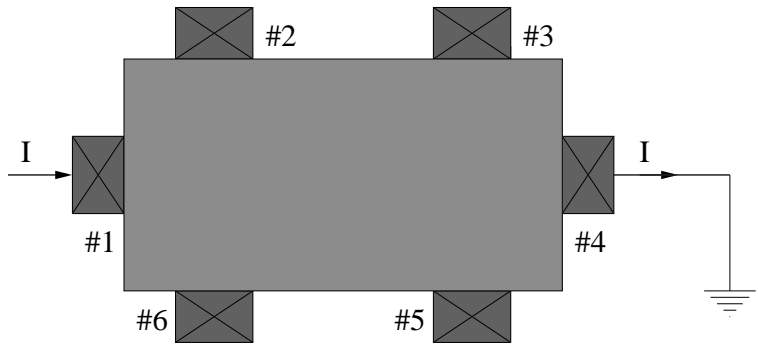}
\caption{a) A schematic of the device used in the experiment on the Quantum Hall effect. 
The source ($\# 1$) and drain ($\# 4$) contacts are used to inject in the system
a steady-state current $I$.
The other probes ($\# 2,\# 3,\# 5, \#6$) are used to measure the
potential drop in the system to obtain the curves in
Fig. \ref{qhe_transport_data.eps}.} 
\label{fig_probe.eps}
\end{center}
\end{figure}

We choose the direction of the electric field as the $\hat z$ axis. In
this direction a magnetic field will be also applied. To reduce the
thermal effects the device is maintained at temperatures below $1$
{\rm K} (the newest experimental setups can reach about $10$ {\rm mK}). The
electron gas is contacted with a multiprobe setup. Two of these
contacts are used to inject a steady-state current and two other
contacts are used to measure the potential drop in the system. A
six-probe setup is shown in Fig. \ref{fig_probe.eps}.

Once the current is fixed we can measure the Hall resistance by measuring the
voltage drop, as for instance, on the probe $4$ and $1$ or the
longitudinal resistance by measuring the potential of the contacts $4$
and $3$ of the device in Fig \ref{fig_probe.eps}.  The results of
these measurements at very high magnetic field and very low
temperature are shown in Fig. \ref{qhe_transport_data.eps}. In that
figure the longitudinal and transversal resistances are reported. By
using the Eq. (\ref{classicalsigmah}) for the Hall resistance and the
Drude part for the longitudinal resistance one expects that the
longitudinal resistance is a constant and the transversal is linearly
dependent on the magnetic field intensity.

From the experimental data we see that this is not true (see Fig.
\ref{qhe_transport_data.eps}). The 
transversal resistance shows clear and large plateaus. In these
plateaus the resistance has a fixed value which is proportional to the
quantum of conductance $e^2/h$ and the constant of proportionality is
a integer or rational number. Outside the plateaus its relation with
${\bf B}$ becomes linear. More strange is the behavior of the
longitudinal resistance. For small value of the magnetic field the
Shubnikov-DeHaas oscillations are seen. However when the amplitude of
these oscillations increases the longitudinal resistance shows an
alternation of high maxima and flat regions where its value is
zero. The plot shown in Fig.
\ref{qhe_transport_data.eps} is a more recent version of the combined result
of von Klitzing \cite{Klitzing1980} on IQHE and Stormer, Gossard and
Tsui
\cite{Tsui1982} on FQHE.

What is paradoxical in these results is that the electron-electron
interactions and the disorder play a fundamental role to obtain it
rather than destroy it. Indeed the results for the conductances in
the previous section can be derived by using only the relativistic
invariance \cite{Girvin1998}, hence it does not matter if the system
is classical or quantum and the electrons form a gas or a solid just
maintaining the translational invariance.

Let us notice another peculiar aspect of this effect. In the region
where the longitudinal resistivity is nearly zero and the transversal one
is quantized, the system behaves like a superconductor ($\rho_{xx}=0$)
and like an insulator ($\sigma_{xx}=0$). The reason for that is the
presence of the finite off-diagonal terms, $\sigma_{xy}$, which
assures the existence of the inverse of the matrix $\hat\sigma$
 and the finite value of the $\rho_{xy}$ resistivity.

\section{The Integer Quantum Hall Effect}
Let us start the theoretical study of the QHE from the Integer Quantum
Hall Effect, this being the first experimentally observed \cite{Klitzing1980}
and the more theoretically understood. In this effect the resistivity $\rho_{xy}$ 
is quantized in terms of the quantum $h/e^2$ and the costant of
proportionality is an integer (measured with a precision of a part over $10^8$).

The very starting point is the quantum motion of a charged particle in
a magnetic field.  As it is well known, the motion of a charged
particle in a magnetic field is obtained from the free Hamiltonian
with the ``minimal substitution"
\be
{\bf p}\to {\bf p}+e{\bf A}
\ee
where ${\bf A}$ is the vector potential\footnote{We choose to indicate
the vector potential as a bidimensional vector. We assume that the
component in the $\hat z$ direction is identically zero.}.  The
presence of an electric potential does not affect such transformation
and the Hamiltonian then reads
\be       
H=\frac{1}{2m}\left({\bf p}+e{\bf A}\right)^2-e\Phi.
\ee

We consider the simplest case of $\Phi\equiv 0$. There are many ways
to solve the Schr\"odinger equation for a single particle. We consider
here a field-theoretical inclined approach that will provide some useful
relations to understand the modern literature. The position and
momentum operators follow the usual commutation relations
\be
\begin{split}
&[x,p_x]=[y,p_y]=i\hbar,\\ &\mbox{others}=0.
\end{split}
\ee
We define the new operators
\be
\begin{split}
&\xi=\frac{1}{eB}\left(p_y+eA_y\right),\\
&\eta=-\frac{1}{eB}\left(p_x+eA_x\right),\\ 
&X=x-\xi,\\ 
&Y=y-\eta,
\end{split}
\ee
which follow the commutation relations
\be
\begin{split}
&[X,Y]=-[\eta,\xi]=-i\ell^2,\\ 
&\mbox{others}=0,
\end{split}
\ee
where we have introduced the magnetic length $\ell^2=\hbar/eB$.  To
obtain these commutation relations we must remember that $A_x$ and
$A_y$ are in general function of both the operators $x$ and $y$. On the
other hand, the rotor of ${\bf A}$ is independent from $x$ and $y$
and hence the vector potential must be a linear function of these
operators and the commutators turn out to be c-numbers.

In terms of these new operators the Hamiltonian takes the simple
form
\be
H=\frac{m}{2}\omega_c^2(\xi^2+\eta^2)
\ee
where we are faced with the problem of a one dimensional harmonic
oscillator in the variables $\eta$ and $\xi$ so that the energy spectrum
is written immediately as
\be
E_n=\left(n+\frac12\right)\hbar \omega_c
\ee
where we have introduced the ciclotron frequency $\omega_c=eB/m$.  The
energy levels take the name of ``Landau Levels" (LL) after the solution
of Landau  of the motion of a particle in a magnetic field \cite{Landau1930}.

To obtain an explicit form for the eigenfunctions we must specify a
gauge. A possible gauge is the Landau gauge
\be
{\bf A}=(By,0)
\ee
and we factorize the eigenfunction as
\be
\psi(x,y)=\frac{e^{ik_x x}}{\sqrt{L}}\chi(y)
\ee
where $\chi$ is a solution of the Schr\"odinger equation
\be
\left[-\frac{\hbar^2}{2m}\frac{d^2}{dy^2}+\frac{m}{2}\omega_c(
 k_x\ell^2-y)^2\right]\chi(y)=E_n\chi(y).
\ee
This is the Schr\"odinger equation of a harmonic oscillator with the
center at $k_x \ell^2$ hence the solution is readily written in terms
of a product of a gaussian and the Hermite polynomials.  Notice that
when evaluated over these functions the mean value of $Y$ depends only
on $k_x$
\be
\langle Y\rangle=k_x \ell^2,
\ee
and because $Y$ does not depend on time we have that in this
representation $Y=k_x \ell^2 \hat 1$.

Now let us turn on an electric field of amplitude $E_y$ in the $y$
direction.  The electric field adds a term which is linear in $y$
hence when we substitute $y=\eta+Y$, we have a term linear in $\eta$
that changes the center of the harmonic oscillator and does not affect
the energy, and a constant term in $Y$ which enters the energy
\be
E_n(k_x,E_y)=\left(n+\frac{1}{2}\right)\hbar \omega_c +\frac{e^2
E_y^2}{2m\omega_c^2}+eE_y \langle Y\rangle.
\ee
From this relation it is easy to calculate the velocities defined as
\be
{\bf v}=\frac{1}{\hbar}\frac{\partial E_n}{\partial {\bf k}}.
\ee
We have
\be
\begin{split}
&v_x=\frac{\ell^2}{\hbar}eE_y,\\ &v_y=0
\end{split}
\ee
and we can calculate the conductivity
\be
\begin{split}
&J_x=-e n v_x=-\frac{e^2\ell^2}{\hbar}n E_y=-\frac{e^2}{h}\nu E_y\\
&J_y=0
\label{jhall}
\end{split}
\ee
where $n$ is the electron density ($N$ is the total number of
electron) and we have defined the {\it filling factor}
\be
\nu=\frac{n}{n_B}=\frac{N}{N_B}=\frac{h n}{eB}.
\label{filling-factor}
\ee
From the Eq. (\ref{jhall}) one can directly compute the conductivity
matrix as
\be
\hat \sigma=\left(
\begin{array}{cc}
0 & -e^2\nu/h
\\ e^2\nu/h & 0
\end{array}
\right).
\label{sigma_hall}
\ee
In the definition of the filling factor (\ref{filling-factor}) we have
also defined the quantity $n_B=eB/h=(2\pi \ell^2)^{-1}$. It is
possible to relate this density to the degeneracy of the Landau
levels. In fact let us consider the case of an infinite system which
is constituted by the replication of a finite size, let us say
$L_x\times L_y=S$, system. The quantum number $k_x$ must be quantized
in this finite size system as $k_x=2\pi j/L_x$ and, on the other hand,
the center of the electron orbit must be contained in such a system
hence we have the upper limit $k_x \ell^2<L_y$. We see then that the
discrete number $j$, which defines the quantization of $k_x$, must be
limited by
\be
j_m=\frac{L_x L_y}{2\pi \ell^2}=\frac{S
B}{\frac{h}{e}}=\frac{\Phi}{\Phi_0}=N_B =n_B S.
\ee   
Notice that this degeneracy is the same for every Landau level.
From this relation we see that the quantity $N_B$ can be interpreted
as the number of magnetic flux quanta contained in the system (we have
used that $\Phi_0=h/e$ is the magnetic flux quantum). Hence the
filling factor can be interpreted as the mean number of flux quanta
carried by the single electron. On the other hand the filling factor
gives information about how many Landau levels are filled, indeed
every Landau level can contain, as maximum, $j_m$ electrons\footnote
{Notice that we consider the electrons fully polarized, hence there is
not the factor $2$ due to the spin.}  and directly from the definition
we have $\nu=N/j_m$.  This simple model seems to reproduce most of the
physics of the IQHE (the form of the conductivity matrix, its
dependence of the universal quantity $e^2/h$ ...), however up to now
the filling factor $\nu$ is a positive real number and there is no way
to introduce a quantization of this quantity.  To introduce a
quantization of the filling factor we need to consider an open system,
i.e. a system where the number of electrons is not fixed but may vary
due to the presence of one or more reservoirs.

Before discussing this situation let us consider the case of the electron
gas confined in a potential $V(y)$ which is almost flat at the center
and bends up at the edge of the
system\footnote{In fact the edges are created by the presence of this
confining potential.}. We also assume that this potential 
is smooth on the scale of the magnetic
length. If we assume that the potential is translationally invariant
in the $x$ direction we can again separate the variables in
Schr\"odinger equation and look for a solution of the form
\be
\Psi(x,y)=\frac{e^{ik_x x}}{\sqrt{L_x}}f_{k_x}(y)
\ee
where $f_k(y)$ will not be the solution of the harmonic
oscillator. Because of the hypothesis that the potential varies
smoothly with respect to the magnetic length scale the function
$f_{k_{x}}$ is centered around $k_{x}\ell^2=Y_{k_x}$ and the energy
$E$ will be again given by the kinetic energy plus a potential term
given by $V(Y_{k_x})$. Hence the group velocities are given by
\be
{\bf v}=\frac{1}{\hbar}\frac{d E_n}{d {\bf k}}
\ee
and again the $v_y$ component will be zero. The energy will depend on
the momentum $k_x$ only in the region of high variation of the
potential $V$ hence the particle will have a non-zero $x$ velocity
only if they are close to the edges of the system. This then defines
two regions, the left and the right edge where the electrons move with
opposite velocities (recall that the energy will decrease when entering
the device, from the left, and increase when leaving the device to the
right). The width of these regions depends on the variation of the
potential: if the potential $V$ varies very sharply these regions are
very small, while if $V$ varies smoothly they are very large.

An example of this behavior is the simple case when the potential
 $V(y)$ is infinite outside the physical edges of the device and
 constant inside (it may be also zero).  Hence its gradient is
 concentrated only in two points. We are then faced with the problem of
 a particle in a high magnetic field confined in a finite region. In the
 Landau gauge this problem is mapped into the problem of a particle
 moving in a 1D well and subjected to the harmonic potential. This
 problem is exactly solvable in terms of special functions (the Kummer
 functions $M$ and $U$ ) \cite{Abramowitz1964} and the condition of
 the confinement gives a relation between the momentum $k_x$ and the
 energy $E_n(k_x)$. On the other hand when the magnetic field is very
 large we can solve the problem in an approximate way. Indeed in this
 hypothesis the magnetic length $\ell$ will be very small and when the
 electron center $k_x \ell^2$ is far away from the boundaries the
 energy is simply given by the harmonic oscillator energy and the ground state
 energy is the corresponding lowest energy level of the harmonic oscillator.
 However if the electron center $k_x \ell^2$ is exactly at one
 boundary the electron wave function must have a node in that point
 and be zero outside the system. This will imply that the ground state
 for this case is the first excited state of the harmonic
 oscillator. We then see that the energy increases continuously when
 the electron center moves from the bulk region, far from the edges,
 towards the boundary regions, close to the edges. A similar behavior
 is verified when the confining potential is smoothly varying.  What
 happens when many Landau levels are filled? We can apply the above
 idea to every Landau level separately (we suppose that the electrons
 are not interacting) and the result is simply that there will be two
 edge states for each Landau level.

Let us now consider the more realistic case of an open system,
where two reservoirs with different chemical potentials can inject
electrons in the gas. We choose to place the reservoirs along the $x$
direction. What is now the total current flowing in the system?
Because the currents at the edges flow in different directions we have
simply
\be
I=\int\limits_{\mu_L}^{\mu_R}d\mu ~e v_x g(\mu)
\ee
where
\be
g(E)=\frac{1}{L}\frac{d j(E)}{dE}=\frac{1}{2\pi \ell^2}\frac{dY}{dE}
\ee 
is the one-dimensional density of states and we have used the fact that $j$ is
related to $k_x$ by $k_x=2\pi j/L$. We use the one-dimensional density
of states because the electrons that are confinated at the edge are
moving in one dimension only.  By using the expression for the
velocity one arrives directly to
\be
I=\frac{e}{h}(\mu_L-\mu_R).
\ee    
This current is the same for every Landau level hence the total
current will be
\be
I_t=N I=\frac{eN}{h}(\mu_L-\mu_R)=\frac{e^2N}{h}(V_L-V_R)
\label{hall-current}
\ee 
where $V_i$ is the potential of the $i$-th edge ($\mu_i=e V_i$). The
integers $N$ count the number of Landau levels that are beyond the
Fermi energy (we are assuming that the chemical potentials $\mu_L$ and
$\mu_R$ are close enough to consider only the linear response
regime). Now we can calculate the conductivity matrix. When the probes
belong to the same edge the current is the same (there is not
backscattering) and one obtains a zero longitudinal resistance. When
the probes belong to different edges the current is given by
(\ref{hall-current}). We then obtain
\be
\hat\sigma=
\left(
\begin{array}{cc}
0 & -\frac{Ne^2}{h}\\
\frac{Ne^2}{h} & 0
\end{array}
\right).
\label{hallcond}
\ee
It is now possible to understand the nature of the plateaus. The
reservoirs will define an energy reference (the Fermi energy) and
moving the magnetic field will vary the frequency $\omega_c$ and the
Landau level energy. Only the Landau levels that are below the Fermi
energy can participate to the current and this defines the number
$N$. When a new Landau level sinks below (or become greater than) the
Fermi energy the number $N$ will increase (decrease) by $1$ and the
conductance can vary only by integers. On the other hand there is a
finite energy gap, given by $\hbar \omega_c$ between two different
Landau Levels.  This finite gap implies the finiteness extension of the
Hall conductance plateaus.

What remains unexplained by this model is why when a new Landau level
sinks below the Fermi energy the longitudinal conductance has a maximum
and the Hall conductance shows a linear behavior. It is mostly
believed that this behavior can be explained by inserting in the above
model the effect of the disorder. The idea is that when a new Landau
level crosses the Fermi level, a path which connects two edges may
appear and this allows for the presence of backscattering and the
longitudinal conductance is not zero. The connection between the two
edges is created by the presence of the impurities that create
localized islands and the electrons can, tunneling from one island to
the other, be backscattered to the other edge. Numerical evidences of
this percolation scheme have been provided but there is not an
accepted model which reproduces the experimental data and allows us to
understand completely this phenomenon.

\section{The Fractional Quantum Hall Effect}

The theory we have presented in the preceding section can explain the
IQHE. However in the 1982 Tsui {\it et al.} showed that in cleaner
sample and at higher magnetic field than that used by von Klitzing a
new series of plateaus in the Hall resistance appears
\cite{Tsui1982}(see Fig. \ref{qhe_transport_data.eps}).  These new
plateaus appear when all the electrons are in the lowest Landau level,
hence there is not the possibility that new edge states can be created
when varying the magnetic field. Another obscure point is that the
transversal conductance is again given by an expression similar to
Eq. (\ref{hallcond}) but with the integer number $N$ substituted by a
rational number.  This rational number is identified with the filling
fraction $\nu$. This effect is now widely known as Fractional Quantum
Hall Effect.

There is not a complete accepted theory for this effect even though many
theoretical ideas have been developed to understand its physics. A review of all
these ideas is outside the scope of this thesis. However the interested reader
 can refer to some beautiful reviews
\cite{DasSarma1996, Chakraborty1988, Girvin1990, Heinonen1998}.

The first observation we can make in dealing with this effect is that
the kinetic energy, since all the electrons reside in the LLL, is a
constant and can be neglected. Hence the electron-electron
interaction, which in the IQHE is treated as a perturbation, plays a
fundamental role. If the interaction cannot be neglected the
correlation between the particles must be taken into account in the
search for the ground state. When we consider the problem of many
electrons usually we first write the single particle wave function and then
the many-particle wave function is given by a Slater
determinant. However when the particles are strongly interacting this
approach fails and we must write a wave-function which takes into
account the correlation. On this track, Laughlin moved in the 1983
\cite{Laughlin1983}. He wrote a trial wave function of the form
\be
\psi=\prod_{i\not = j}(z_i-z_j)^m \exp\left(-\sum_i \frac{|z_i|^2}{4\ell^2}\right),
\ee      
where $m$ is an odd integer\footnote{The request that $m$ is an odd
integer is due to the anti-symmetry related to the Pauli exclusion
principle.} and $z_j=x_j+i y_j$ is a complex variable which describes
the position of the $j-th$ particle.  He proved by using numerical
calculation that this wave-function minimizes the interaction energy.
This minimization can be easily understood by observing that in the
Laughlin wave-function the multiplicity of the zeros is greater than
that of the fermion (this case is recovered by setting $m=1$).
Laughlin was also able to show that the state described by this
wave-function is incompressible, hence there is a gap in the energy spectrum.
The presence of the gap is necessary to understand the finite
extension of the plateaus as in the case of the IQHE.

Laughlin pointed out also that the quasi-particle e\-xci\-ta\-tions of this
ground state have fractional charges given by $e/m$.  The direct
observation of such fractional charges has captured the major experimental
efforts and was achie\-ved in the 1997 by several experimental groups
with different techniques \cite{Maasilta1997, Saminadayar1997,
de-Picciotto1997}.

The Laughlin's proposal can explain the presence of the plateaus at
filling fraction given by a rational number which is the inverse of an
odd number. However a more recent theory extended the model capturing
also a series of possible values of the filling fractions. The
starting point of these theories was the pioneering work of Jain
\cite{Jain1989} which pointed out that the FQHE can be explained as
the IQHE phase of the ground-state excitations. These quasi-particles,
called Composite Fermions, carry a certain number of magnetic flux
quanta per particle and see a residual magnetic field. Under certain
conditions these fermions can develop a IQHE phase. With this idea
Jain was able to predict a series of filling fraction given by
\be
\nu=\frac{p}{p m\pm 1}
\ee
where $p$ is an odd integer and $m$ is an integer. 

An earlier approach to the generalization of the allowed values of $\nu$ is the hierarchical 
approach developed by Haldane \cite{Haldane1983} and Halperin 
\cite{Halperin1984}.
The general idea of this approach goes as follows. We consider a given FQHE
state in its ground state. When we change the magnetic field we add
quasi-particles to the system.  These quasi-particles move in the
magnetic field determined by the difference of the external magnetic
field and the magnetic field carried by the particles which form the
ground state. When we increase again the magnetic field the
quasi-particles develop a new Quantum
Hall phase. The new filling fraction is now given by a complex
fraction composed by the integer numbers which identify the starting
state and the new integer which identify the new developed phase. A
complete review of these ideas can be found in Wen \cite{Wen1995}. A
major aspect of this theory is that a series of excitations starts to
develop and a given filling fraction $\nu=m/p$ might be the result of
many of these steps. This in turn implies that many modes can propagate
in the system. This can be the basis to understand some recent
experimental results \cite{Roddaro2002, Griffiths2000, Comforti2002}

A major shortcoming of this theory is that its starting point is the
phenomenological relation (\ref{phe-qhe}) (see Ref. \cite{Wen1995} and
references therein) where the presence of a developed Quantum Hall phase
has been assumed implicitely.  Up to now, at the best of our knowledge,
 no one was able to derive this
equation from the fundamental properties of the electrons in the
magnetic field confined in a finite size device.
Even in the Composite Fermions
approach, one of the most widely used theories to study the FQHE, 
it is not clear how to derive, starting from the correlated wave function, 
the presence of a gap which can justify the quantization of the Hall conductance. 

\section{Edge dynamics}
In a series of beautiful papers Wen shows that the edge dynamics can
be described by using a so called Chiral Luttinger Liquid (\chill)
\cite{Wen1990,Wen1991a,Wen1991b}. 
We want to discuss briefly this theory without entering in the
details. In the following chapters we will compare this theory with
our model.

The Wen's starting point is the phenomenological equation
(\ref{phe-qhe}).
%report 
%\be
%J^\mu=\sigma_H \epsilon^{\mu}_{\phantom{\nu}\nu\lambda} \partial^\nu
%A^\lambda.
%\ee
This equation is the requirement that a QHE is fully developed and we
are in one of the plateau of the Hall conductance. It relates the
response of the system (the current $J^\mu$) to the external
perturbation (the potential $A^\mu$) in a non-trivial way.  We can
define a Chern-Simons field $a^\mu$ by
\begin{equation}
J^\mu=\sigma_H \epsilon^{\mu}_{\phantom{\nu}\nu\lambda} \partial^\nu
a^\lambda
\ee
and obtain the Eq. (\ref{phe-qhe}) as the equation of motion of the
density lagrangian
\be
{\cal L}=\frac{\nu e^2}{2h}\epsilon^{\mu\lambda\eta}\tilde
A_\mu\partial_\lambda \tilde A_\eta.
\ee
where $\tilde A_\mu=a_\mu-A_\mu$ and $a_\mu$ is the dynamical field.

As one can easily verify this lagrangian is not invariant with respect
to a gauge transformation in a space with boundaries.  To make the
action gauge invariant Wen assumes that the lagrangian is written for
the bulk operators and does not take into account the effects of the
edges. Taking into account this contribution adds a term whose form
can be derived from the gauge invariance and can be expressed in terms
of current-current correlation functions.  Wen was also able to show
by assuming the locality of the theory that the excitations must have
a gapless and linear spectrum. In particular, if one defines the
current in the momentum space
\be
J_k^{\alpha}=\int d\sigma
\frac{1}{\sqrt{L}}e^{ik\sigma}J^{\alpha}(\sigma),
\ee
where $\alpha=(0,1)$ indicates the temporal ($0$) or spatial ($1$)
component of the current vector and $\sigma$ parameterizes the
boundary, then one gets
\be
\begin{split}
&[H,J_k^\alpha]=ck J_k^\alpha,\\ &[J_k^+,J_{k'}^+]=\frac{\nu
e^2}{h}\delta_{k+k',0},\\ &[J_k^+,J_{k'}^-]=[J_k^-,J_{k'}^-]=0,
\end{split}
\ee
where by definition $J^\pm=1/2(J^0\pm J^\sigma/c)$.  In these
equations we have assumed that $c$ is the velocity of the boundary
excitations and that they are chiral waves.  From these equations it
is also easy to show that the density fluctuations localized at the
edge follow the commutation rules
\be
[J_k^0,J_{k'}^0]=\frac{\nu e^2}{h}\delta_{k+k',0}.
\ee
Wen now shows that one can ``bosonize" the theory. The bosonization
technique is a powerful idea to solve the problem of interacting
one-dimensional electrons model or Tomonaga-Luttinger model. The main
observation is that in the low-energy regime, the densities of a one
dimensional electron system are boson operators. Hence instead of solving
the problem for many interacting electrons with fermion properties one
can solve the problem of the density modes which have bosonic properties
\cite{Haldane1981}.

We define a boson field $\phi$ with density lagrangian
\be
{\cal L}=\frac{1}{2}[(\partial_0 \phi)^2-(\partial_\sigma \phi)^2]
\ee
and we require that this field is chiral, i.e. it satisfies the
constraint
\be
(\partial_0 -\partial_\sigma)\phi=0.
\ee
The equation of motion for this field is
\be
(\partial_0 -\partial_\sigma)(\partial_0+\partial_\sigma)\phi=0
\ee
which can be solved by assuming that
\be
\phi=\phi_0+\tilde \phi_0+p_\phi(t+\sigma)+\tilde p_\phi(t-\sigma)
+\sum_{n\not =0}\frac{1}{n}\left(\alpha_n
 e^{-in(t-\sigma)}+\tilde\alpha_n e^{-in(t+\sigma)}\right)
\ee
where $\phi_0,~\tilde \phi_0,~p_\phi$, and $\tilde p_\phi$ define the
zero wavelength mode. The conjugate momentum
\be
\pi=\frac{\delta{\cal L}}{\delta\partial_0\phi}
=\partial_0\phi=p_\phi+\tilde p_\phi+\sum_{n\not =0}\left(\alpha_n
 e^{-in(t-\sigma)}+\tilde\alpha_n
 e^{-in(t+\sigma)}\right)=\phi_L+\phi_R
\ee
must satisfy the commutation rule $[\phi,\pi]=1$ hence we can derive
the commutation rules of the operators in the expansion for $\phi$
\be
\begin{split}
&[\alpha_n,\alpha_n]=[\tilde
\alpha_n,\tilde\alpha_m]=n\delta_{n,-m},\\
&[\phi_0,p_\phi]=[\tilde\phi_0,\tilde{p}_\phi]=i,\\ &\mbox{others}=0.
\end{split}
\label{boson-commutators}
\ee
It is possible to use the chirality condition to eliminate the field
$\tilde \phi$, $\tilde p_\phi$ and $\tilde \alpha_n$. The hamiltonian
is diagonalized
\be
H=\frac{1}{2}p_\phi^2+\sum_{n>0}\alpha_n^\dagger\alpha_n
\ee
where we have neglected the constant terms.

By defining
\be
J^\alpha=\frac{\sqrt{\nu}}{2\pi}\epsilon^{\alpha\beta}\partial_{\beta}\phi_L,
\ee
where $\phi_L$ denotes the modes which satisfy the chirality
condition, we can verify, starting from the boson algebra, that these
operators satisfy the algebra of the edge excitation. Hence we have
obtained a boson representation for the density fluctuations
operators.  To fully define all the algebra we need also the operators
for the quasi-particles.  Indeed the density $J^\alpha$ describes the
edge density fluctuation with respect to an equilibrium density
$\rho_0$. We look for an operator that can change such equilibrium
density. To obtain a form for this operator we define the equilibrium
charge
\be
Q=e\int d\sigma J^0(\sigma)
\ee
and the quasi-particle operator
\be
[\Psi,Q]=e^*\Psi
\label{psi-def}
\ee
where $e^*$ can be different fron the electron charge.
In fact in a physical system the quantum charge is the
electron charge.  However, here, we are postulating that the
quasi-particle charge may not be the electron charge. We have seen in
the preceding section that the ``charge fractionalization" is useful
to understand the phenomenology of the FQHE.

In the chiral boson theory the charged operators have a form
\be
\Psi=:e^{i \gamma\phi_L}:
\ee
where the symbol $:\ldots :$ denotes the normal ordering and $\gamma$
is a constant. This constant can be determined by the commutation
relation (\ref{psi-def}) by using the boson commutators
(\ref{boson-commutators}) obtaining
\be
\gamma=\frac{e^*}{e}\frac{1}{\sqrt{\nu}}.
\ee
Require that the operator $\Psi^\dagger$ creates a unitary charge
corresponds to choose $\gamma\sqrt{\nu}=1$.  The other requirememt that
one can make is that the operator $\Psi$ satisfies the usual fermion
(anti-)commutation relations. By using the Haussdorf lemma
\be
e^A e^B=e^{[A,B]}e^B e^A
\ee
we obtain $\gamma^2=1/\nu={\rm odd~integer}$. Notice that this last
relation says to us that if the condition $\nu=1/{\rm odd~integer}$ is
not fulfilled hence the fermion creation operator must be composed by
many different branches.

One can now calculate the fermion correlation function $G$ defined as
\be
G=\langle \Psi^\dagger(t,\sigma)\Psi(0,0)\rangle.
\ee
By using again the Haussdorf lemma we obtain, in the limit of large
system size, $\sigma/L\ll 1$
\be
G=\left(\frac{i}{t+\sigma}\right)^\frac{1}{\nu}.
\ee

\section{Transport through a constriction}

It is possible to use the formulation for the edge dynamics we have
depicted in the preceding sections to understand the effect of
tunneling between the edges \cite{Wen1991b}. Such a phenomenon is
present when two edges are close enough to have a non zero
superposition of the quasi-particle wavefunction. The inclusion of
tunneling in the edge model is done ``by hand" in the sense that there
is not a completely accepted model to derive the tunneling amplitude
$\Gamma$ from the electron or quasi-particle properties. In the
hamiltonian for the edges we then add a term proportional to the
product of two distinct quasi-particle operators
\be
H_T=\Gamma \Psi_L^\dagger \Psi_R+\Gamma^* \Psi_R^\dagger \Psi_L.
\ee 
In the framework of the linear response theory one can calculate the
tunneling current and the tunneling conductance which changes the Hall
conductance in a non-universal way. The result of this approach is a
non linear behavior of the current as a function of the tunneling
voltage, the bias difference between the edges, i.e. the Hall voltage,
and of the temperature \cite{Wen1991b}. In particular Wen obtained
that
\be
I\propto V^{\alpha-1}
\ee
where $\alpha=2/\nu$. A similar relation holds for the temperature
dependence of the tunneling current
\be
I\propto T^{\alpha-1}.
\ee
These relations stem directly from the power-law behavior of the
correlation function $G$ of the quasi-particle.

The non-linear relation between the current and tunneling voltage are
expected also for the Luttinger liquid model \cite{Haldane1981}.
However the intrinsic difficulties in obtaining a very clean
one-dimensional electron liquid reduce the possibility of the
experimental observation of these results. In this sense the Wen's
papers open the possibility of a verification of these result on the
edge of a quantum Hall liquid which, as we have briefly discussed, forms
a very clean one-dimensional Luttinger liquid.

Similar results were extended later by Kane and Fisher, by using a
Renormalization Group (RG) approach, to the effect of the impurity
which mediates the tunneling \cite{Kane1995}. In particular, they
showed that any impurity between two edges under the RG analysis
implies a strong coupling i.e.  the edges will eventually close, the tunneling
current diverges and the Hall bar separates in two distinct regions
(this phenomenon is known as ``overlap catastrophe").

In this thesis we will generalize this approach in many ways. The
first point we want to make is the generalization of the approach to
the edge dynamics eliminating the limitation of incompressibility of
the Hall liquid in the bulk. This is done by a hydrodynamical approach
to the density fluctuation in the QHE and by the projection in the
Lowest Landau Level.  The projection on the LLL simplifies the
Hamiltonian for the system quenching the kinetic energy but, on the
other hand, introduces a non trivial quantum commutator between the
density fluctuations and restores the full hardness of the problem. The
hydrodynamical approach eliminates the fluctuations present at the
scale of the magnetic length. This allows us to fully describe the
system by using only the density fluctuations neglecting the real
electrons and allows us to consider only the low energy excitations of
the system. Similar approaches were used to study the tunneling 
of electron from a Fermi liquid to the edge of a fractional Quantum Hall liquid \cite{Conti1998,Conti1998a}.

On the other hand there are not conceptual difficulties in inserting
in our model the intra- and inter-edge interactions. In this way we
arrive to a problem similar to a Luttinger liquid model where an edge
maps to a chiral branch of the liquid. However our edges are still
interacting while in the Luttinger liquid model the reduction to
chiral waves is done by eliminating the interactions.

The scheme we have depicted before and that we will discuss in the
following brings us to the Hamiltonian for the density
fluctuations. We bosonize these fields and obtain an equation of
motion for the amplitude of the fluctuations. We choose a situable
form of the intra- and inter-edge interaction and solve the equation
of motion in many different edges profiles. In particular we are
interested to the case when a constriction, which brings the
edges to stay close, is present. It is possible to solve this case by using a
scattering approach, i.e by defining the transmission and reflection
coefficients for the wave which impinges on the constriction. We will
show that the constriction does not change the low frequency (long
wavelength) behavior of the conductance thus recovering the linear relation
between the Hall conductance and the filling factor\footnote{Notice
that in our model the filling factor is not quantized at all, hence we
are unable to recover the quantized Hall conductance.}. To fully
understand some new experimental results \cite{Roddaro2002} we will
introduce the tunneling between the edges. We restrict the tunneling
only to the region of the constriction, hence we use the quasi-particle
operators that are given by the solution of the equation of motion
when the constriction is present. We develop a perturbative approach
starting from the equation of motion for our operators and use it to
calculate the tunneling conductance. This quantity, which modifies the
Hall conductance, is a non linear function of the tunneling voltage
and of the temperature as shown by Wen\cite{Wen1991b}.  However the
constriction introduces an energy scale. The characteristic time $t_0$
is given by the time that an edge excitation needs to travel through
the constriction. This time introduces the energy scale $h/t_0$. We
will show that the response functions of the system have different
behaviors if the energy is above or below this energy threshold and
this will affect the tunneling resistance.

Recent experimental results, reported by Roddaro {\it et al.}
\cite{Roddaro2002}, show that the tunneling conductance for low bias
voltage has a different behavior when compared with the Wen's results.
In particular they report that in their experimental setup at
relatively high temperature the conductance shows a large maximum for
zero bias voltage and two deep minima when the voltage is
increased. When the temperature is lowered two main aspects
appear. The zero bias maximum initially start to increase as expected
but beyond a temperature near about $400~ \rm{mK}$ the conductance
starts to develop a large zero bias minimum. Beyond this temperature
also a strong asymmetry starts to appear. From the discussion with the
experimentalist we known that the appearance of the central deep in
the conductance is not related to the presence of localized
impurities.

Our model seems to capture some of the aspects of these results. In
particular the presence of the deep minima seems to be related to the
presence of the constriction and the comparison with the experiments
can give information on $t_0$ and $e^*$. We predict also the presence
of small oscillations on the tunneling resistance at large bias.  The
frequency of these oscillations is given by
\be
\Delta V_T=\frac{h}{e^* t_0}=\frac{h c_1\cosh 2\theta_1}{e^* d\cosh 2\theta_2}
\ee
where $c_1$ is the velocity of the modes out of the constriction and
$\theta_1$ and $\theta_2$ are related to the intra- and inter- edge
interactions inside and outside the constriction.

The thesis is organized as follows.  In this first chapter we have
discussed the general phenomenology of the QHE and the Luttinger
Liquid model. To do that we moved from the Classical Hall effect,
given a look to the main experimental results of von Klitzing
\cite{Klitzing1980} and Tsui, Stormer, and Gossard \cite{Tsui1982} and
discussed the theoretical result for the IQHE and the FQHE. We have
also briefly reviewed the Wen's theory for the edge dynamics and then
discussed our main results.

In the second chapter we will concentrate on the derivation of our
model and of the equation of motions. We will discuss its properties
in the Hilbert space and solve the equation of motion in some simple
cases that will prove to be useful in the following.

In the third chapter we will discuss the problem of defining and
calculating the transport properties in these devices. We give a
definition of the conductance and show how we recover, in the simple
cases studied in the second chapter, the ideal Quantum Hall
conductance.

In the fourth chapter we will abandon the limitations of a
translational invariant system and discuss the presence of the
tunneling between the edges.  In this case we show that the tunneling
gives a non-universal correction to the Quantum Hall conductance. We
also compare our model with those present in the literature.

Finally, in the last chapter, we will discuss the comparison
with some experimental results and the future perspectives of
this research, and sumarize our main conclusions.

The results presented in this thesis have been submitted for pubblication 
in the Physical Review B journal. A preprint of this paper is available \cite{DAgosta2003}.

\chapter{The model for the Quantum Hall bar}
We are interested in the study of a two-dimensional electron gas
confined in a finite region in the presence of a high magnetic
field.  We will show that
the description of the electron gas in terms of two interacting chiral
Luttinger Liquid follows from some semi-classical assumptions we will
discuss in the following. We want to point out also that we do not 
require that the system is in a Quantum Hall phase.

The ground state of the electron gas is characterized by a density
$\rho_0({\bf r})$ which takes into account the electron-electron
interaction and the action of the external potential which confines
the system. We consider the excitations of this system as density
fluctuations $\dr({\bf r})$ near the ``equilibrium" density
$\rho_0({\bf r})$.  The dynamics of these fluctuations is determined by
the Hamiltonian
\be
H=\frac12 \int d{\bf r}d{\bf r'}~ \dr({\bf r})V({\bf r-r'})\dr({\bf
r'})
\label{hamiltonian}
\ee
and by the commutation relations for the fields $\dr$. In considering
the Hamiltonian (\ref{hamiltonian}) we have projected all the operators
in the lowest Landau level.  Such a projection quenches the kinetic
energy (it becomes simply a constant) but, as we will discuss later,
introduces non-trivial commutation relations for the field $\dr$.

In the definition of the Hamiltonian (see eq. (\ref{hamiltonian})) we
have inserted an interaction potential $V({\bf r-r'})$ between two
density fluctuations at different points.  This electron-electron
potential has a coulomb origin. In the following, however, we will
assume a simple form for this potential.
% and in the conclusions we will
%discuss the possibility to generalize our model to a more realistic
%potential.

\section{The Luttinger Liquid model}
In this section we will briefly review the solution of the Luttinger
Liquid model because we will use a similar approach to the solution of
our model and this analysis will be useful when we will compare our
model with others present in the literature.  We do not derive the
Luttinger model starting from the physical properties of the
interacting electrons in one dimension. The interested reader can
refer to the seminal paper of Haldane \cite{Haldane1981} or to the
book of Mahan \cite{Mahan1981}.

The Luttinger Liquid model describes the dynamics of a one dimensional
electron liquid. The Fermi surface of such a system is composed by 
two points $\pm k_F$ then with the same energy there are two species
of excitations, the left and right movers defined by the sign of their
momentum. We make the association $Left=L=+1$ and $Right=R=-1$. In the
following we will consider also a spinorial notation where the upper
component will be always related to the left movers.

When the electrons are not interacting the Hamiltonian operator is (we
use $\hbar=c=1$)
\be
H_0=v_F\int dx (\rho_R^2+\rho_L^2)
\ee 
where $v_F$ is the Fermi velocity and $\rhor$ ($\rhol$) is the density
operator of the right (left) movers \cite{Haldane1981, Mahan1981} 
\footnote{This Hamiltonian differs by a constant from the usual definition
one can find in the references.}.
The density operators follow the
commutation relations
\be
[\rhoa(x),\rhob(x')]=-i\sigmaz \partial_x \delta(x-x'),
\ee
where $\alpha$ and $\beta$ assume the values $R$ or $L$.
These relations can be derived by starting from the definition of the
density operators in terms of the electron creation and annihilation
operators and their anti-commutation rules. 

It is customary to define
the operators
\be
\partial_x \thetaa\equiv -\alpha \rhoa.
\ee
These new operators
will be useful in the following, when we will introduce the interaction, to
diagonalize the total Hamiltonian. It easy to show that $\rhoa$ and
$\thetaa$ are conjugate field
\be
[\rhoa(x),\thetab(x')]=-i\delta_{\alpha,\beta}\delta(x-x').
\ee

When we consider the interaction we add to the free Hamiltonian $H_0$
the term
\be
\begin{split}
H_i=&V_1\int dx (\rhor^2+\rhol^2) +V_2 \int dx
(\rhor\rhol+\rhol\rhor)\\ =&V_1\int dx
((\partial_x\thetar)^2+(\partial_x\thetal)^2) +V_2 \int dx
(\partial_x\thetar\partial_x\thetal+\partial_x
\thetal\partial_x\thetar)
\end{split}
\ee
where $V_1$ and $V_2$ are the limit of vanishing momenta of the
Coulomb interaction between the right and left movers. 
$V_1$ is the interaction between electrons that belong to 
the same species, while
$V_2$ is the inter-species interaction. Notice
that the interaction preserves the momentum.

The first step towards the solution of this problem is the
diagonalization of the total Hamiltonian $H=H_0+H_i$. We can
accomplish this task with the linear transformation
\be
\begin{split}
&\thetan=\thetar+\thetal,\\ &\thetaj=\thetar-\thetal,
\end{split}
\ee
and obtain the new Hamiltonian as
\be
H=\frac{v}{2}\int dx \left[g(\partial_x
\thetan)^2+\frac{1}{g}(\partial_x\thetaj)^2\right]
\ee
where we have defined
\be
\begin{split}
v&=v_F
\sqrt{\left(1+\frac{V_1}{v_F}\right)^2-\left(\frac{V_2}{v_F}\right)^2},\\
g&=\sqrt{\frac{v_F+V_1+V_2}{v_F+V_1-V_2}}.
\end{split}
\ee
The Hamiltonian is separated but the new field $\thetan$ and $\thetaj$
are conjugate fields, indeed their commutation relations are
\begin{equation}
[\partial_x
\thetan(x),\thetaj(x')]=[\partial_x\thetaj(x),\thetan(x')]=
-2i\delta(x-x').
\ee
Notice also that these fields are not chiral modes. The decomposition of
the problem in terms of non-interacting chiral modes is not yet complete.

To obtain a set of independent fields we consider the new
transformation on the fields $\thetan$ and $\thetaj$
\be
\begin{split}
\phi_R=g\thetan+\thetaj,\\
\phi_L=g\thetan-\thetaj.
\end{split}
\ee
The Hamiltonian in terms of these fields is now
\be
H=\frac{v}{8g}\int dx \left[(\partial_x
\phi_R)^2+(\partial_x\phi_L)^2\right]
\label{diagonalhamiltonian}
\ee
and we have
\be
\left[\partial_x\phi_\alpha(x),\phi_\beta(x')\right]=-4ig 
\sigmaz \delta(x-x').
\label{diagonalcommutationrelation}
\ee
The equation of motion for the fields $\phi$ are
\be
\begin{split}
&\dot\phi_R=v\partial_x\phi_R\\ &\dot\phi_L=-v\partial_x\phi_L
\end{split}
\ee
hence these modes are chiral.

We finally can express the density operator in terms of these chiral
modes
\be
\begin{split}
\rhor(x)=\frac{g+1}{4g}\partial_x\phi_R-\frac{g-1}{4g}\partial_x\phi_L,\\
\rhol(x)=\frac{g-1}{4g}\partial_x\phi_R-\frac{g+1}{4g}\partial_x\phi_L,
\label{rhochiral}
\end{split}
\ee
and obtain the total density operator $\rho=\rhor+\rhol$ as
\be
\rho=\frac12 \partial_x (\phi_R-\phi_L)=\frac{1}{2v}\partial_t(\phi_R+\phi_L).
\ee
The total current is defined by using the continuity equation
$\dot\rho=-\partial_x I$ and then
\be
I=-\frac{v}{2}\partial_x(\phi_R+\phi_L)=-\frac12\partial_t(\phi_R-\phi_L).
\ee

When we consider the chiral Luttinger liquid ($\chi$LL) we fix one of
the fields $\phi$ to zero and consider the dynamics of the other field
given by the Hamiltonian (\ref{diagonalhamiltonian}) and the
commutation relation (\ref{diagonalcommutationrelation}). Notice that
we have started with two interacting chiral modes and after a
non-canonical transformation we have obtained two non-interacting
chiral modes with non-trivial commutation relation. As it is seen from
Eq. (\ref{rhochiral}), the chiral modes $\phi$ are a linear
combination of the starting interacting chiral modes $\theta_R$ and
$\theta_L$.

\section{Lowest Landau level projection and the hydrodynamical approximation}
The Hamiltonian ($\ref{hamiltonian}$) alone cannot determine the
dynamics of the fields $\dr$. To do that we need to give either the
equation of motion or the commutation rules for these fields. To
calculate the commutation relations we project all our field on to the
lowest Landau Level. The theoretical approach to this projection and
its physical implications are discussed in a more detailed way in many
reviews or articles (see as an example Ref. \cite{Girvin1986}). In the
following we will use it in fixing the wave functions of the single
electron.

In order to emphasize the role of the two main physical
approximations, i.e., the LLL projection and the hydrodynamical
approximation, it is convenient to start from the second quantized
form of the density fluctuation
\be \delta\hat\rho({\bf r})=\sum_{k\not =h}\hat c_k^\dagger \hat c_h
\varphi_k^* ({\bf r})\varphi_h({\bf r})\label{a1.1} \ee
from which we have
\be
\begin{split}
[\delta\hat \rho({\bf r}),\delta\hat \rho({\bf
r}')]=&\sum_{k\not=h,m\not=l}\left(\hat c^\dagger_k
\hat c_l \delta_{h,m}-\hat c^\dagger_h \hat c_m
\delta_{k,l}\right)\\
&\times\varphi_k^*({\bf r})\varphi_h({\bf r})\varphi_m^* ({\bf
r}')\varphi_l({\bf r}').
\label{commutation-1}
\end{split}
\ee 
The indices $k$ and $h$ in Eq.(\ref{a1.1}) label states in the LLL
within the Landau gauge ${\bf A}=(By,0,0)$. The condition $k\neq h$
excludes the equilibrium contribution to the density $\rho_0 ({\bf
r})$.  In the spirit of the hydrodynamic approach, the latter is
related to the ground state expectation value \be \hat c^\dagger_k
\hat c_l \to
\langle \hat c^\dagger_k \hat c_k\rangle \delta_{k,l}=n(k)\delta_{k,l}
\label{a1.3}
\end{equation}
via
\be
\rho_0(y)=\int\limits_{-\infty}^\infty
\frac{dk}{2\pi}~\frac{n(k)}{\sqrt{\pi
l^2}}\exp\left[-\frac{(y+kl^2)^2}{l^2}\right].
\label{a1.4}\ee 
Here $n(k)$ is the occupation number for the state with momentum $k$
and in the homogeneous case, $n(k)=n_0$, the evaluation of the gaussian
integral gives
\be
\rho_0=\frac{n_0}{2\pi l^2} \Rightarrow n_0=2\pi l^2\rho_0=\nu 
\label{a1.5}
\ee
where $\nu$ is the filling factor.

The commutator between two different density fluctuations can be
derived from the well known result \cite{Girvin1986}
\be
[\rho({\bf q}),\rho({\bf k})]=\left(e^{k^*q\ell^2 /2}-e^{-kq^*\ell^2
/2}
\right)\rho(\bf{k+q}).
\ee
When we consider the long wavelength density fluctuation we expand to
the leading order in $k\ell$ and $q\ell$ and transform to real space to
obtain
\be
[\rho({\bf r}),\rho({\bf r'})]\simeq i\ell^2
\epsilon_{ij}\partial_i\rho_0({\bf r})\partial_j\delta({\bf r-r'}).
\label{commutation1}
\ee
Notice that we have substituted the density operator $\rho$ with its
equilibrium expectation value $\rho_0({\bf r})$. This approximation is
justified as long as we are interested to the linear dynamics of small
fluctuations about the equilibrium state.

From these commutation relations and from the Hamiltonian
(\ref{hamiltonian}) we can derive the equation of motion of the
density fluctuation
\be
\partial_t\delta\rho({\bf r},t)=\ell^2 (\partial_i
\rho_0({\bf r}))\epsilon_{ij}\partial_i \int d{\bf r'} V({\bf r},{\bf r'})
\delta\rho({\bf r'},t).
\label{generalmotion}
\ee
By this equation it is evident that the density fluctuation are
localized in the region of variation of the equilibrium density
$\rho_0$. The regions of maximum variation of $\rho_0$ are located
near the edges hence the density fluctuation are confined near the
edges of the Hall bar. Because this model for the edge dynamics agrees
with that we find considering the large magnetic field limit of the
hydrodynamical Euler equations \cite{Aleiner1994} we call our approach
``hydrodynamical".

Attached to a point of every edge we consider a local right handed
system of coordinates where the variable $y$ indicates the region of
variation of the equilibrium density. To be more precise we consider
the device reported in the Fig. \ref{trans-probe.eps}.
\begin{figure}[!ht]
\begin{center}
\includegraphics[clip,width=7cm]{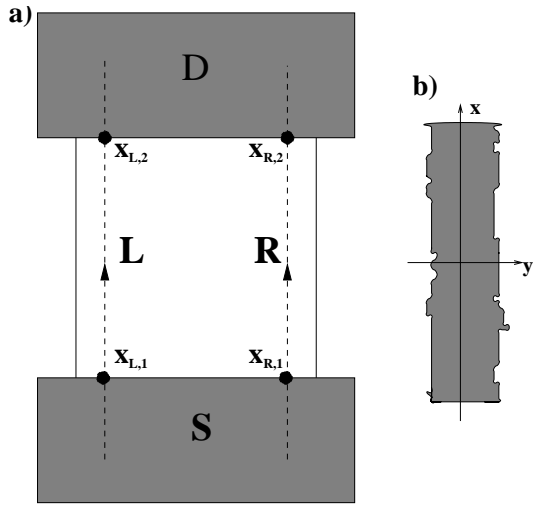}
\caption{{\bf a)} The scheme of a simple Quantum Hall bar contacted with two 
reservoirs. If we neglect the effect of the disorder, the edges runs
parallels and translationally invariant along the $x$ direction.  The
$y$ axis is chosen to be orthogonal to the edges. {\bf b)} A magnification
of the edge region. We integrate with respect the $y$ direction to
obtain a field varying only in the $x$ direction.}
\label{trans-probe.eps}
\end{center}
\end{figure}
In this figure we draw a simple schematization of a Quantum Hall bar connected
to two reservoirs (the drain (D) and the source (S)) bar where the
physical edges are considered as parallels and translationally
invariants.  The coordinate $y$ is then orthogonal to the edge while
the $x$ coordinate run along the edge.

The observation about the localization of the density fluctuation
allows us to greatly simplify the problem. Indeed rather than solve
the complete two-dimensional problem determined by the equation of
motion (\ref{generalmotion}), which implies the exact knowledge of the
equilibrium density $\rho_0$ \cite{Aleiner1994}, we integrate over the
$y$ direction, starting from the edge to well inside the bulk when we
consider the fluctuation localized on the left edge and viceversa for
the right edge, and obtain a one dimensional density fluctuation. In
this way we define the operators
\be
\dra(x)=\int\limits_\alpha dy~\dr({\bf r})
\label{dra}
\ee
where the index $\alpha$ identifies the edge to which the density
fluctuation belongs. These new operators must satisfy the commutator
law (\ref{commutation1}) when we integrate with respect to the $y$ and
$y'$ variables, following the prescription in (\ref{dra}). It is
possible to show that the result of this operation is
\be
[\dra(x),\drb(x')]=-i \frac{\nu(x)}{2\pi}\sigmaz \partial_x
 \delta(x-x').
\label{edgecommutationrelations}
\ee 
In the appendix \ref{commutationrel} we give an alternative derivation
of this equation starting directly from the operators $\dra$ in the
case of sharp and smooth edges.  We have also allowed for the
possibility that the filling fraction $\nu$ is dependent on the
position $x$ along the edge\footnote{This can account, as an example,
for the situation where two Hall liquids with different filling
factors, are connected.}. In the following, for simplicity, we will
drop this degree of freedom and consider a uniform filling
fraction. 

The Hamiltonian for the operators $\dra$ is derived from the
Hamiltonian (\ref{hamiltonian}) by an integration. To do that we
suppose that the interaction potential is slowly varying on the
dimension of the edge i.e. we approximate
\be 
V({\bf r-r'})\simeq V_{\alpha,\beta}(x,x')
\ee
with
\be
V_{\alpha,\beta}(x,x')=V(x,y_\alpha ;x',y_\beta)
\ee
where we have allowed for the possibility that the two density
fluctuations belong to different edges. With this approximation we get
(repeated indexes are summed over)
\be
H=\frac12 \int~ dx dx'~ \dra(x)V_{\alpha,\beta}(x,x')\drb(x').
\label{edgehamiltonian}
\ee
This Hamiltonian together with the commutation relations
(\ref{edgecommutationrelations}) determines the equation of motion for
the $\dra$ operator
\be
\partial_t \dra(x)=-\frac{\nu(x)\sigmaz}{2\pi}\int\limits_{-\infty}^\infty
dx' \partial_x V_{\beta,\gamma}(x,x')\delta\rho_\gamma(x').
\label{drequationofmotion}
\ee
This is our first major result. The solution of this equation
describes the excitations of our system and will be the fundamental tool
to obtain the response functions.

The commutation relations (\ref{edgecommutationrelations}) when $\nu$
is constant are the reproduction of the Kac-Moody current algebra
(\ref{diagonalcommutationrelation}) for the density field in the
\chilll model. Notice however that our derivation of the \chilll model
has nothing to do with the QHE. Indeed it depends only on the high
magnetic field limit (projection on the LLL) and on the
coarse-graining (hydrodynamical approximation). Our derivation is
therefore valid for every real value of the filling fraction whereas
the QHE occurs only for certain value of $\nu$.  We want to point out
also that in our model every edge exhibits a \chilll behavior even if
the interactions are turned off, and that it is
not possible to map the problem of two interacting QH edges to one
Luttinger Liquid model.    

\section{The equation of motion}
Let us now discuss the solutions of the equation of motion 
(\ref{drequationofmotion}). To do that it is convenient to define the
field $\phia$ such that
\be
\dra(x,t)=\partial_x \phia(x,t).
\ee
These fields satisfy the commutation relations
\be
[\phia(x),\phib(x')]=i\frac{\nu}{4\pi}\sigmaz \sign(x-x').
\ee
In the following, we will consider also the function $\varphi_\alpha(x)$ which
are the solutions of the equation
\be
i\omega \sigmaz \partial_x \varphi_\beta(x)=\frac{\nu}{2\pi}
\int\limits_{-\infty}^\infty dx'~\partial_x 
V_{\alpha,\beta}(x,x')\partial_{x'}\varphi_\beta(x')
\label{varphiequationofmotion}
\ee
and we will show that the field $\phia$ can be expressed in terms of the functions $\varphi_\alpha$ and the operators $b$ which turn out to have a Bose 
statistics.

To complete this task, we need to study the properties of the solutions
of the equation (\ref{varphiequationofmotion}). The first point we want 
to make is that these solutions form a complete and orthonormal basis of the 
Hilbert space. More precisely we will show that:
\begin{itemize}
\item all the frequencies $\omega_n$ are real and we label these states such that
$\omega_n>0$ if and only if $n>0$;
\item the eigenfunctions $\varphi_{n,\alpha}$ form a complete basis in the 
Hilbert space with the completeness relation
\be
\sum_n \sign(\omega_n)\varphi_{n,\alpha}(x)\varphi^*_{n,\beta}(x')\partial_{x'_\beta}
=\sigmaz \delta(x-x')
\ee
and the orthonormality condition
\be
\sum_\alpha \sigmaz \int dx~ 
\varphi^*_{n,\alpha}(x)\partial_x\varphi_{m,\beta}(x)=\sign(\omega_n) 
\delta_{n,m};
\ee
\item the eigenfunction $\varphi_{n,\alpha}(x)$ is doubly degenerate.
\end{itemize}
The proof of these properties is detailed in the appendix \ref{eigenvalueproperties}. 

Having defined a basis for the Hilbert space we can develop the field $\phia$ and
 $\dra$ in this basis. Indeed for every (spinorial) function $f_\alpha (x)$ we have
\begin{equation}
\begin{split}
f_\alpha (x)=&\sum_\beta\int dx'~ \delta_{\beta,\alpha} \delta(x-x') f_\gamma (x')\\
=&\sum_{n} \varphi_{n,\alpha}(x)f(n)
\end{split}
\end{equation}
where we have defined
\be
f(n)=-i\sign(\omega_n) \int dx'~\varphi_{n,\alpha}(x')\sigmaz \partial_{x'}f_\beta (x').
\ee
This result implies that we can define the operator $\phi(n)$ as
\be
\dra(x)=\sum_n \phi(n)\varphi_{n,\alpha} (x),
\ee	
where $\phi(n)$ turns out to satisfy the commutation relation
\be
[\phi(n),\phi^\dagger(n')]=\frac{\nu}{2\pi} \sign(\omega_n) \delta_{n,n'},
\ee
and this suggests to define the operators $b$ such that
\be
\phi(n)=\sqrt{\frac{\nu}{2\pi}}\left(b_n \theta(n)+b_{-n}^\dagger\theta(-n)\right).
\ee
It is easy to show that we have $[b_n^\dagger,b_n]=1$ thus these operators
follow the Bose statistics. 
We have then proved that we can expand the $\dra$ operator as
\be
\dra(x)=\sqrt{\frac{\nu}{2\pi}}\sum_{n>0} \left(b_n\partial_x \varphi_{n,\alpha}(x)
+b_n^\dagger \partial_x \varphi^\dagger_{n,\alpha}(x)\right)
\ee
and we obtain for the Hamiltonian
\be
H=\sum_{n>0}\omega_n\left(b^\dagger_n b_n+\frac12\right).
\label{diagonalbhamiltonian}
\ee

We define the current by starting from the continuity equation. The current in
the edge $\alpha$ is, by definition, given by
\be
\partial_x I_\alpha(x)=e\partial_t{\dra}(x)
\ee
and, up to a constant, we identify
\be
I_\alpha(x)=e\partial_t \phi_\alpha(x).
\ee
We have then related all the quantities we want to calculate to the solutions
of the equation of motion (\ref{varphiequationofmotion}) and to the 
dynamics of the boson operators which is determined by the Hamiltonian
(\ref{diagonalbhamiltonian}). 

\section{Solutions of the equation of motion in simple cases}
It is now instructive to solve the equation of motion for the function $\varphi_\beta$ 
is some simple cases, namely the case of two translationally invariant edges and 
when  a constriction is present.

\subsection{Translationally invariant case}
The translationally invariant case is the simplest we can consider and we have for the potential the form
\be
V_{\alpha,\beta}(x,x')=V_{\alpha,\beta}(x-x')=
\left(
\begin{array}{cc}
V_1(x-x') & V_2(x-x')\\
V_2(x-x') & V_1(x-x')
\end{array}
\right)
\ee
where $V_1$ ($V_2$) is the intra-(inter-) edge interaction\footnote{Recall that
the upper component corresponds to the left edge while the lower component to the
right edge.}.
We seek for the solution of the equation (\ref{varphiequationofmotion}) in terms
of plane waves, so we choose the form $\varphi_\alpha(x)=\varphi_\alpha(k)e^{ikx}$.
This leads us to the $2\times2$ eigenvalue problem
\be
i\omega
\left(
\begin{array}{cc}
1 & 0\\
0 & -1
\end{array}
\right) 
\left(
\begin{array}{c}
\varphi_L\\
\varphi_R
\end{array}
\right)
=\frac{ik\nu}{2\pi}
\left(
\begin{array}{cc}
V_1(k) & V_2(k)\\
V_2(k) & V_1(k)
\end{array}
\right)
\left(
\begin{array}{c}
\varphi_L\\
\varphi_R
\end{array}
\right),
\ee
where $V_1(k)$ and $V_2(k)$ are the Fourier transform of $V_1(x)$ and $V_2(x)$ 
respectively.
The eigenvalues are given by
\be
\omega_k=\pm|k|\frac{\nu}{2\pi}\sqrt{V_1^2(k)-V_2^2(k)}=\pm c |k|
\ee
where $c$ is the sound velocity.
Notice that for each positive frequency there are two counterpropagating modes:
the ``up-moving" solution is
\be
\varphi^u_{k,\alpha}(x)=\frac{e^{ikx}}{\sqrt{kL}}
\left(
\begin{array}{c}
u_k\\
-v_k
\end{array}
\right)
\label{upmoving}
\ee
and the ``down-moving" solution is
 \be
\varphi^d_{k,\alpha}(x)=\frac{e^{-ikx}}{\sqrt{kL}}
\left(
\begin{array}{c}
v_k\\
-u_k
\end{array}
\right).
\label{downmoving}
\ee
In these expression we have assumed $k>0$ and we have introduced the
edge length $L$. This length is assumed to be arbitrary large (as usual when 
dealing with plane-wave) and will not enter the physical results. On the other hand
the presence of the factor $1\sqrt{k}$ is imposed by the orthonormality condition.
The orthonormality condition also fixes $u_k^2-v_k^2=1$ hence we can parameterize 
\be
\begin{split}
u_k=\cosh(\theta_k),\\
v_k=\sinh(\theta_k),
\end{split}
\label{amplitude}
\ee
where the ``mixing angle" $\theta_k$ is given by
\be
\tanh 2\theta_k=\frac{V_2(k)}{V_1(k)}.
\ee

Let us discuss briefly these results. If we have two parallel non-interacting edges
we have $\theta_k\equiv 0$ for every $k$ thus $u_k\equiv 1$ and $v_k\equiv 0$.
This implies that the ``up-moving" modes are fully concentrated on the left 
edge while the ``down-moving" modes are concentrated only on the right. In this 
sense the left and right modes are well defined concepts and a good basis to
discuss the properties of the system. Moreover
we have a linear energy spectrum of these excitations with the 
edge velocity determined by the intra-edge interaction. 

When the edges are
interacting the modes propagates both on the left and on the right edge. The
concept of left edge mode is ill-defined and the good basis in this case is
the up-moving and down-moving modes. We will show, when discussing the 
transport, that the presence of the inter-edge interaction will not change
the linear relation between the conductance and the filling fraction $\nu$ 
when one consider the limit of low-energy (which corresponds to the limit
$k\to 0$) excitation.  We want also point out that we have recovered the
standard expression for the dispersion of the edge waves in the ordinary (non-chiral)
Luttinger liquid model and that the $\chi$LL persists even if the interaction
potential $V_2$ is turned off. This is due to the anomalous commutation
relation (\ref{edgecommutationrelations}) between the density fluctuations 
on the same edge.

\subsection{A conservation law}
At this point it is interesting to discuss the existence of a conserved quantity 
for the equation of motion (\ref{varphiequationofmotion}). We assume for
the interaction potential the form
 $V_{\alpha,\beta}(x-x')=V_{\alpha,\beta}(x)\delta(x-x')$. It is now possible to
show that the quantity
\be
\phia^\dagger(x)\sigmaz \phib(x)=\phi_L^\dagger(x)\phi_L(x)- \phi_R^\dagger(x)\phi_R(x)
\ee
is conserved
\be
\partial_x \left(\phia^\dagger(x)\sigmaz \phib(x)\right)=0.
\label{conserved1}
\ee
The proof of the existence of this conservation law rests on the assumption
of the existence of the $V_{\alpha,\beta}$ matrix inverse\footnote{For clear physical assumption the matrix $V_{\alpha,\beta}$ must be symmetric.} for every value of
$x$. Within this assumption, we can consider the equation of motions
for the field $\phia$ and its complex conjugate
\be
\begin{split}
i\omega \phia(x)&=\frac{\nu}{2\pi}\sigmaz V_{\beta,\gamma}(x)\partial_x \phi_\gamma(x),\\
-i\omega \phia^\dagger(x)&=\frac{\nu}{2\pi} \partial_x \phi_\gamma^\dagger(x)V_{\gamma,\beta}(x)\sigma^z_{\beta,\alpha}.
\end{split}
\ee
By taking the matrix $V_{\alpha,\beta}$ to the left-hand side of these equations
and then multiplying the first (second) equation on the left (right) by $\phib^\dagger\sigma^z_{\beta,\alpha}$ ($\sigmaz\phia$) and summing the results
we obtain the conservation law. Notice that, because $\phia$ and $\varphi_\alpha$
follow the same equation of motion, this conservation law must be verified
also by $\varphi_\alpha$.

The physical interpretation of this conservation law is interesting. 
It is possible to connect this law with the continuity equation and the
conservation of the total charge and current.  
We have defined the current using
the continuity equation
%\begin{eqnarray}
%\partial_x  I_\alpha(x)&=&e\partial_t\delta\rho_\alpha(x)\\
%&=&-ie\omega \partial_x \phi_\alpha(x)
%\end{eqnarray}
and up to a constant we have 
\be 
 I_\alpha(x)=-ie\omega
\phi_\alpha(x) 
\ee 
and we can express (\ref{conserved1}) in the form
%\footnote{The current is a real operator.}
\be
\begin{split}
0&=\partial_x\frac{1}{e^2\omega^2}\left( I_L^2(x)-
I_R^2(x)\right)
 \label{conserved2}.
\end{split}
\ee 
To appreciate the meaning of the above conservation law, let us
consider the solution of the equation of motion. For simplicity we
consider the translationally invariant case. The left and right
density fluctuations are given by
\be
\begin{split}
\delta\rho_L(x)&=-ie\sum_{q>0}\sqrt{\frac{\nu}{2 \pi q}}
q\left[e^{iqx}\left(u  b_{u,q}-v 
b_{d,q}^\dagger\right)-e^{-iqx}\left(-v  b_{d,q}+u 
b_{u,q}^\dagger\right)\right],\\
\delta\rho_R(x)&=-ie\sum_{q>0}\sqrt{\frac{\nu}{2 \pi q}}
q\left[e^{iqx}\left(-v  b_{u,q}+u 
b_{d,q}^\dagger\right)-e^{-iqx}\left(u  b_{d,q}-v 
b_{u,q}^\dagger\right)\right]
\end{split}
\ee
and the currents ($cq=\omega_q$)
\be
\begin{split}
 I_L(x)&=-iec\sum_{q>0}\sqrt{\frac{\nu}{2 \pi q}}
q\left[e^{iqx}\left(u  b_{u,q}+v 
b_{d,q}^\dagger\right)-e^{-iqx}\left(v  b_{d,q}+u 
b_{u,q}^\dagger\right)\right],\\ 
I_R(x)&=-iec\sum_{q>0}\sqrt{\frac{\nu}{2 \pi q}} q\left[-e^{iqx}\left(u
 b_{u,q}+v  b_{d,q}^\dagger\right)+e^{-iqx}\left(v 
b_{d,q}+u  b_{u,q}^\dagger\right)\right].
\end{split}
\ee

Now we define
\begin{eqnarray}
&&\delta\rho_u(x)=-ie\sum_{q>0}\sqrt{\frac{\nu}{2 \pi
    q}}q\left[e^{iqx} b_{u,q}-e^{-iqx}
    b_{u,q}^\dagger\right],\\
    &&\delta\rho_d(x)=-ie\sum_{q>0}\sqrt{\frac{\nu}{2 \pi
    q}}q\left[e^{iqx} b_{d,q}^\dagger-e^{-iqx} b_{d,q}\right]
\end{eqnarray}
for the up and down moving fluctuations densities and again from the
continuity equation
\begin{eqnarray}
&& I_u(x)=-iec\sum_{q>0}\sqrt{\frac{\nu}{2 \pi q}}
    q\left[e^{iqx} b_{u,q}-e^{-iqx} b_{u,q}^\dagger\right],\\
&& I_d(x)=-iec\sum_{q>0}\sqrt{\frac{\nu}{2 \pi
    q}}q\left[e^{iqx} b_{d,q}^\dagger-e^{-iqx} b_{d,q}\right].
\end{eqnarray}
In terms of the up and down moving density fluctuations we have the
simple relations
\begin{eqnarray}
&&c(\delta\rho_u+\delta\rho_d)=c\delta\rho= I_u-I_d,\\ 
&&c(\delta\rho_u-\delta\rho_d)= I_u+ I_d= I,
\end{eqnarray}
so that the conservation of the total charge and the total current
also implies the conservation of the product \be ( I_u-
I_d)( I_u+ I_d). \ee Next we observe that
\begin{eqnarray}
&&\delta\rho_L(x)=u\delta\rho_u(x)-v\delta\rho_d(x),\\
&&\delta\rho_R(x)=u\delta\rho_d(x)-v\delta\rho_u(x),
\end{eqnarray}
and \be \left(
\begin{array}{l}
 I_L(x) \\  I_R(x)
\end{array}
\right) = \left(
\begin{array}{cc}
u & -v \\ -v & u
\end{array}
\right) \left(
\begin{array}{l}
 I_u(x) \\  I_d(x)
\end{array}
\right). \ee As a result we get \be  I_L^2- I_R^2\equiv 
I_u^2- I_d^2. \ee This follows by writing the rotation matrix
between ($ I_R$, $ I_L$) and ($ I_u$, $ I_d$) in terms
of Pauli matrices and observing that
\begin{eqnarray}
(u\sigma_0-v\sigma^y)\sigma^z(u\sigma_0-v\sigma^y) =\sigma^z.
\end{eqnarray}
Then we have reduced the conservation law written in terms of the left
and right currents to the product of two conserved quantities for the
up and down currents. The conservation of the currents in the up and
down basis follows from the diagonal form of the Hamiltonian in such
basis and the decomposition $ I_u^2- I_d^2=( I_u-
I_d)( I_u+ I_d)$ follows from the commutation rules for the up
and down moving boson operators.

\subsection{Non-translationally invariant case}

We now consider the effect of the presence of a constriction which
breaks the translational symmetry. This constriction
can be created by depleting a portion of the sample by applying a voltage
to the metallic gate fabricated on top of the mesa. When a finite $k$ wave impinges
on the constriction it can be partially reflected and transmitted. How this 
will affect the conductance is determined by the $k\to 0$ limit of the 
reflection coefficient. If this limit is zero there will be no correction to the 
ideal Hall conductance.  

We model the presence of the constriction by considering a piece-wise 
inter-edge potential. This choice is based on the assumptions that the 
interaction potentials have a Coulomb origin. When
the edges are forced by the constriction to stay close the mean distance 
between the density fluctuations is lesser than outside thus
the inter-edge interaction is greater inside the constriction than outside.  
We suppose also that the region when the inter-edge potential switches  
from the value outside the constriction to the value inside the constriction
can be neglected. 
The situations we want to consider in this section are plotted in Fig. 
\ref{fig_constriction.eps} where a constriction is localized in a finite region 
of the Hall bar (panel a) or the constriction extends until the drain (panel b).
\begin{figure}[!ht]
\begin{center}
\includegraphics[width=7cm,clip]{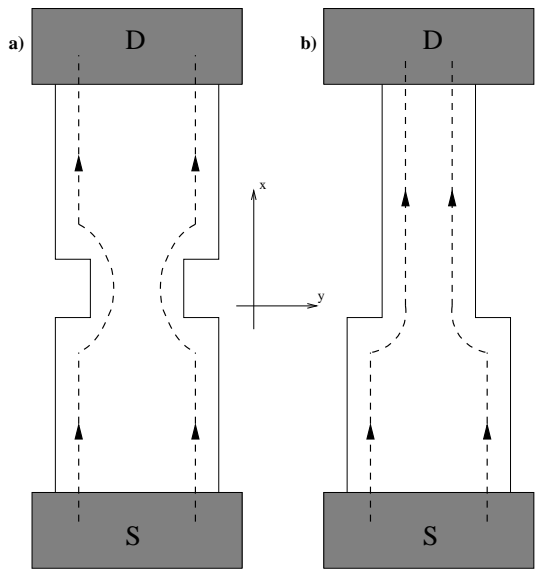}
\caption{The two types of constriction we want to consider. On the left the
constriction is localized in a finite region while on the right it is semi-infinite.}
\label{fig_constriction.eps}
\end{center}
\end{figure}

We start by considering the case of a finite size constriction. We assume that the
system is symmetric with respect to the $x=0$ point and that the length of the 
constriction is $d$.
To keep the analysis simple, we assume that the interaction potentials $V_1$ and
$V_2$ are short ranged on the scale of the density fluctuations: this in particular
implies that only points at the same value of the variable $x$ interact and 
$V_{\alpha,\beta}$ has the form
\be
V_{\alpha,\beta}=
\left(
\begin{array}{cc}
V_1 & V_2(x)\\
V_2(x) & V_1
\end{array}
\right)
\ee
where 
\be
V_2(x)=
\left\{
\begin{array}{cc}
V_{2,1}&x<-d/2\\
V_{2,2}&|x|<d/2\\
V_{2,3}&x>d/2
\end{array}
\right. .
\ee
 The region labelled $1$, $2$
and $3$ are implicitely defined in the piece-wise form of the potential $V_2$. 
We seek for a piece-wise solution. As is standard in scattering theory we label
the full solution with the quantum numbers of the incident wave. An ``up-moving"
solution will then have the form
\be
\tilde\varphi_{k_1}^u=
\left\{
\begin{array}{lc}
\varphi_{k_1}^u(x)+r^u \varphi^d_{k_1}(x) & x<-d/2\\
A^u\varphi_{k_2}^u(x)+B^u\varphi^d_{k_2}(x) & |x|<d/2\\
t^u \varphi_{k_3}^u(x)& x>d/2
\end{array}
\right. .
\ee
The wave vectors in the regions $2$ and $3$ are determined in terms of the incident
momentum $k_1$ by the conservation of the energy
\be
c_1k_1=c_2k_2=c_3k_3
\label{soundspeeds}
\ee
where $c_1$, $c_2$ and $c_3$ are the sound velocities in the three regions. In a 
similar way one can construct the solution for the ``down-moving" solution
\be
\tilde\varphi_{k_3}^d=
\left\{
\begin{array}{lc}
t^d\varphi_{k_3}^d(x)& x<-d/2\\
A^d\varphi_{k_2}^d(x)+B^d\varphi^u_{k_2}(x) & |x|<d/2\\
\varphi_{k_3}^d(x)+r^d \varphi_{k_3}^u(x)& x>d/2
\end{array}
\right. .
\ee
In these expressions the spinorial functions $\varphi^u_k(x)$ and $\varphi^d_k(x)$
are given by the Eqs. (\ref{upmoving}) and (\ref{downmoving}) where the index
 $\alpha$ has been dropped for the sake of notation's simplicity. 

The matching conditions are dictated by the physical requirement that there is 
not energy accumulation at the interfaces. This is equivalent to the requirement
of the continuity of the solution at the points $x=\pm d/2$ and gives four conditions
to determine the coefficients $A$, $B$, $t$, and $r$ for the up- and down-moving
wave functions\footnote{We must recall that the equation of motion for the 
spinorial wave functions $\varphi$ contains only the first derivatives with
respect to the time and position hence the request of the continuity 
at one point is sufficient to fully determine the solution for the scattering problem.}. 

The solution of the set of the equations obtained by imposing the matching 
conditions can be obtained in a straightforward way. We get, 
after having expressed the results in terms of the mixing angles,
 \be
\begin{split}
t^u
&=\frac{e^{-i\frac{k_1+k_3}{2}d}}{\cos(k_2d)\cosh(\theta_1-\theta_3)-i
\sin(k_2d)\cosh(2\theta_2-\theta_1-\theta_3)}\sqrt{\frac{c_1}{c_3}},\\
r^u &=-\frac{\cos(k_2d)\sinh(\theta_1-\theta_3)+i \sin(k_2
d)\sinh(2\theta_2-\theta_1-\theta_3)}
{\cos(k_2d)\cosh(\theta_1-\theta_3)-i
\sin(k_2d)\cosh(2\theta_2-\theta_1-\theta_3)}e^{-ik_1d},\\ A^u
&=\frac{\cosh(\theta_3-\theta_2)e^{-i\frac{k_1+k_2}{2}d}}
{\cos(k_2d)\cosh(\theta_1-\theta_3)-i
\sin(k_2d)\cosh(2\theta_2-\theta_1-\theta_3)}\sqrt{\frac{c_1}{c_2}},\\
B^u &=\frac{\sinh(\theta_3-\theta_2)e^{i\frac{k_2-k_1}{2}d}}
{\cos(k_2d)\cosh(\theta_1-\theta_3)-i
\sin(k_2d)\cosh(2\theta_2-\theta_1-\theta_3)}\sqrt{\frac{c_1}{c_2}}.
\label{solution}
\end{split}
\ee
The coefficients for the down-moving solution can be obtained by the substitutions
\be
t^u\to t^d;~r^u\to r^d;~A^u\to B^d;~B^u\to A^d,
\ee
and 
\be
\theta_1\to \theta_3;~ \theta_3 \to \theta_1;~ c_1 \to c_3.
\ee
It is easy to verify that the wave function (\ref{upmoving}) with the coefficients
determined by (\ref{solution}) 
satisfies the conservation law (\ref{conserved1}).

We recover the case of a semi-infinite constriction by considering the limit
$d\to 0$ in the expressions for the transmission and reflection coefficients 
\cite{Oreg1995},
\be
\begin{split}
t^u&=\frac{1}{\cosh(\theta_1-\theta_3)}\sqrt{\frac{c_1}{c_2}},\\
r^u&=-\frac{\sinh(\theta_1-\theta_3)}{\cosh(\theta_1-\theta_3)}.
\end{split}
\ee
If one defines the interaction renormalized  filling factor $\nu_i=\nu e^{-2\theta_i}$
for the various region then it is possible to rewrite the above results 
as
\be
\begin{split}
r^u=&\frac{\nu_1-\nu_3}{\nu_1+\nu_3},\\
t^u=&\frac{2\nu_3}{\nu_1+\nu_3}\sqrt{\frac{c_1}{c_3}}.
\end{split}
\ee
These expressions remain valid when we consider the case of two regions
with different filling factors \cite{Chklovskii1998}.
In the following we will consider the symmetric case $\theta_1=\theta_3$ i.e. the 
interaction is symmetric with respect to the center of the constriction. From the
expressions (\ref{solution}) we get
\be
\begin{split}
t^u
&=\frac{e^{-i kd}}{\cos(k_2d)-i
\sin(k_2d)\cosh(2\theta_2-2\theta_1)},\\
r^u &=-\frac{i \sin(k_2
d)\sinh(2\theta_2-2\theta_1)}
{\cos(k_2d)-i
\sin(k_2d)\cosh(2\theta_2-2\theta_1)}e^{-ikd},\\
A^u&=\frac{\cosh(\theta_1-\theta_2)e^{-i\frac{k_1+k_2}{2}d}}
{\cos(k_2d)-i
\sin(k_2d)\cosh(2\theta_2-2\theta_1)}\sqrt{\frac{c_1}{c_2}},\\
B^u &=\frac{\sinh(\theta_1-\theta_2)e^{i\frac{k_2-k_1}{2}d}}
{\cos(k_2d)-i
\sin(k_2d)\cosh(2\theta_2-2\theta_1)}\sqrt{\frac{c_1}{c_2}}.
\label{symmetricsolution}
\end{split}
\ee
We will use these expressions to calculate the Hall conductance in the presence
of a constriction. 

The generalizations to the case when many constrictions are present or a constriction connects regions with different filling factors are straightforward.

\section{The tunneling Hamiltonian}
We want now to consider the effect of the tunneling between different edges. The
physical origin of tunneling lies in the fact that the electron quasi-particles are not completely localized in one or the other edge i.e. the density matrix $\rho({\bf r},{\bf r'})$ has a finite value even if ${\bf r}\not ={\bf r'}$.

The physics of the tunneling is obviously lost in the hydrodynamical 
approximation. We need then to insert the tunneling by hand in our 
Hamiltonian. To do that we need to define a quasi-particle creation operator 
which adds a
quasi-particle with charge $e^*$ (not necessarily equal to the electron 
charge $e$) at the point $x$ of the edge $\alpha$. 
This is accomplished by requiring that this operator
 $\Psi^\dagger_\alpha(x)$ satisfies the commutation relation with the 
quasi-particle density
\be
[\Psi^\dagger_\alpha(x),\drb(x')]=-\frac{e^*}{e}\delta_{\alpha,\beta}
\delta(x-x')\Psi^\dagger_{\alpha}(x).
\label{psicommutation2}
\ee
At the best of our knowledge there is not exist a general theory
which predict the correct value of $e^*$ for arbitrary value of the filling factor. For 
certain values of the filling factor, as an example those given by $\nu=1/(2m+1)$
where $m$ is an integer, it is believed that $e^*=\nu e$. Two approaches
are then possible. One can fix $e^*=\nu e$ and then makes a comparison 
with the experimental results. On the other hand  it is possible to treat $e^*$ as a phenomenological parameter determined by
the confrontation with the experiment. 

The equation (\ref{psicommutation2}) allows us to express the quasi-particle
creation operator in terms of the boson operator via
\be
\begin{split}
\Psi^\dagger_\alpha(x)=U^\dagger_\alpha \exp\left[-i\frac{e^*}{e}\sqrt{\frac{2\pi}{\nu}}
\sigmaz \sum_{n>0} \varphi_{n,\beta}^*(x)b^\dagger_n\right]\\
\times\exp\left[-i\frac{e^*}{e}\sqrt{\frac{2\pi}{\nu}}
\sigmaz \sum_{n>0} \varphi_{n,\beta}(x)b_n\right]
\end{split}
\label{psiboson}
\ee
where the unitary fermion operator\footnote{We have omitted a normalization constant which depends on a short-range cut-off. We choose this normalization constant such that this operator is dimensionless.} $U^\dagger_\alpha$ commutes with all the boson operators and increases the total charge $Q_\alpha$ by $-e^*$ 
\be
[U_\alpha^\dagger,Q_\beta]=e^*\delta_{\alpha,\beta}U^\dagger_\alpha
\ee
where the total charge operator is defined as
\be
Q_\alpha=\int\limits_{-\infty}^\infty dx~\dra(x).
\ee 
The solution
of the equation (\ref{psicommutation2}) can be obtained by the observation
that the commutation relation can be converted to a differential equation
by using the different representations
\be
\begin{split}
&b\to \frac{\partial}{\partial b^\dagger},\\
&b^\dagger\to -\frac{\partial}{\partial b},
\end{split}
\ee
which in turn imply that
\be
\begin{split}
[b,f(b^\dagger)]=&\frac{\partial f(b^\dagger)}{\partial b^\dagger},\\
[b^\dagger,g(b)]=&-\frac{\partial g(b)}{\partial b}.
\end{split}
\ee
The operator $U_\alpha$ is then the ``arbitrary constant" in the solution 
of the differential equation and its properties can be deduced from the physical
insight that it must increases the total charge $Q_\alpha$. We have also written
the quasi-particle creation operator in a normal ordered way: this will avoid some
complications with the normalization \cite{Luther1974}.

In terms of the quasi-particle operators, the tunneling between the
edges coupled by a constriction at $x=0$ is described by the Hamiltonian
\be
H_T=\Gamma :\Psi_L^\dagger(0)\Psi_R(0):
+\Gamma^* :\Psi_R^\dagger(0)\Psi_L(0):
\ee
where $\Gamma$ is the (phenomenological) tunneling amplitude and the
symbol $:\ldots :$ indicates the normal ordering. 
The complete Hamiltonian then reads
\be
H=H_0+H_T=\sum_{n>0}\hbar \omega_n b^\dagger_n b_n+\Gamma :\Psi_L^\dagger(0)\Psi_R(0):
+\Gamma^* :\Psi_R^\dagger(0)\Psi_L(0):
\ee
where now the tunneling Hamiltonian can be viewed as an interaction Hamiltonian
for the bosons and now the total charge in a given edge is not a constant of motion:
its time derivative defines a tunnel current operator $I_T$ as follows
\be
\begin{split}
I_T&=\frac{i}{\hbar}[H_T,Q_L]\\ 
&=i\frac{e^*}{\hbar}(\Gamma:\Psi_L^\dagger(0)\Psi_R(0):
-\Gamma^* :\Psi_R^\dagger(0)\Psi_L(0):).
\end{split}
\ee

In the following chapter, when we will deal with the formulation
of the transport and derive the relations between the current and voltage, 
we will describe a method to calculate this tunneling 
current by defining an exact boson propagator and developing a perturbative 
scheme to evaluate its expectation value.

We must point out also that we do not specify any statistics for the quasi-particle
operator. Wen was able to show that the quasi-particle operator follows 
a Fermi algebra if and only if $\nu$ is the inverse of an odd integer. 
If $\nu$ is the inverse of an even integer the quasi-particle follows a 
Bose algebra. In the intermediate case the quasi-particle has not 
a definite statistics. We do not restrict the range of variation of
the filling fraction and then we do not have real fermion operator for the 
quasi-particle. We will show in the following that the restriction to
fermion operator is not necessary and we can fully develop
a perturbative approach to the tunneling.

\section{The multi-probe setup}

In the previous sections of this chapter we have introduced our model for
the quantum Hall bar and the way to describe the edges and their excitations.
For the sake of simplicity in introducing the notation and the various concepts
we have treated the case of two parallel edges with
only two probes, namely the drain and the source. 
However as can be seen 
from the scheme of the experimental setup in Fig.~\ref{fig_probe.eps} we need to
consider a more complicated situation where many probes, edges and contacts are
present. To do that we consider the schematic in the Fig.~\ref{fig_probe_theo.eps}
where we have introduced six different probes, two of them are used to inject
a steady-state current in the system (the D and S probes in the figure) and 
the others are used to measure either the Hall or the longitudinal voltage. The presence of the constriction is also included. 
\begin{figure}[ht!]
\begin{center}
\includegraphics[width=7cm,clip]{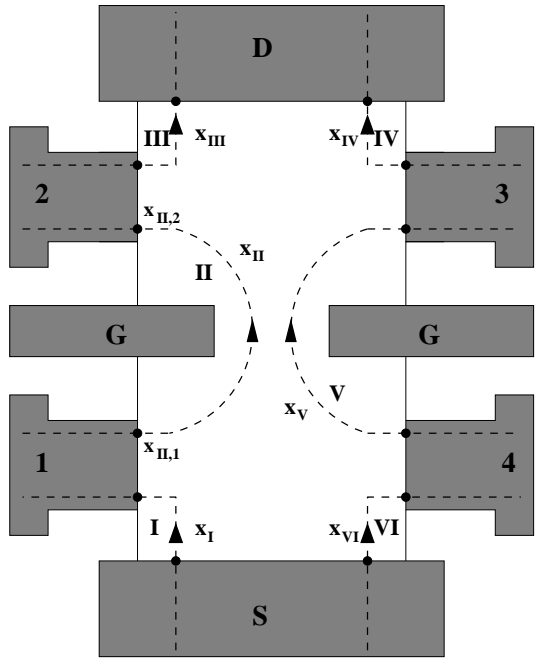}
\caption{A schematization of the experimental device where many edges and 
probes are represented. The gates (G) create a depletion zone forbidden for 
the electrons and then force the edges to stay close.} 
\label{fig_probe_theo.eps}
\end{center}
\end{figure}

We choose the following notation. Every edge is identified by an arabic figure,
while the probes are indicated by roman figures. The point where an 
edge exits the device is determined by the notation $x_{\alpha,m}$ where
$\alpha$ is the arabic figure which refers to the edge, while $m$ identifies the
probe. Generally a point belonging to a given edge is identified by its position $x_\alpha$ with respect to the edge. The number $x_\alpha$ varies in
$[-\infty,\infty]$ but we consider that when the electron leaves the device
the correlation vanishes exponentially with the increase of the position. 

We need to change also the definition of the density and quasi-particle operators.
The density fluctuation that belongs to the edge $\alpha$ is now indicated
by $\delta\rho(x_\alpha)$ while the quasi-particle creation operator is 
$\Psi^\dagger(x_\alpha)$. The simplest way to think at this change of notation
is to attach the index $\alpha$ directly to the variable $x$ rather than to the
operator. The usefulness of such notation will be proved when we will deal,
in the next chapter, with the calculation of the transport properties. 
When we consider the delta function of the position, namely the quantity
$\delta(x_\alpha-x_\beta)$, we identify it with the product of a delta function
relative to the edge index times the delta function of the position
\be
\delta(x_\alpha-x_\beta)\equiv\delta_{\alpha,\beta}\delta(x-x').
\ee

The intra-edge interaction
is substantially not affected by this change of notation. For the inter-edge
interaction we consider that our device is specular with respect to the $x$ axis.
In such a way when we consider the interaction potential $V_2(x_\alpha-x'_\beta)$
we consider two point with the same abscissa belonging to two different edges.

\chapter{The formulation of the transport}
In this chapter we derive the expression for the conductance
matrix in terms of the correlation function of the quasi-particles. In
the transport theory we need to calculate the current induced into the
reservoir $i$ due to a change in the potential $V_j$ of the reservoir
$j$. If we consider only the linear relation between the change in the
potential $\delta V_j$ and the induced current $\delta I_i$ we have
\be
\delta I_i=\sum_j G_{ij}({\bf V})\delta V_j.
\label{defconductance}
\ee
As it is recalled in this definition, the conductance matrix $G_{ij}$ is 
a function of the potential of all the reservoirs. In the
following, to simplify the notation, this
dependence will be understood even if not indicated explicitely.  
We choose the convention that a current is defined positive
when it enters a reservoir (leaving the system) and negative
viceversa.  The matrix element $G_{ij}$ are also subjected to the
physical conditions
\be
\sum_i G_{ij}=\sum_j G_{ij}=0
\ee
which represent the gauge invariance and the charge conservation,
respectively.  These expressions determine the value of the diagonal
elements $G_{ii}$ once the off-diagonal elements are known. Notice
also that in general there is not other way to calculate these
diagonal elements.  The rest of the thesis will deal with the
calculation of the conductance matrix in various cases study.

\section{The conductance matrix}
We want now to express the conductance matrix in terms of the boson
correlation function.

The starting point is the definition of the current of the edge
$\alpha$ in terms of the displacement field $\phi$. As we have
discussed in the previous chapter this relation can be derived from
the continuity equation which gives
\be
I(x_\alpha)=e\partial_t \phi(x_\alpha).
\ee

The current in the terminal $i$ is then the algebraic summation of
all the currents that enter or leave this terminal. We define the
function
\be
\xi_{\alpha,i}=\left\{
\begin{array}{cc}
+1 &\mbox{if~}\alpha\mbox{~enters~}i\\ -1
&\mbox{if~}\alpha\mbox{~leaves~}i\\ 0 & \mbox{otherwise}
\end{array}
\right.
\label{contactfunction}
\ee
such that the current on the terminal $i$ is given by
\be
I_i=\sum_{\alpha}\xi_{\alpha,i}I(x_{\alpha,i}).
\ee
This relation and the function $\xi$ represent the mathematical formulation
of our convention on the sign of the currents flowing in the system.
In general this is the sum of the two different currents, one leaving
the terminal and the other entering it. We choose, as the versus of the
total current, the versus from the source to the drain in the two edges. This choice
fixes the ``contact function" $\xi_{\alpha,i}$ and, as we will see, the sign of the
potential drops at the contacts.

By remembering that in the linear response theory the voltage is
coupled to the charge density, we have
\be
\delta I(x_\alpha,\omega)
=i\frac{e^2}{\hbar}\sum_\beta \int\limits_{-\infty}^\infty dx'_\beta
\delta V(x'_\beta) \int\limits_0^\infty dt \langle [\partial_t \phi(x_\alpha,t),\partial_{\xb '}
\phi(\xb ')]\rangle e^{i(\omega+i\eta)t}
\ee
where $\omega$ is the frequency of the external potential and $\langle
\ldots \rangle$ is the equilibrium average. We are interested in the
static limit hence in the following expressions the limit
$\omega\to 0$, when not explicitely indicated, is understood.

After an integration by parts with respect to the time, we arrive at
\begin{eqnarray}  \label{linearresponse2}
\delta I(x_\alpha)=i\frac{e^2}{h} \sum_\beta \int
dx'_\beta~[\omega \partial_{x'_\beta}D(x_\alpha,x'_\beta;\omega)]
\delta V(x'_\beta),
\end{eqnarray}
where
\begin{equation}
\label{defD}
D(x_\alpha,x'_\beta;\omega) \equiv -2 \pi i \int\limits_0^\infty
\langle [\hat \phi(x_\alpha,t),\hat \phi(x'_\beta)]\rangle
e^{i(\omega+i\eta)t}dt
\end{equation}
is the retarded displacement-field propagator, whose explicit
expression in terms of ``phonon eigenfunctions" is
\be
\label{propagator}
D(x_\alpha,x'_\beta;\omega)= \nu \sum_{n>0} \left[
\frac{\varphi_{n}(x_\alpha)\varphi_{n}^{*}(x'_\beta)}{\omega-\omega_n+i\eta}
-
\frac{\varphi_{n}^{*}(x_\alpha)\varphi_{n}(x'_\beta)}{\omega+\omega_n+i\eta}
\right]. 
\ee
In performing the integration by parts we have used the fact that
\be
\partial_{\xb'}
D(\xa,\xb';t=0^+)=-2\pi[\phi(\xa),\partial_{\xb'}\phi(\xb')]\propto
\delta(\xa-\xb')
\ee
vanishes unless $\xa=\xb'$. We will show in the
following that the evaluation of this function on the point
$x_{\alpha,i}$ is related to the conductance. However, as we have
discussed briefly before, the conductance matrix elements we can
calculate are the off-diagonal ones only. Hence in doing these calculations
we deal only with the case $\xa\not =\xb'$ and this validates our
procedure of integrating by parts.

We need now a model for the external potentials which are applied to
the reservoirs. We model only the potentials in the region
outside the device. To do that we consider that the potential is uniform in
the portion of the edge that runs inside the reservoirs, drops to
zero at the contact points between the leads and the device, and
is zero inside the device.
Thus we choose an external potential which satisfies
\be
\partial_{\xa} \delta V(\xa)=\sum_j \xi_{\alpha,j}\delta(\xa-x_{\alpha,j}) \delta V_j
\ee
where the ``contact function" was defined in the
Eq.~(\ref{contactfunction}).  This expression for the external
potential makes very easy to calculate the conductance matrix. Indeed
after an integration by parts in the Eq.~(\ref{linearresponse2}),
carried out on the assumption that the correlation function decay
exponentially for $x\to\pm\infty$, we obtain the expression of the
current arriving at the reservoir $i$ via the edge channel $\alpha$
due to a potential variation applied to the edge channel $\beta$ by
the reservoir $j$ ($i\not =j$)
\be
\begin{split}
\delta I_i&=\sum_j G_{ij}\delta V_j\\
&=\sum_j\left(-\frac{ie^2}{\hbar}\sum_{\alpha,\beta}
\xi_{\alpha,i}\xi_{\beta,j}
\lim_{\omega\to 0} \omega D(x_{\alpha,i},x_{\beta,j},\omega)\right) \delta V_j.
\end{split}
\label{conductances}
\ee
We have thus obtained an expression for the conductance matrix in
terms of the boson correlation function.

As an application we want to calculate this function in the case of
two translationally invariant edges. In this case the relation between
the energy and the momentum is $\omega=kc$ where $c$ is the phonon
velocity and the wave functions are given by the expressions
(\ref{upmoving}) and (\ref{downmoving}).  To ensure that the wave
function vanishes in the limit $|x|\to \infty$ we shift, in the
argument of the complex exponential, $k\to k+ i\eta\sign(x)$ in the
(\ref{upmoving}) and $k\to k-i\eta\sign(x)$ in the (\ref{downmoving}).
After plugging in these wavefunctions in the expression for the
correlation function we reduce the sum over $k>0$ to the integral over
the real axis in $k$. We readily see that only the poles $ck=\pm
(\omega+i\eta)$ contribute to the integral thus giving
\be
\begin{split}
-i\lim_{\omega\to 0}\omega D(\xa,\xb',t)=&
\nu
\Theta (x_\alpha-x'_\beta) \left ( 
\begin{array}{cc} 
u^2 & -uv \\ -uv & v^2
\end{array} 
\right)\\
&+\nu \Theta (x'_\beta-x_\alpha) \left (
\begin{array}{cc} 
v^2 & -uv \\ -uv & u^2
\end{array} 
\right)
\end{split}
\label{simplepropagator}
\ee   
where the functions $u$ and $v$ are the limits $k\to 0$ of the functions
$u_k$ and $v_k$ we have defined in the (\ref{amplitude}).  In the case
of non-interacting edges we have $u=1$ and $v=0$, thus we have upward
propagation on the left edge and downward propagation on the right
edge. This makes the conductance $G_{ij}$ vanishes unless the
reservoir $j$ is not ``upstream" the reservoir $i$ according to the
definition of an ideal Quantum Hall system. 

When the edges are interacting, from the direct application of the
Eqs. (\ref{conductances}) and (\ref{simplepropagator}), 
we obtain  
\be
G_{21}=G_{12}=\nu\frac{e^2}{h}e^{-2\theta}.
\ee
Notice that in this simple case an edge contacts both the
reservoirs.
%As pointed out by Wen \cite{Wen1991b} 
%this relation does not change the 
%exact quantization of the Hall conductance because
%the interaction modifies, in the same way, the edge potentials,
%thus giving again 
%\be
%I=\nu\frac{e^2}{h}(V_S-V_D).
%\ee
One might wonder how this result modifies the relation between the 
current and the source and drain potentials. As pointed out by Wen  \cite{Wen1991b}, when the interaction is present, the edge potentials are 
a linear combination of the source and drain potentials with the coefficients 
determined by the mixing angle $\theta$.
By taking into account
this relation we arrive at 
\be
I=\nu \frac{e^2}{h}(V_S-V_D)
\ee
and this implies that 
a measure of the Quantum Hall conductance is insensitive to the 
presence of a translationally invariant inter-edge interaction.
%must recover the ideal result.
 
A similar calculation can be developed for the case of a constriction by using the 
reflection and transmission coefficients in Eq.~(\ref{solution}).  
In this case the calculation is complicated by the piece-wise form
of the wave functions. To keep the analysis as simple as possible
we consider the case when four reservoirs are attached just 
above and below the constriction and 
the interaction is present only inside the 
constriction region (i.e. we fix $\theta_1=\theta_3=0$). 
By moving from the wave functions (\ref{upmoving}) and (\ref{downmoving}) we have obtained fixing, as an example, the 
first reservoir
\be
\begin{split}
&G_{21}=\nu\frac{e^2}{h} \frac{1}{2\pi i}\int\limits_{-\infty}^\infty
dk \frac{e^{ik(x_1-x_2)}}{k-\omega/c-i0^+}t_k^u,\\
&G_{31}=0,\\
&G_{41}=\nu\frac{e^2}{h} \frac{1}{2\pi i}\int\limits_{-\infty}^\infty
dk \frac{e^{-ik(x_1+x_4)}}{k-\omega/c-i0^+}r_k^u.
\end{split}
\ee
With $\theta_1=\theta_3=0$ the reflection and transmission coefficients
read
\be
\begin{split}
t_k^u
&=\frac{e^{-i k_1 d}}{\cos(k_2d)-i
\sin(k_2d)\cosh(2\theta_2)},\\
r_k^u &=-\frac{i \sin(k_2d)\sinh(2\theta_2)}
{\cos(k_2d)-i
\sin(k_2d)\cosh(2\theta_2)}e^{-ik_1d},
\end{split}
\ee   
where we have $c_1 k_1=c_2 k_2$. In performing the integration 
to calculate the conductances we observe that we must close
both the integrals in the upper half complex plane. Since $t_k$ and
$r_k$ have only poles inside the integration path
in the lower half complex plane, the only 
pole is at $k=\omega/c+i\eta$ and we obtain
\be
\begin{split}
&G_{21}=\nu \frac{e^2}{h}t^u(\omega/c),\\
&G_{41}=\nu \frac{e^2}{h}r^u(\omega/c).
\end{split}
\ee
The ideal Quantum Hall conductance is recovered by observing that
in the limit of vanishing frequency $\omega \to 0$ we have $t^u \to 1$ and $r^u\to 0$. We are then arrived to the conclusion that the constriction and the interaction do not modify the ideal Quantum
Hall conductance. Indeed one must observe that the interaction
does not allow for the passage of one quasi-particle from 
one edge to the other. On the other hand the constriction,
in the limit of small energy, becomes fully transparent thus recovering
the translationally invariant results.  

\section{The tunneling conductance}
When we consider the tunneling, we know that the full Hamiltonian cannot be
diagonalizable. Thus we develop a perturbative scheme to calculate the
tunneling conductance i.e. the correction to the ideal Quantum Hall conductance induced by the presence of the tunneling. We choose to develop this theory
starting from the equations of motion. Such an approach does not rest
on any assumptions about the statistics of the quasi-particles and, then, can be
applied to the case when $\nu$ is not the inverse of an odd number.

As we have pointed out the presence of the tunneling adds to the free Hamiltonian an interaction
potential between the bosons operators. The task 
at hand is then to calculate the corrections, due to this interaction, to the
displacement field $\phi$. We show that this correction can be exactly 
expressed in terms of a tunneling current propagator and we provide 
a perturbative evaluation of the latter.

We introduce now some compact notation. We define the phonon spinorial
operator
\be
B_n^i\equiv\left(
\begin{array}{c}
b_n\\
b_n^\dagger
\end{array}
\right)
\ee
where the index $i\in\{1,2\}$ (the upper part corresponds to $1$), and the 
associated phonon propagator
\be
D_{n,n'}^{i,j}(t)\equiv -\frac{i}{\hbar}\Theta(t)
\langle[B^i_n(t),B_{n'}^{j \dagger}]\rangle.
\ee
We define also the spinorial eigenfunction
\be
\varphi^i_n(\xa)\equiv
\left(\begin{array}{c}
\varphi_n(\xa)\\
\varphi_n(\xa)^*
\end{array}\right).
\ee

With these definitions the phonon field propagator can be written as
\be
\begin{split}
D(\xa,\xb',t)&=\hbar\nu\sum_{i,j}\sum_{\{n,n'\}>0} \varphi^i_n(\xa) D^{i,j}_{n,n'}(t) \varphi^j_{n'}(\xb').
\end{split}
\ee
In the following we will suppress the summation symbols and the sums over
repeated indices will be understood.
The phonon operator satisfies the equation of motion 
\be
i\partial_t B_n^i=\Omega_n^{i,j} B_n^j-\frac{Y^i_n}{e}I_T
\ee
where we have defined
\be 
\Omega_n^{i,j}=
\left(
\begin{array}{cc}
\omega_n & 0\\
0 & -\omega_n
\end{array}
\right),
~~
Y^i_n=\left(
\begin{array}{c}
\gamma_n\\
-\gamma^*_n
\end{array}
\right),
\ee
and
\be
\gamma_n=\sqrt{\frac{2\pi}{\nu}}\sum_\alpha \varphi^*_n(0_\alpha).
\ee

It is now straightforward to verify that the phonon propagator satisfies the equation of motion
\be
(i\partial_t \delta_{i,l}-\Omega^{i,l}_n)D^{l,j}_{n,n'}=\frac{(-1)^i}{\hbar} \delta_{i,j}\delta_{n,n'}\delta(t)-\frac{Y^i_n}{e}{\cal G}^j_{n'}(t)
\label{dmotionseq}
\ee
where the auxiliary operator 
\be
{\cal G}^i_{n}(t)=-\frac{i}{\hbar}\Theta(t)\left\langle\left[I_T(t),B_n^{j \dagger}\right]\right\rangle
\ee
in turn satisfies
\be
(i\partial_t \delta_{i,j}-\Omega^{i,j}_n){\cal G}^j_{n}=-\frac{Y^{i*}_n}{e}\tilde M_T(t)
\label{gcurrentprop}
\ee
with 
\be
\tilde M_T(t)=M_T(t)-\frac{{e^*}^2}{\hbar^2}\langle H_T \rangle\delta(t)
\ee
where
\be
M_T(t)=-\frac{i}{\hbar}\Theta(t)\left\langle\left[I_T(t),I_T\right]\right\rangle
\ee
is the tunneling current propagator. We will show later that this propagator can be expressed in terms
of the quasi-particle correlation functions. 

The Eqs. (\ref{dmotionseq}) and (\ref{gcurrentprop}) can be readily solved
by Fourier transformation to obtain 
\be
\begin{split}
D_{n.n'}^{i,j}=&[D^{(0)}]^{i,j}_{n,n'}(\omega)\\
&+\frac{\hbar^2}{e^2}[D^{(0)}]^{i,l}_{n,{n_1}}(\omega)
Y_{n_1}^l \tilde M_T(\omega) (Y_{n_2}^m)^*
\left([D^{(0)}]^{m,j}_{{n_2},n'}(\omega)\right)^*
\end{split}
\ee
where $[D^{(0)}]^{i,j}_{n,n'}(\omega)$ is the noninteracting phonon propagator.
Because $\tilde M_T$ is related to the quasi-particle correlation functions,
we have then expressed the phonon propagator in terms of the tunneling
current propagator. The noninteracting phonon propagator is determined
by the Fourier transformation of the operator
\be
[D^{(0)}]^{i,j}_{n,n'}(t)=\frac{(-1)^i}{\hbar}\delta_{i,j}\delta_{n,n'}\delta(t) (i\partial_t\delta_{i,j}-\Omega_n^{i,j})^{-1}.
\ee
Notice that there are no approximations on the 
calculation of the tunneling amplitude. 
Thus this set of equations constitutes an exact approach to the problem
of the tunneling.

Let us denote by
$G^{(0)}_{ij}$ the conductance obtained in the absence of tunneling
and by 
\be 
\label{deltaGdef} 
\delta G_{ij}=G_{ij} - G^{(0)}_{ij} 
\ee
the correction due to the tunneling.  
By starting from the definition (\ref{conductances}), with some straightforward
manipulations, one arrives at
\begin{equation}
\label{deltaG}
\begin{split}
\delta G_{ij} =& -\frac {i}{\nu^2} \lim_{\omega \to 0} \sum_{\alpha_i
\gamma}\xi_{\alpha i} [D^{(0)}](x_{\alpha i},0_\gamma;\omega)\\
&\times [\omega\tilde M_T(\omega)]\sum_{ \delta
\beta_j}[D^{(0)}]^*(0_\delta,x_{\beta j}';\omega)\xi_{\beta j}
\end{split}
\ee 
where the indices $\gamma$ and $\delta$ run over the two edges
that are coupled by tunneling at $x=0$, and the Green's function of
the noninteracting displacement field, $[D^{(0)}]_{\alpha
\beta}(x,x';\omega)$, is given by Eq.~(\ref{propagator}).

As a concrete example, let us consider a four-terminal geometry, as may be
obtained from Fig.~\ref{fig_probe_theo.eps} by considering only the terminals
from $1$ to $4$.  Let us assume for simplicity that the mixing angle $\theta$ is
independent of $x$. Then from Eq.~(\ref{simplepropagator}) we
immediately get
\be
\begin{split}
\sum_{\gamma} [D^{(0)}](x_\alpha,0_\gamma;\omega) =& \sum_{\gamma}
[D^{(0)}](0_\gamma,-x_\alpha;\omega)\\ =& \frac{i
\nu}{\omega}e^{-\theta}\left[\Theta(x)
\left(
\begin{array}{c}
u\\
-v
\end{array}
\right)
+
\Theta(-x) \left(
\begin{array}{c}
-v\\ ~u
\end{array}
\right)
\right]
\end{split}
\ee 
where the upper (lower) component refers to the left (right) edge
and $u=\cosh\theta$, $v=\sinh\theta$.

Substituting this in Eq.~(\ref{deltaG}) we find 
\be 
\label{linearresponse4}
\delta G_{ij}= \sum_{\alpha_i \beta_j} \delta
G_{\alpha_i \beta_j}(x_i,x_j)\xi_{\alpha i}\xi_{\beta j},
\ee 
where
\begin{equation}
\label{deltaG4}
\begin{split}
\delta G_{\alpha \beta}(x,x') =& -i e^{-2 \theta} \lim_{\omega \to
0}\frac{\tilde M_T(\omega)}{\omega} 
\left\{ 
\Theta(x)\Theta(-x')
\left(
\begin{array}{cc}
u^2 & -uv\\
-uv & v^2
\end{array} 
\right)\right.\\
&+\Theta(-x)\Theta(x')
\left(
\begin{array}{cc}
v^2 & -uv\\
-uv & u^2
\end{array} 
\right)
+ \Theta(x)\Theta(x') 
\left(
\begin{array}{cc}
-uv & u^2\\
v^2 & -uv
\end{array}
 \right)\\
&+\left. 
\Theta(-x)\Theta(-x')
\left(
\begin{array}{cc}
-uv & v^2\\
 u^2 & -uv
\end{array}
\right)
\right\}.
\end{split}
\ee 
Putting this in Eq.~(\ref{linearresponse4}) and noting that
$\xi_{\alpha 1}=\xi_{\beta 4}=1$, $\xi_{\alpha 2}=\xi_{\beta 3}=-1$
(with the labels $i$, $j$ as specified in the figure) we finally
obtain the correction to the Landauer-B\"uttiker conductances of the
ideal system: 
\be
\label{deltaGideal} 
\delta G_{ij}= i
e^{-2\theta}\lim_{\omega \to 0} \frac{{\tilde M}_T(\omega)}{\omega}
\left(\begin{array}{cccc} 
uv&v^2&-uv&-v^2\\
u^2&uv&-u^2&-uv\\
-uv&-v^2&uv&v^2\\ 
-u^2&-uv&u^2&uv
\end{array}\right). 
\ee
In the following for simplicity of notation we define the function
\be
g_T\equiv i\lim_{\omega\to 0} \frac{M_T(\omega)}{\omega}.
\ee

We consider the specific setup of
Fig. \ref{fig_probe_theo.eps} which represent 
an ideal representation of the experimental setup proposed
by Roddaro {\it et al.} \cite{Roddaro2002}.
The resistance $R_{xx}$ of the quantum point contact is measured
between terminals 3 and 4 
\be
R_{xx}=\frac{V_4-V_3}{I}, 
\ee 
where $I$ is the source-to-drain
current. By considering that the constriction does not affect the
source and drain probes, the full conductance matrix reads 
\be
\label{gij} 
G_{ij}=\frac{e^2}{h}\nu\left(
\begin{array}{cccccc}
1  &  0 & 0 		     & 0 		& 0 		      & -1               \\ 
0  &  1 & 0 		     & -1 		& 0 		      & 0	       \\ 
-1 &  0 & 1+\delta g_{11} & \delta g_{12}     & \delta g_{13}      & \delta g_{14}\\ 
0  &  0 &-1+\delta g_{21}& 1+\delta g_{22} & \delta g_{23}      & \delta g_{24}\\
0  & -1 & \delta g_{31}    & \delta g_{32}     & 1+\delta g_{33}  & \delta g_{34}\\
0  &  0 & \delta g_{41}    & \delta g_{42}     & -1+\delta g_{43} & 1+\delta g_{44}
\end{array}
\right),  
\ee 
where the indices $i$, $j$ run over $\{S,D,1,2,3,4\}$
and the right bottom submatrix is given by $\delta g_{ij} = \frac{h
\delta G_{ij}}{\nu e^2}$. In the next chapter we will evaluate 
the $g_T$ in a perturbative way where the
small parameter is the tunneling amplitude $\Gamma$: we will show 
that the leading term in $\delta G_{ij}$ is proportional to $|\Gamma|^2$ hence 
it must constitute a small correction to the ideal Quantum Hall conductance. 
In the Appendix \ref{Landauer-Buttiker} we 
discuss briefly the Landauer-B\"uttiker formalism and give
some insights about the calculation of this matrix.
As it is customary in the experimental
setup we fix $V_D=0$, $I_S=-I_D=-I$ and $I_1=I_2=I_3=I_4=0$.
With these constraints, the equation~(\ref{defconductance}) can be
easily solved\footnote{Notice that the gauge invariance and 
the charge conservation constraints make singular this matrix. We need then
to eliminate a row and a column before proceed with the standard technique
to solve the linear system.}, and to the lowest non vanishing order in $g_T$ we get
\be 
R_{xx}=\frac{h^2}{\nu^2 e^4} e^{-2\theta}(u+v)u~
g_T=\frac{h^2}{\nu^2 e^4} e^{-\theta}\cosh{\theta}~ g_T.
\label{tunnel-resistor}
\ee 
Notice that this perturbative result is valid only so long as
$R_{xx}$ is much smaller that $\frac{h}{e^2}$: The tunneling amplitude
$\Gamma$ must be sufficiently small to justify this perturbative approach.

\section{The tunneling amplitude}

In the previous section we have connected the conductance matrix to the 
current-current correlation function $\tilde M_T$. The next step is to
connect this function with the quasi-particle correlation function.
In this step we introduce a perturbative approach and develop all
the terms to the second order in the tunneling amplitude $\Gamma$. 

In $\tilde M_T$ two terms are present. The first one is the current-current
correlation function proportional to $\langle [I_T(t),I_T]\rangle$. Because
the tunneling current $I_T$ is proportional to $|\Gamma |$ this correlation function
is already second order in the tunneling amplitude. 
The second term is the average on the ground state
of the 
operator $H_T$. This average must be calculated to the second
order in $|\Gamma|$. 
By using the Kubo formula \cite{Mahan1981} we arrive at
\be
\langle H_T(t)\rangle=\frac{i}{\hbar}\int\limits_{-\infty}^t dt'~\langle
[H_T(t'),H_T(t)]\rangle .
\ee 

To simplify the notation we define the operator
\be
A(t)=:\Psi_L^\dagger(0,t) \Psi_R(0,t):
\ee
and then write the tunneling Hamiltonian as
\be
H_T(t)=\Gamma A(t)+\Gamma^* A^\dagger(t)
\ee
and the tunneling current as
\be
I_T(t)=\frac{ie^*}{\hbar}(\Gamma A(t)-\Gamma^* A^\dagger(t))
\ee
where we have used the fact that the tunneling takes place at the 
point $x=0$.
In this notation we have then
\be
\langle [H_T(t'),H_T(t)]\rangle =|\Gamma|^2 (\langle [A(t),A^\dagger(t')]\rangle+
\langle [A^\dagger(t),A(t')]\rangle)
\ee
and 
\be
\langle[I_T(t),I_T(0)]\rangle=\frac{{e^*}^2}{\hbar^2}|\Gamma|^2
(\langle[A(t),A^\dagger(0)]\rangle+\langle[A^\dagger(t),A(0)]\rangle).
\ee
Notice that in doing these averages the anomalous means that involved 
two $A$ or two $A^\dagger$ operator have been dropped. This is 
justified by the presence of the operator $U_\alpha$ in the definition
of the quasi-particle operator. Indeed, being unitary operators, we have
$U_R^\dagger U_R=U_L^\dagger U_L\equiv 1$ while the averages of the
others product of two $U$ operators is zero. We have restored here
the left $L$ and right $R$ notation for the sake of simplicity. Indeed in the
constriction only two edges are present (we do not have any contact inside
the constriction). Notice also that we have chosen the coordinate systems in 
these edges to have the $x=0$ point at the same height of the sample.  

The simple relation between a commutator and
its hermitian conjugate
\be
[A,B]^\dagger=-[A^\dagger , B^\dagger],
\ee 
if we define the correlation functions
\be
\begin{split}
&G_-(t;t')=\langle A(t')A^\dagger(t)\rangle,\\
&G_+(t;t')=\langle A^\dagger(t)A(t')\rangle,
\end{split}
\ee
allows us to recast the average of the tunneling Hamiltonians commutator as
\be
\begin{split}
\langle[H_T(t'),H_T(t)]\rangle =&|\Gamma|^2\left(
\langle [A^\dagger(t),A(t')]\rangle-\langle [A^\dagger(t),A(t')]\rangle^\dagger\right)\\
=&2i|\Gamma|^2 \mbox{Im}(G_-(t';t)-G_+(t';t))\\
=&4i|\Gamma|^2\mbox{Im}G_-(t';t)
\end{split}
\ee
where we have used the relation $G_+(t';t)=G_-^*(t';t)$ that we will prove in the
following chapter. The tunneling current correlation function can be rewritten as
\be
\langle [I_T(t),I_T]\rangle=-4i\frac{{e^*}^2}{\hbar}|\Gamma |^2\mbox{Im}G_-(t;0).
\ee

Because by definition we have
\be
\tilde M_T(\omega)=M_T(\omega)-\frac{{e^*}^2}{\hbar^2}\\
\int\limits_{-\infty}^\infty dt~\delta(t) \langle H_T(t)\rangle e^{i\omega t}
\ee
we must calculate the Fourier transform of the tunneling correlation function and
of the second term in the right hand side.
To calculate this second term we use the Dirac $\delta$ to perform the 
integration on time and we have
\be
\begin{split}
\int\limits_{-\infty}^\infty dt~\delta(t) \langle H_T(t)\rangle e^{i\omega t}&=
-4\frac{|\Gamma |^2}{\hbar} \int\limits_{-\infty}^0dt'~\mbox{Im}G_-(t';0)\\
&=-4\frac{|\Gamma |^2}{\hbar}\int\limits_{0}^\infty dt'~\mbox{Im}G_-(-t';0)\\
&=4\frac{|\Gamma |^2}{\hbar}\int\limits_{0}^\infty dt'~\mbox{Im}G_-(t';0)
\end{split}
\ee
where we have used the fact $G_-(t;0)=G_+^*(t;0)=G_+(-t;0)$ we will prove in 
the next chapter.
It is now easy to obtain the relation between $\tilde M_T(\omega)$
and the function $G_-(t)$. We get
\be
\begin{split}
\lim_{\omega\to 0}\frac{\tilde M_T(\omega)}{\omega}=&
-\frac{4{e^*}^2|\Gamma |^2}{\hbar^3} \left[ \lim_{\omega\to 0}\int\limits_{0}^\infty
dt~\frac{\cos(\omega t)-1}{\omega} \mbox{Im}G_-(t)\right.\\
&\left.+i\lim_{\omega\to 0}
\int\limits_{0}^\infty dt~\frac{\sin(\omega t)}{\omega}\mbox{Im}G_-(t)\right]\\
=& -\frac{4i {e^*}^2|\Gamma |^2}{\hbar^3}\int\limits_0^\infty dt~t\mbox{Im}G_-(t).
\end{split}
\label{mtilde}
\ee

The presence of a bias voltage $V_\alpha$ on the edge
$\alpha$ modifies the time evolution of the corresponding
quasi-particle operator 
\be 
\hat \Psi^\dagger (x_\alpha,t) \to
\hat \Psi^\dagger (x_\alpha,t)e^{- i\frac{e^* V_\alpha
t}{\hbar}}.
\ee  
The underlying physical assumption is, of course, that
each edge is in equilibrium with a reservoir, from which it generates,
at potential $V_\alpha$.
Under this assumption, the bias voltage dependence of the conductances
can be calculated with no additional effort. Indeed the function $G_-$ can be
easily expressed in terms of the quasi-particle creation and annihilation operators.
From the definition we have
\be
\begin{split}
G_-(t;t')&=\langle A(t')A^\dagger(t) \rangle\\
&=\langle :\Psi_L^\dagger(0,t') \Psi_R(0,t')::\Psi_R^\dagger(0,t) \Psi_L(0,t'):\rangle
\end{split}
\ee
thus this correlation function is modified by a phase factor due to the presence
of the finite edge voltages. Notice that the same phase factor modifies the function 
$G_+$.

With arguments
similar to those used to derive the Eq. (\ref{mtilde}) we have
\be
\begin{split}
\langle [H_T(t'),H_T(t)]\rangle &=4i\cos(\omega_T t')\mbox{Im}G_-(t';t),\\
\langle [I_T(t),I_T]\rangle &=-\frac{4i{e^*}^2}{\hbar^3}\cos(\omega_T
 t)\mbox{Im}G_-(0;t),
\end{split}
\ee
and 
\be
\begin{split}
g_T(\omega_T)\equiv i \lim_{\omega\to 0}\frac{\tilde M_T(\omega)}{\omega} =&
\frac{4 {e^*}^2|\Gamma |^2}{\hbar^3}\int\limits_0^\infty 
dt~t\cos(\omega_T t)\mbox{Im}G_-(t;0)\\
=& \frac{4 {e^*}^2|\Gamma |^2}{\hbar^3}\frac{\partial}{\partial \omega_T}\int\limits_0^\infty dt~\sin(\omega_T t)\mbox{Im}G_-(t;0)\\
=& \frac{4 {e^*}^2|\Gamma |^2}{\hbar^3}\frac{\partial}{\partial \omega_T}\mbox{Im}\int\limits_0^\infty dt~ e^{i\omega_T t}\mbox{Im}G_-(t;0),
\end{split}
\label{tunnelingamplitude}
\ee
where we have defined the frequency
\be
\omega_T=\frac{{e^*}V_T}{\hbar}.
\ee
We evaluate the tunneling voltage, the voltage across the quantum point contact,
by
\be
V_T=\frac{V_1-V_2}{2}+\frac{V_3-V_4}{2}
\ee
and it is possible to prove that $V_T=V_H$ where $V_H$ is the Hall voltage.
The amplitude $g_T(\omega_T)$ 
must be plugged in the Eq. (\ref{deltaGideal}) to calculate
the correction to the ideal conductance. The result of this operation
is the main topic of the next chapter.

\chapter{The transport properties in the presence of the constriction}
In this chapter we apply the formalism we have developed in the
previous chapter to the case when the tunneling takes place inside 
the constriction.

We need to calculate the retarded response function $G_-(\xa,\xb',t)$
defined as
\be
G_-(x,t;x',t')= \left\langle :\Psi_L^\dagger(x',t') \Psi_R(x',t'):
:\Psi_R^\dagger(x,t)\Psi_L(x,t): \right\rangle.
\ee
To do this calculation we use the Haussdorf lemma
\be
e^Ae^B=e^Be^A e^{[A,B]}
\ee
and the expression of the operator $\Psi$ in terms of the boson 
operators as defined in Eq. (\ref{psiboson}).

We start by considering the zero temperature limit and obtain an
approximated form of this correlation function. The presence of
the constriction introduces the time scale, $t_0=d/c$, i.e. 
the time a density fluctuation needs to travel through
the constriction. We can then write the correlation function
as the sum of two different contributions, one for $t<t_0$ and the other 
for $t>t_0$. This allows us to perform the Fourier transform we need
to calculate the current. 

The non-zero temperature case is then recovered by a conformal
transformation \cite{Shankar1990} that allows us to obtain
the form of the correlation function directly from the zero temperature
expression. Also in this case the correlation function can be separated in
two different contributions for short and long times. 

\section{General results}
By plugging in the definition of the $G_-(x,t;x',t')$ correlation function the general
form of the quasi-particle operator $\Psi^\dagger(\xa)$ 
given by the Eq. (\ref{psiboson}) we get 
\be
\begin{split}
G_-(x,t;x',t')=& \exp\left[\frac{2\pi{e^*}^2}{\nu e^2}\sum_{
n>0}\left(
\varphi_{nR}(x)+\varphi_{nL}(x)\right)\right.\\
&\left.\times\left(\varphi_{nR}^{*}(x')+\varphi_{nL}^{*}(x')\right)
e^{i\omega_k(t-t')} \right]
\end{split}
\label{gminuscomplete}
\ee 
while when we perform the same calculation on $G_+(x,t;x',t')$ we have
\be
\begin{split}
G_+(x,t;x',t')=& \exp\left[\frac{2\pi{e^*}^2}{\nu e^2}\sum_{
n>0}\left(
\varphi_{nR}^*(x)+\varphi_{nL}^*(x)\right)\right.\\
&\left.\times\left(\varphi_{kR}(x')+\varphi_{kL}(x')\right)
e^{-i\omega_k(t-t')} \right].
\end{split}
\ee 
To obtain these relations we have used the fact that the temperature is zero. 
In this case the average of the boson operator exponentials on the ground
state gives
\be
\langle e^{\alpha b^\dagger_n}e^{\beta b_n}\rangle =1
\ee
where $\alpha$ and $\beta$ are arbitrary coefficients. We will see that in the
case of non-zero temperature this average is a function of the temperature
via the Bose distribution.

From these expressions we get immediately the relations
\be
G_-(x,t;x',t')=G_+^*(x,t;x',t')
\ee
and if we fix $x=x'=0$ we have also
\be
G_-(0,t;0,0)\equiv G_-(t)= G_+(-t).
\ee
This validates our results obtained in the previous chapter about the relation 
between the tunneling amplitude and the $G_-(t)$ correlation function.
Notice that now and in the following we use the fact that the times $t$ and
$t'$ appear always in the combination $t-t'$ to set $t'\equiv 0$. 
When we fix $x=x'=0$ we obtain
\be
\begin{split}
G_-(t)=& \exp\left[\frac{2\pi{e^*}^2}{\nu e^2}\sum_{
n>0}\left|
\varphi_{nR}(0)+\varphi_{nL}(0)\right|^2 e^{i\omega_k t} \right].
\end{split}
\label{gminuse}
\ee

To calculate the effects of the constriction we substitute in the Eq.~(\ref{gminuscomplete})
the wave functions $\varphi$ determined in this case. The result
of this operation is 
\be
\begin{split}
G_-(x,t;x,&t')=\exp \left[\frac{2\pi{e^*}^2}{L\nu e^2}e^{-2\theta_2}
\sum_{k_2>0}\frac{e^{ik_2 c_2 (t-t')}}{k_2}\right.\\
&\left.\phantom{\frac{{e^*}^2}{L}}
\times \left(|A^u e^{ik_2x}-B^ue^{-ik_2x}|^2+|A^d
e^{-ik_2x}-B^de^{ik_2x}|^2\right)\right]
\end{split}
\ee 
where the coefficients $A^u$, $B^u$, $A^d$ and $B^d$ are given by
(\ref{solution}). By assuming that the tunneling is localized at the point
$x=x'=0$ and that the system is symmetric with respect to the center
of the constriction, i.e. we assume $\theta_1=\theta_3$, and 
substituting the expression (\ref{solution}) in this equation 
we obtain our key result
\be 
\label{finalG}
\begin{split}
G_-(t)=&\exp \left[\frac{4\pi{e^*}^2}{L\nu
e^2}\frac{\cosh(2\theta_2)}{\cosh(2\theta_1)}
e^{-2\theta_2}\sum_{k_2>0}\frac{e^{ik_2 c_2 t}}{k_2}
\right.\\
&\times\left.
\left(\frac{\cosh(2\theta_{12})-\sinh(2\theta_{12})\cos(k_2d)}
{1+2\sinh^2(\theta_{12})\sin^2(k_2d)}\right)\right]
\end{split}  
\ee 
where we have defined
$\theta_{12}=\theta_1-\theta_2$.

\section{The translationally invariant system}
Let us discuss, starting from the Eq.~(\ref{finalG}) the case of translationally invariant edges. To do that we fix $\theta_1=\theta_2=\theta$. With this position
the correlation function (\ref{finalG}) greatly simplifies to give
\be 
\label{weng}
G_-(t)=\exp \left[\frac{4\pi{e^*}^2}{L\nu
e^2}
e^{-2\theta}\sum_{k>0}\frac{e^{ik c t}}{k}
\right].
\ee 
To calculate this expression we use the well known analytical results
\begin{equation}
\begin{split}
&\sum_{n=1}^\infty \frac{\cos(n q)}{n}=-\frac12 \ln(2-2\cos(q))\\
&\sum_{n=1}^\infty \frac{\sin(n q)}{n}=\frac12 (\pi-q)
\end{split}
\end{equation}
which can be summarized in 
\be
\sum_{n=1}^\infty \frac{e^{inq}}{n}=-\ln \left(1-e^{iq}\right).
\ee
If $q$ is small compared to $1$ we have the approximated results
\begin{equation}
\begin{split}
\sum_{n=1}^\infty \frac{\cos(n q)}{n}\simeq-\ln(q), ~\sum_{n=1}^\infty
\frac{\sin(n q)}{n}\simeq\frac{\pi}2, ~\sum_{n=1}^\infty
\frac{e^{i n q}}{n}\simeq -\ln(iq).
\end{split}
\end{equation}
To calculate the expression~(\ref{weng}) we write the momentum
$k$ in terms of a integer $j$ defined by $k=2\pi j /L$ and then
evaluate the series in $j$\footnote{Notice that the lower limit $k>0$ forces the 
sum on $j$ to start
from $1$.} to obtain 
\be
G_-(t)=\exp\left[-\frac{2{e^*}^2}{\nu
e^2}e^{-2\theta}\ln\left(1-e^{\frac{2\pi i c
t}{L}-\delta}\right)\right] 
\ee 
and in the limit of large system size
$ct/L\ll 1$ we have 
\be 
G_-(t)=\left(\delta-\frac{2\pi
ict}{L}\right)^{-\frac{2}{\nu}\frac{{e^*}^2}{e^2}e^{-2\theta}} 
\ee
where $\delta$ assures the convergence of the series even when $t\to
0$. This function is the propagator for the Luttinger Liquid model,
with the anomalous exponent
$\frac{2}{\nu}\frac{{e^*}^2}{e^2}e^{-2\theta}$. Notice that if we
assume $e^*=\nu e$ we get for this exponent
\be 
2\nu
e^{-2\theta}=2\tilde\nu.
\ee  
In this case the tunneling differential
conductance at zero temperature is predicted to have a power law
behavior with exponent given by $2({e^*/e})^2/\tilde\nu-2$.
In fact to obtain the relation between the tunneling current and the 
tunneling voltage we must evaluate the Eq.~(\ref{tunnelingamplitude}). 
To perform this calculation we introduce a Heaviside function, $\Theta(t)$,
to extend the integral over the real axis and then we evaluate the Fourier transform
by calculating the convolution between the Fourier transforms of the $G_\pm(t)$ 
and of the $\Theta(t)$. Some details on the calculation of Fourier transform of the $G_\pm(t)$ are discussed in the Appendix~\ref{calculationofg}.
The result of these operations is the tunneling amplitude
\be
\label{wengtunnelingt0}
\begin{split}
g_T(\omega_T)=&\frac{2\pi |\Gamma|^2 {e^*}^2}{\hbar^3 \Gamma(\alpha)}
\left(\frac{L}{2\pi c}\right)^{\alpha}
\frac{d}{d\omega_T} 
|\omega_T|^{\alpha-1} 
\sgn{\omega_T}
\end{split}
\ee
where we have defined 
\be
\alpha=\frac{2}{\nu}\frac{{e^*}^2}{e^2}e^{-\theta}.
\ee
The Eq.~(\ref{wengtunnelingt0}) reproduces the result of Wen for the tunneling at zero temperature 
in the translationally invariant case \cite{Wen1991b}. We want to point
out that in the calculation of Wen a ultraviolet cut-off determined by the
magnetic length scale appears. This cut-off is necessary if one evaluates the sum 
in the correlation function as an integral to guarantee the convergence. 
In our case however an infrared cut-off
determined by the sample length $L$ appears. The finiteness of this length assures
the convergence of the series in the $G_-$ correlation function eliminating the singularity at $k\to 0$. 

When we consider the finite temperature case we must evaluate the 
boson thermal average $\langle b^\dagger_n b_n \rangle$ which appears 
in the argument of the exponential in the definition of the $G_-$ correlation 
function. The result of this evaluation is that the correlation function
at finite temperature can be obtained from that at zero temperature by
the conformal transformation \cite{Shankar1990}
\be
\label{conftransf} \left(\delta\pm i t\right) \to \frac{\sin[\pi
T(\delta\pm it)]}{\pi T}~. 
\ee 
Notice that we are using units in
which $\hbar = k_B=1$, where $k_B$ is the Boltzmann constant.  The
correct physical dimensions are restored via the substitution $T\to
k_B T/\hbar$ and this is understood in the following equations.
With this substitution the evaluation of the $g_T$ function follows the same 
steps we have discussed in the zero temperature case. In this case however
one must evaluate the integral
\be
\int_{-\infty}^\infty dt~e^{i\omega t} \frac{(\pi T t)^\alpha}{\sinh^\alpha (\pi tT)}.
\ee
The evaluation of this integral is discussed in the Appendix~\ref{calculationofg}.
We have thus obtained
\be
\begin{split}
g_T(\omega_T,T)=&2^{\alpha+1}\frac{{e^*}^2|\Gamma|^2}{\hbar^3} \left(\frac{L}{2\pi c}\right)^{\alpha}(\pi T)^{\alpha-1}
\sin\left(\frac{\pi\alpha}{2}\right)\\
&\times\frac{\partial}{\partial\omega_T} \mathrm{Im}
B\left(\frac{\alpha}{2}-i\frac{\omega_T}{2\pi
T},1-\alpha\right)
\end{split}
%\label{final-gt}
\ee
where $B$ is the Euler Beta function defined by \cite{Abramowitz1964}
\be
B(\alpha,\beta)=\frac{\Gamma(\alpha)\Gamma(\beta)}{\Gamma(\alpha+\beta)}.
\ee
With some straightforward manipulations on the $B$ function we can
recover the result of Wen \cite{Wen1991b} for the finite temperature
tunneling conductance.

\section{The effect of the constriction}
We consider now the effect of the constriction which 
breaks the translational invariance. To do that we must evaluate
the sum which compares in the $G_-$ correlation function in the Eq.~(\ref{finalG}).
We define the quantity 
\begin{equation}
\begin{split}
S_-(t,d)=&\frac{2\pi}{L}\frac{\cosh(2\theta_2)}{\cosh(2\theta_1)}\\
&\times \sum_{k_2>0}\frac{e^{ik_2 c_2t}}{k_2}%\\
\left(\frac{\cosh(2\theta_{12})-\sinh(2\theta_{12}) \cos(k_2d)}
{1+2\sinh^2(\theta_{12})\sin^2(k_2d)}\right)
\end{split}
\label{Sminus}
\end{equation}
and in the following we will evaluate it numerically.
We then separates its real and imaginary 
parts
\begin{equation}
\begin{split}
ReS_-(t,d)=&\sum_{n=1}^\infty\frac{\cosh(2\theta_{12})-\sinh(2\theta_{12})
\cos(2\pi dn/L_2)}{1+2\sinh^2\theta_{12}\sin^2(2\pi dn/L_2)}\\
&\times\frac{\cos(2\pi n c_2t/L_2)}{n},\\
ImS_-(t,d)=&\sum_{n=1}^\infty\frac{\cosh(2\theta_{12})-\sinh(2\theta_{12})\cos(
2\pi dn/L_2)}{1+2\sinh^2\theta_{12}\sin^2(2\pi dn/L_2)}\\ &\times
\frac{\sin(2\pi c_2tn/L_2)}{n}. 
\end{split}
\label{sum}
\end{equation}
Notice that in these functions we have defined the integer
$n$ such that $k_2=2\pi n/L_2$ and the length $L_2$, defined by
\be
L_2=L\frac{\cosh(\theta_1)}{\cosh(\theta_2)},
\label{l2}
\ee
which takes into account the different propagation velocities of the 
wave function in the regions $1$ and $2$. The convergence of the
series is guaranteed by the oscillatory behavior of the trigonometric 
functions.

The constriction introduces the characteristic time $t_0=d/c_2$ which 
is the time an edge wave needs to travel through the constriction. 
We discuss the two limits of $t\gg t_0$ and $t\ll t_0$. 
In both these two limits
we recover an expression for the $G_-$ correlation function
which is similar to the translationally invariant case and this allows us
to obtain the tunneling conductance. 

First let us consider the $d\to 0$ regime. From the Eq.~(\ref{sum}) we get
\begin{equation}
\begin{split}
ReS_-(t,d\to 0)=&e^{-2\theta_{12}}\\ 
&\times \left[-\frac12
\ln\left(2-2\cos\left(\frac{2\pi c_2t}{L_2}\right)\right)\right]\\
\simeq & -e^{-2\theta_{12}}\ln(2\pi c_2t/L_2),\\ 
ImS_-(t,d\to 0)=&e^{-2\theta_{12}}\left(\frac{\pi}2 (1-4 c_2t/L_2)\right)\\ 
\simeq &e^{-2\theta_{12}}\frac{\pi}2. 
\label{sum2}
\end{split}
\end{equation}
By plugging in these results in the expression for the $G_-$ correlation function
we get
\be 
G_{-,d\to0}(t)=\left(\delta-\frac{2\pi ic_2t}{L_2}\right)^
{-\frac{2}{\nu}\frac{{e^*}^2}{e^2} e^{-2\theta_1}}.
\label{gdzero}
\ee
From the definition (\ref{l2}) and from the energy conservation 
$k_1 c_1=k_2c_2$ we have also the relation
\be
\frac{L_2}{c_2}=\frac{L}{c_1}.
\ee
With this result we recover the same correlation function of the 
translationally invariant case when the constriction {\it is not present}. 

In the other limit $d\to \infty$ we have substituted in the two
functions in the Eq.~(\ref{sum}) the averaged values, 
$\langle \cos(k_2d)\rangle=0$, $\langle
\sin^2(k_2d)\rangle=1/2$ obtaining
\begin{equation}
\begin{split}
ReS_-(t,d\to \infty)=&(1+\tanh^2(\theta_{12}))\\ &\times
\left[-\frac12 \ln\left(2-2\cos\left(\frac{2\pi
c_2t}{L_2}\right)\right)\right]\\ \simeq &
-(1+\tanh^2(\theta_{12}))\ln(2\pi c_2t/L_2) ,\\ ImS_-(t,d\to
\infty)=&(1+\tanh^2(\theta_{12}))\left(\frac{\pi}2 (1-2
c_2t/L_2)\right)\\ \simeq &
(1+\tanh^2(\theta_{12}))\frac{\pi}2. 
\label{sum3}
\end{split}
\end{equation}
Again when we consider the function $G_-(t)$ we get a power law of $t$
\be G_{-,d\to\infty}(t)=\left(\delta-\frac{2\pi
ic_2t}{L_2}\right)^{-\frac{2}{\nu}\frac{{e^*}^2}{e^2} e^{-2\theta_2}
(1+\tanh^2(\theta_{12}))}.
\label{gdinfinity}
\end{equation}
In this case the presence of the constriction affects the exponent of
this correlation function and can change the behavior of the tunneling
amplitude.

The two limits of short and long times (with respect to $t_0$) are clearly
visible in the numerical evaluation of the function $S_-(t,d)$ as shown in
the Fig.~\ref{sumnew.eps}.
\begin{figure}[!ht]
\begin{center}
\includegraphics[clip,width=10cm]{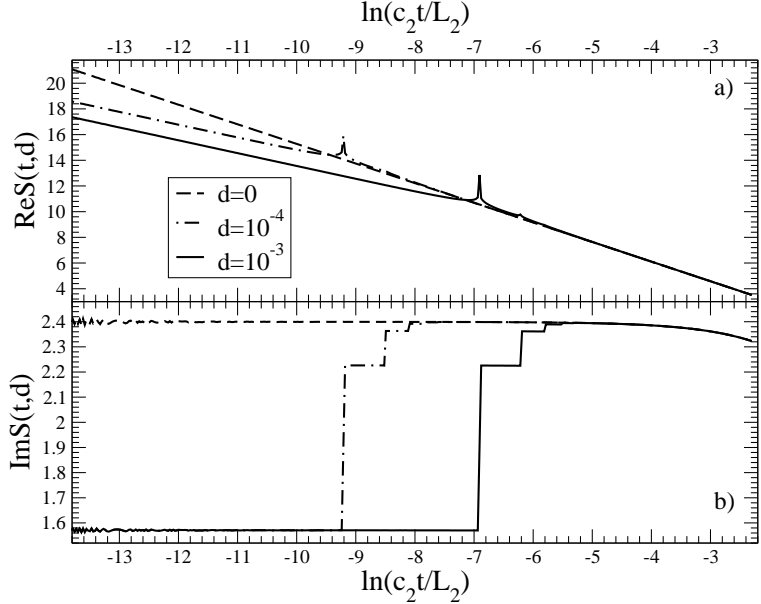}
\caption{{\bf a)} Plot of $ReS_-(t,d)$ vs. $\ln(c_2t/L_2)$ for various values
of $d$. Observe the two different regimes for $c_2t>d$ and $c_2t<d$.
The two slopes agree very well with the approximated result of
Eq.(\ref{Sminus}). We have chosen $\exp(\theta_{12})=1.5275$ in this
calculation.
{\bf b)} Plot of $ImS(t,d)$ vs. $\ln(c_2t/L_2)$ for various values of $d$. We
used the same parameters as a). The values for
small and large $c_2t/L_2$ agree well with the expected results (see
Eqs.(\ref{sum2}) and (\ref{sum3})).  The downward curvature at large
times arises from the finite size of the system used in the numerical
calculation and disappears in the limit of large system size.}
\label{sumnew.eps}
\end{center}
\end{figure}
In the numerical calculation we have fixed the value of $d$ and $\theta_{12}$ 
and then varied the value of $c_2 t$. 
As it is seen from the Fig.~\ref{sumnew.eps}
the two limits we have discussed are reached for $c_2 t\gg d$ and $c_2 t\ll d$.
In the Fig.~\ref{sumnew.eps} we report the behavior of the functions $ReS$ and
$ImS$ vs. the time for some values of $d$ and fixed $\theta_{12}$. Notice that
we have restricted our calculation to the case $c_2t/L_2\ll 1$ and $d/L\ll 1$.
The agreement of the calculated expressions and the approximated results
(\ref{sum2}) and (\ref{sum3}) is very good.

Hence from now on we approximate the $S_-(t,d)$ function as the sum
of the long and short time behaviors
\be 
%\label{Sminus}
\begin{split}
S_-(t,d)=&\Theta(t-t_0)S_-(t,d\to 0)\\ &+\Theta(t_0-t)(S_-(t,d\to
\infty)-\Delta_-)
\end{split}
\ee 
where the $\Delta_-$ function assures the continuity of $S_-$ at the point
$t=t_0$. The two functions $S_-(t,d\to 0)$ and $S_-(t,d\to\infty)$ are given
by the limits for long and short times of the function $ReS$ and $ImS$, 
respectively.

With this approximation we have separated the calculation of the tunneling
correlation function $G_-$ in the calculation of the short and long times 
behaviors. As we have seen these behaviors are similar to the translationally
invariant case. We
then expect that the low energy behavior (which corresponds to the low
bias voltage region) of the conductance will be dominated by the long
time part of $S_-(t,d)$.  Vice-versa, the response to a high bias
voltage will be dominated by the short time behavior of $S_-(t,d)$.
Within this approximation the $G_-(t,d)$ reads
\be 
\label{approxG}
\begin{split}
G_-(t) =\Theta(t-t_0)G_{-,d\to 0}(t)+\Theta(t_0-t)G_{-,d\to
\infty}(t) \exp\left[-\frac{2{e^*}^2}{e^2\nu}e^{-2\theta_2}\Delta_-\right]
\end{split}
\ee
where the exponential factor in the second term in the right hand side 
stems from the presence of the constant $\Delta_-$.

We have thus obtained the expression for the $G_-$ correlation function 
when the constriction is present. The next step is to evaluate the 
function $g_T(\omega_T)$. In this case we must evaluate an 
integrals of the form
\be
\int_{t_0}^\infty dt~e^{i\omega t}(\delta\pm it)^{-\alpha}.
\ee
In the Appendix~\ref{calculationofg} we discuss some
details about this calculation. 
Our result for the $g_T(\omega_T)$ function is
\be
\label{gtunnelingt0}
\begin{split}
g_T(\omega_T)=&\left(\frac{4|\Gamma|^2 {e^*}^2 t_0}{\hbar^3}\right)
\left(\frac{L_2}{2\pi c_2 t_0}\right)^{\alpha}\sin\left(\frac{\pi
\alpha}{2}\right)\\ 
&\times\frac{d}{d\omega_T}\left[ |\omega_T t_0|^{\alpha-1}
\left(\cos\left(\frac{\pi\alpha}{2}\right)~\sgn{\omega_T t_0}
\mathrm{Re} \Gamma(1-\alpha,-i \omega_T t_0)\right. \right.\\
&+\left.\sin\left(\frac{\pi\alpha}{2}\right)
\mathrm{Im}\Gamma(1-\alpha,-i\omega_T t_0)\right)\\ 
&+|\omega_T t_0|^{\beta-1} 
\left(\cos\left(\frac{\pi\beta}{2}\right)\sgn{\omega_T t_0}
(\Gamma(1-\beta)\right.\\
&-\mathrm{Re}\Gamma(1-\beta,-i\omega_T t_0))\\
&\left.\left.-\sin\left(\frac{\pi\beta}{2}\right)\mathrm{Im}\Gamma(1-\beta,-i\omega_T
t_0)\right)\right]\\
\end{split}
\ee
where we have defined 
\be
\begin{split}
%\omega_T=&\frac{e^*}{\hbar}V_T,\\ 
\alpha=&\frac{2}{\nu}
\frac{{e^*}^2}{e^2}e^{-2\theta_1},\\
\beta=&\frac{2}{\nu}\frac{{e^*}^2}{e^2}
e^{-2\theta_2}(1+\tanh^2(\theta_{12}))
\end{split}
\ee 
and $\Gamma(z_1,z_2)$ is the incomplete $\Gamma$ function
\cite{Abramowitz1964}. Notice that in the limit $\theta_{12}\to 0$ we 
have $\alpha=\beta$ and we recover the result of the translationally
invariant case. Notice also that this function is written as the sum
of two distinct contributions which came from the long and short time
regimes. However the $g_T$ function does not show
two distinct behaviors for small and large frequency as 
is shown in Fig.~\ref{conductance-t0.eps} where we plot the longitudinal
resistance $R_{xx}$, defined in the Eq.~(\ref{tunnel-resistor}), for various 
values of the mixing angle $\theta_2$ inside the constriction. In fact
the Fourier transformations of both the long and short time regimes have 
contributions in the whole range of the frequency values. 
\begin{figure}[!ht]
\begin{center}
\includegraphics[clip,width=12cm]{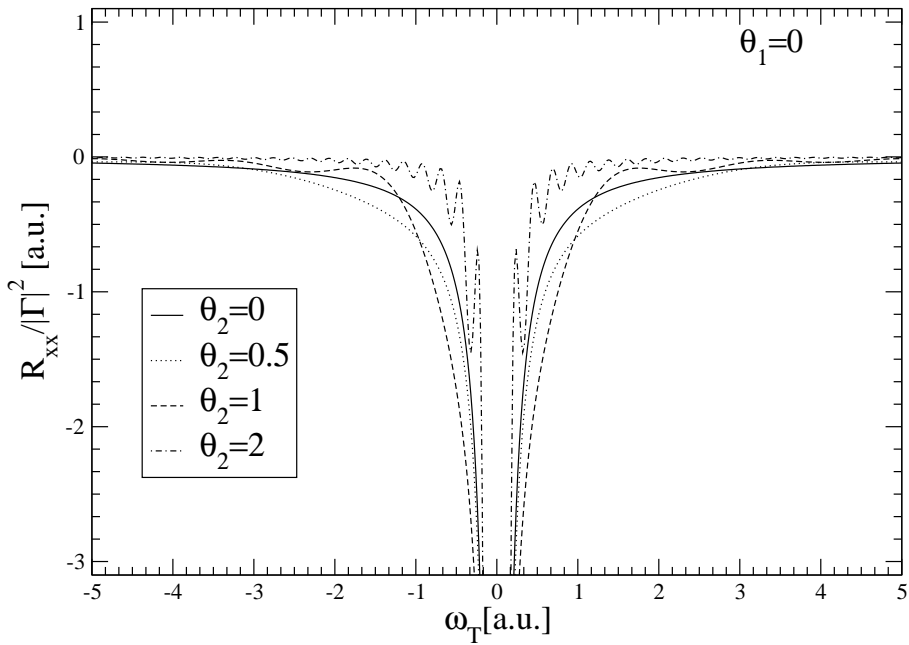}
\caption{Plot of the resistance $R_{xx}/|\Gamma|^2$ given by
Eq.~(\ref{tunnel-resistor}) with the $g_T$ calculated in
Eq.~(\ref{gtunnelingt0}) for various values of $\theta_2$ at fixed
$\theta_1=0$.  The oscillations at large bias voltage becomes more and
more pronounced with increasing $\theta_2$.  
}
\label{conductance-t0.eps}
\end{center}
\end{figure} 

In Fig.~\ref{conductance-t0.eps} we plot $R_{xx}(\omega_T)$ in the
case that the inter-edge interaction is confined to the region of the
constriction (i.e., we set $\theta_1 = \theta_3=0$ and let $\theta_2$
assume several different values).  Experimentally, $\theta_2$ can be
increased by narrowing the constriction by the application of a gate
potential. When $\theta_2 = 0$ there is no interaction and $R_{xx}$
diverges as $V_T^{\alpha-2}$ at low bias.

This low-bias behavior does not change upon increasing $\theta_2$
because the long time behavior is dominated by the exponent $\alpha$
which does not depend on $\theta_2$.  At larger bias voltage, however,
the plot of $R_{xx}$ shows oscillations, which become more pronounced
with increasing $\theta_2$. We can express the period of these
oscillations in terms of the physical parameters of the theory 
\be
\Delta V_T=\frac{h}{e^* t_0}=\frac{h c_1}{e^*d}\frac{\cosh
2\theta_1}{\cosh 2\theta_2}~.  
\label{deltavt}
\ee 
The frequency of the oscillations
increases with increasing $\theta_2$ as it is apparent in
Fig. \ref{conductance-t0.eps}.

%The finite temperature behavior of the tunneling resistance can be
%derived from the zero-temperature behavior of the same quantity by
%means of the conformal transformation \cite{Shankar1990} 
%\be
%\label{conftransf} \left(\delta\pm i t\right) \to \frac{\sin[\pi
%T(\delta\pm it)]}{\pi T}~. 
%\ee 
%Notice that we are using units in
%which $\hbar = k_B=1$, where $k_B$ is the Boltzmann constant.  The
%correct physical dimensions are restored via the substitution $T\to
%k_B T/\hbar$ and this is understood in the Eqs. (\ref{final-gt}) 
%and (\ref{finalgt_wen}) below.  
Making the conformal transformation~(\ref{conftransf}) in
Eqs.~(\ref{gdzero}) and ~(\ref{gdinfinity}), and substituting in
Eqs.~(\ref{Sminus}) and (\ref{approxG}) we obtain, after lengthy
calculations, the result for $g_T$ in the case of finite temperature
\be
\begin{split}
g_T(\omega_T,T)=&4\frac{{e^*}^2|\Gamma|^2}{\hbar^3}\left(\frac{L_2}{2\pi
c_2}\right) \left(\frac{L_2T}{2c_2}\right)^{\alpha-1}
\frac{\sin\left(\frac{\pi\alpha}{2}\right)}{\sinh^\alpha(\pi T t_0)}\\
&\times\frac{\partial}{\partial\omega_T} \mathrm{Im}
\left\{2^{\beta-1}
\sinh^\beta(\pi T t_0) B\left(\frac{\beta}{2}-i\frac{\omega_T}{2\pi
T},1-\beta\right)   
\right.\\ 
&+\frac{e^{i\omega_T t_0}}{\alpha-i\frac{\omega_T}{\pi T}}
F\left(\alpha,1;1+\frac{\alpha}{2}-i\frac{\omega_T}{2\pi T};\frac{1}{1-e^{2\pi Tt_0}}\right)\\
&\left.-\frac{e^{i\omega_T t_0}}{\beta-i\frac{\omega_T}{\pi
T}} F\left(\beta,1;1+\frac{\beta}{2}-i\frac{\omega_T}{2\pi
T};\frac{1}{1-e^{2\pi T t_0}}\right) \right\},
\end{split}
\label{final-gt}
\ee
where $F$ is the hypergeometric function of four arguments (also
indicated as ${_2}F_1$) and $B$ the Euler beta function
\cite{Abramowitz1964}.  In the case $\theta_1=\theta_2$ we have
$\alpha=\beta$, the first and third term cancel against each other and
we recover Wen's result. 
%\be
%\begin{split}
%g_T(\omega_T)=&4\frac{{e^*}^2|\Gamma|^2}{\hbar^3}\left(\frac{L_2}{2\pi
%c_2}\right)\left(\frac{L_2 T}{c_2}
%\right)^{\alpha-1}\sin\left(\frac{\alpha\pi}{2}\right) \\
%&\times\frac{d}{d\omega_T}\mathrm{Im}
%B\left(\frac{\alpha}{2}-i\frac{\omega_T}{2\pi T},1-\alpha\right).
%\end{split}
%\label{finalgt_wen}
%\ee

In Fig. \ref{fig_tnot0.eps}(a-d)
\begin{figure}[!ht]
\begin{center}
\includegraphics[clip,width=12cm]{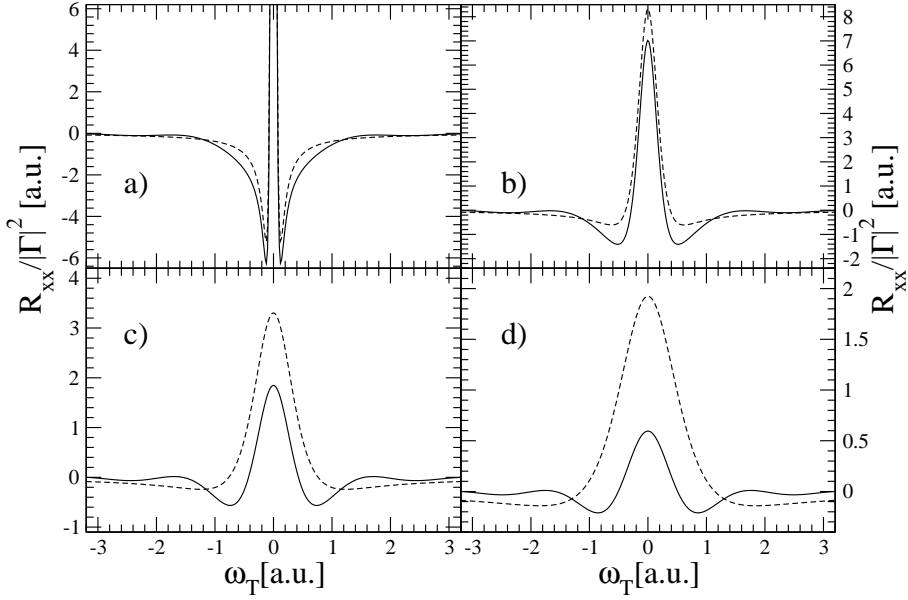}
\caption{Plot of the differential resistance $R_{xx}/|\Gamma|^2$ vs.  $\omega_T$
for a system with inter-edge interaction within the constriction
(continuous line, $\theta_1=0$, $\theta_2=1$) and without inter-edge
interaction (dashed line, $\theta_1=\theta_2=0$).  The four curves
correspond to different temperatures: $\pi T d/c_1=0.1$ (a),
$0.5$ (b), $1$ (c), and $1.5$ (d).}
\label{fig_tnot0.eps}
\end{center}
\end{figure}
we plot the differential resistance $R_{xx}$ vs. bias voltage for a
system {\it without} inter-edge interaction (dashed line --
$\theta_2=\theta_1=0)$ and {\it with} inter-edge interaction (solid
line -- $\theta_1=0$, $\theta_2=1$) for different values of $\pi d
T/c_1 =0.1, 0.5,1$, and $1.5$.  The non vanishing value of $\theta_2$
within the constriction induces oscillations in the $R_{xx}$ vs.
$\omega_T$ relation with the same period as in the zero temperature
case. However, we now have a maximum at zero bias voltage and two
minima at finite bias voltage.  This behavior is due to the fact that
the temperature introduces a new energy scale.  When the $e^*V > k_BT$
we are essentially in the zero temperature case and the resistance
$R_{xx}$ decreases with decreasing bias voltage (see
Fig. \ref{conductance-t0.eps}). But, when the $e^*V<k_BT$ the
resistance turns around and begins to increase, reaching a maximum at
zero bias.  This behavior implies the presence of two minima located
at bias voltages of the order of magnitude of $k_BT/e^*$: these are
clearly seen in Fig. \ref{fig_tnot0.eps}.  The finite value of
$R_{xx}$ at zero bias (independent of $V_T$ to first order) indicates
that the constriction is behaving like an ohmic resistor in this
regime, even though the resistance is strongly temperature-dependent.

The presence of a constriction adds another energy scale in the
problem, associated with the inverse of the characteristic time $t_0$.
For temperatures smaller than $\hbar/t_0$ the low bias behavior is
dominated by the same exponent $\alpha$ (cf. Fig. \ref{fig_tnot0.eps}
(a,b)) irrespective of whether the inter-edge interaction is present
or not.  When the temperature, instead, is greater than $\hbar/t_0$
the exponent $\beta$, which depends on the strength of the interaction
within the constriction, controls the behavior of $R_{xx}$
(cf. Fig. \ref{fig_tnot0.eps}(c,d)). As a consequence the minima at
finite bias are generally deeper and shift to lower voltages.

The effect of the constriction depends quantitatively on both the
inter-edge interaction parameter $\theta_2$ and the temperature.  To
appreciate this we plot in Fig. \ref{fig_tnot02.eps} the differential
resistance $R_{xx}$ for different values of the inter-edge interaction
and the temperature. More specifically we have plotted $R_{xx}$
without interactions ($\theta_2=\theta_1=0)$ and with interactions
within the constriction ($\theta_1=0$, $\theta_2=0.2$) for $\pi d
T/c_1 =0.5, 1, 5$, and $10$.
\begin{figure}[!ht]
\begin{center}
\includegraphics[clip,width=12cm]{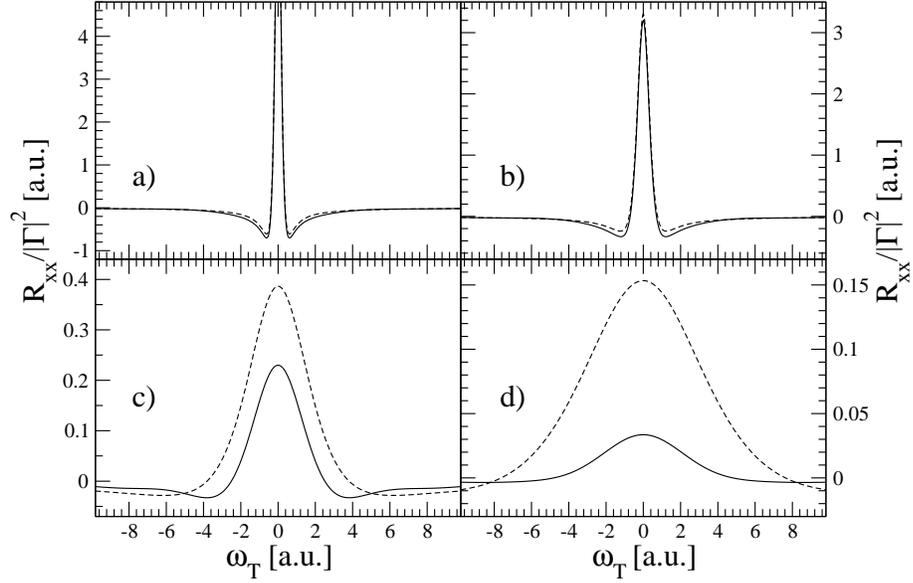}
\caption{Plot of the differential resistance $R_{xx}/|\Gamma|^2$ vs.  frequency
with and without inter-edge interaction within the constriction.
Solid line -- $\theta_1=0$, $\theta_2=0.2$; Dashed line --
$\theta_1=\theta_2=0$.  Temperatures are $\pi d T/c_1 =0.5$ (a),
$1$ (b), $5$ (c), and $10$ (d).}
\label{fig_tnot02.eps}
\end{center}
\end{figure}
We notice that the effect of the inter-edge interaction disappears at
sufficiently low temperature, since it is always the long times
exponent $\alpha$ that matters in that regime.  The effect of the
interaction shows up upon increasing the temperature above the
crossover energy $\hbar/t_0$: the latter decreases with increasing
$\theta_2$.  Such a trend is clearly seen by comparing
Figs. \ref{fig_tnot0.eps} and \ref{fig_tnot02.eps}.

%
%Finally we would like to comment about recent measurements of
%tunneling characteristics through a constriction \cite{Roddaro2002} in
%the weak inter-edge tunneling regime at high magnetic field.  At
%relatively high temperatures ($T > 400 mK$) the experiment clearly
%shows the emergence of a zero bias peak in the differential
%longitudinal resistance which is qualitatively consistent with the
%results presented above.  The experiment also shows well defined
%minima at finite bias voltage, which, according to the previous
%discussions may reveal the effect of the constriction. In fact, the
%system without inter-edge interactions never shows deep minima in this
%temperature range.
%
%At lower temperatures, on the other hand, the experiment shows a
%completely different behavior which is not qualitatively consistent
%with the present theory irrespective of the presence of inter-edge
%interactions. Strong tunneling effects\cite{Fendley1995},
%which can be treated by the thermodynamic Bethe Ansatz, are not likely
%to explain the unexpected {\it decrease} in $R_{xx}$ that is seen at
%these temperatures.  This clearly suggests that a different physical
%mechanism comes into play at these temperatures and some additional
%physical input is needed. One could, for instance, speculate that,
%within the constriction, the hydrodynamic approximation may be too
%crude and better treatment of the edge structure may be required.
%This is, however, outside the scope of the present work.

\chapter{Conclusions}
In this chapter we compare our results with the theory
of Wen and the measurements
made by a group in Pisa showing that our model can explain  
some of the 
experimental features that are missed by the previously developed theories.
After that we would like to discuss a few future research lines we can follow to further 
investigate these systems. Finally, we will state our conclusions.   

\section{Comparison with the experiments}
The group of F. Beltram, V. Pellegrini, and
S. Roddaro in Pisa has
performed a series of measurements of the 
tunneling current due to quasi-particles tunneling between two edges 
belonging to the same Quantum Hall liquid \cite{Roddaro2002}. 
In such a system, it is
believed that the tunneling particles are the bulk quasi-particles 
with fractional charge. The results of Wen \cite{Wen1991b}, 
and Kane and Fisher \cite{Kane1995}, predict that the tunneling
conductance $G(T,V)$ must show a power law behavior both in voltage
and in temperature,
\be
G(0,V)\sim V^\alpha;~~~~G(T,0)\sim T^\alpha
\ee
with the same exponent $\alpha$ determined by the
filling factor
\be
\alpha=2 \nu-2,
\ee
and, because in these theories $\nu<1$, we have $\alpha<0$. 
The divergence of the conductance at very low bias is removed
by the presence of a finite temperature, thus recovering a linear relation
between the current and the voltage. Notice that the expected
divergence is $G(0,V_T)\to -\infty$ for $V_T\to 0$ then
the effect of the temperature is to generate a zero bias maximum and
two minima at finite bias.   
In the Fig. \ref{roddaroexperiment.eps} we report the experimental
data for the measure of the tunneling conductance \cite{Roddaro2002}.
The magnetic field is determined by fixing the filling factor to $\nu=1/3$ while 
the temperature is varied from $900$ mK to $30$ mK. 
\begin{figure}[!ht]
\begin{center}
\includegraphics[width=12cm,clip]{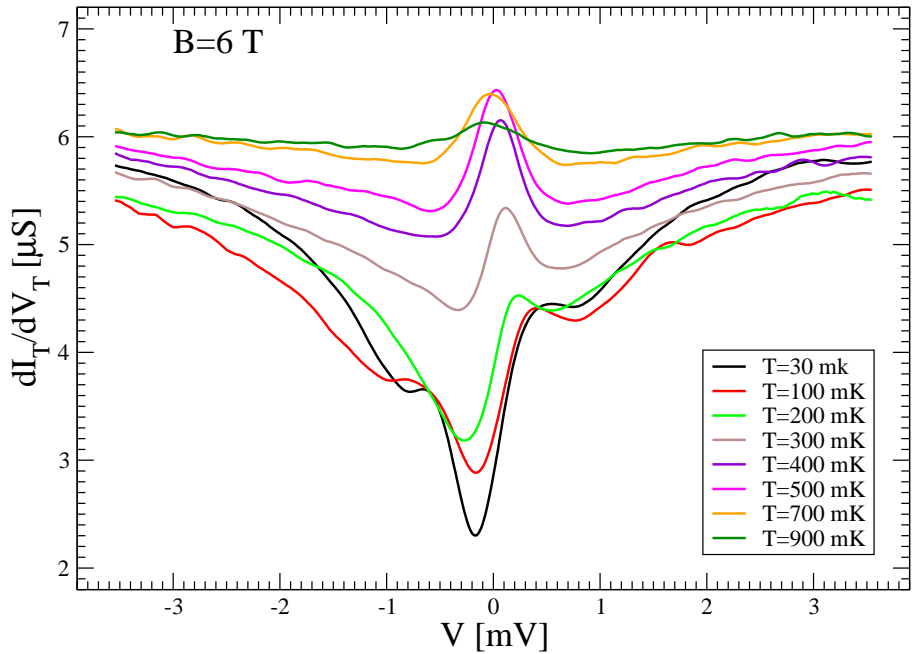}
\caption{Plot of the differential conductance vs. the tunneling voltage for
various temperatures at fixed magnetic field. The magnetic field and the
experimental device fix $\nu=1/3$. Data from \cite{Roddaro2002}.}
\label{roddaroexperiment.eps}
\end{center}
\end{figure}

For high temperatures ($900-400$ mK), the central maximum 
is clearly seen. Also two deep minima are present. The general 
behavior is in qualitative good agreement with the expected theoretical behaviors 
(see Figs.~\ref{fig_tnot0.eps} and \ref{fig_tnot02.eps}). In this range of 
temperatures is also evident an asymmetry for the change of the 
sign of the tunneling potential. The origin of this asymmetry, which becomes more
and more evident lowering the temperature, is not completely understood.
There is the possibility, as discussions with the experimentalists have 
pointed out,
that it can be an artifact of the experimental setup.  

For lower temperatures ($400-30$ mK) the structure 
of the response function
changes dramatically. 
The asymmetry becomes more evident, the central maximum
disappears and a deep central minimum starts to develop. 
The origin of this central minimum is completely unknown and 
we expect that, if it is a genuine experimental result, we must
plug in some other physical insights in our model to explain it.

For high temperatures and in the region of the central maximum, both our model
and that of Wen can be compared with the experimental results. 
To do that we need to fit the experimental data with the theoretical
curves to fix the free parameters: the translationally invariant interaction
angle for Wen or the inside and outside mixing angles in our model.
We choose to fit the data by using the least square minima technique.
We have then looked at the minima of the sum of the squared difference of the
experimental and theoretical curves to fix the parameters.
\begin{figure}[!ht]
\begin{center}
\includegraphics[width=12cm,clip]{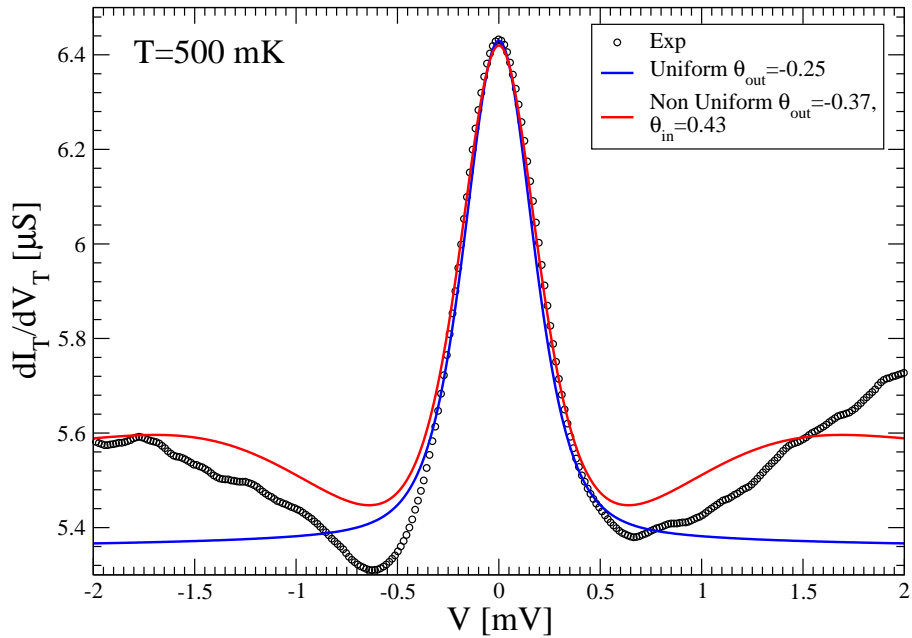}
\caption{Plot of the experimental and theoretical results. The filling fraction
$\nu=1/3$ and the temperature is fixed at $T=500$ mK. We have
fixed the total amplitude of the theoretical curves because we do not 
known the value of the tunneling amplitude $|\Gamma |$ and partially 
removed the asymmetry by moving the central maximum at $V_T=0$.
}
\label{plot_fit_data.eps}
\end{center}
\end{figure}

In Fig.~\ref{plot_fit_data.eps} we report the curves for the longitudinal
conductance at $T=500~\mbox{mK}$ obtained after having determined the best fit parameters. 
From this plot is clearly seen that both the Wen's theory and our model
recover the structure of the central maximum. However the Wen's theory
does not recover neither the deep minima that are present at $|V|\sim 0.7$ mV 
nor the rise of the conductance at larger voltage. 

We must point out that both the models predict either an attractive interaction
in the region 
outside the constriction or a larger than $e^*=e/3$ quasi-particle charge. 
In fact we use as fit parameters the exponents of the tunneling
conductance that tie together the quasi-particle charge $e^*$ and
the outside interaction parameters in products like $(e^*/e)^2 e^{-2\theta}$
and thus it is not possible to obtain from our fit direct 
measurements of the quasi-particle charge and of the mixing angle $\theta$, 
separately. Similar fits with the other experimental curves in the range 
$900$-$400$ mK have shown an increase of the term  $(e^*/e)^2 e^{-2\theta}$ when lowering the temperature. This dependence is not understood and is not
predicted neither by our model nor by the theory of Wen.   
On the other hand it is not possible to extract from the experimental data a measure of the oscillation period as determined by the Eq. (\ref{deltavt}).
  
\section{Perspectives}
The results obtained so far allow us to consider some possible developments.
In particular two arguments are attracting our attention. The first is the study
of the tunneling in the edges of the Quantum Hall liquid in more detail.
Indeed, when we have dealt with the problem of the tunneling into an edge, the problem
of the fine structure of the edge was not considered at all. However has been 
pointed out by several authors that, even for the Laughlin state where $\nu=1/m$ 
with $m$ an odd integer, the edge can undergo a reconstruction \cite{Chklovskii1992, Chamon1994, Wan2002, Rosenow2002,Joglekar2003} 
and acquire a complex structure. This reconstruction can affect the tunneling 
amplitudes by modifying the form of the quasi-particle creation and annihilation
operators. One way to take into account this effect can be the inclusion in the tunneling Hamiltonian of other terms that describe the possibility that 
several species of quasi-particles can propagate into the edge.
On this research line lies also the result of Eric Yang {\it et al.} \cite{EricYang2002}: 
The quasi-particle species that can undergo the tunneling between two edges 
depend on the separation width of the edges. In particular, if one considers
the tunneling between two Laughlin states, when the edges are far-off 
the tunneling quasi-particles are the fractionally charged Laughlin quasi-particles.
However, when the edges are close enough, other quasi-particles with different
charges can do the tunneling. This research line can thus clarify the properties 
of the tunneling quasi-particles and their relation with the bulk quasi-particle.

The other argument we want to explore is a more detailed study of the effect
of the constriction on the transport properties of the Quantum Hall bar. 
In fact we have not yet considered the possibility that the constriction, 
behaving like a semi-opaque barrier, has some tunneling resonances that 
will enhance the longitudinal conductance even when the magnetic field
is not present. A possible way to attack this problem is to solve the quantum
mechanical problem of the tunneling amplitudes of a constriction in a bidimensional 
system with zero magnetic field. 
We expect that these amplitudes depend on the energy of the 
impinging wave. When the magnetic field is not zero one can use
the basis developed for the tunneling in the previous case to 
calculate the new transmission and reflection coefficients. 
This research can also open the way to a more accurate study of the 
dependence of the tunneling amplitude $\Gamma$ as a function of the physical
parameters of the impinging quasi-particles.
 
\section{Conclusions}
The main subject of this thesis has been the study of the transport
properties of a two dimensional electron gas in the presence
of a high magnetic field. 
When the magnetic field
is very strong the kinetic energy is quenched and
the dynamics of the interacting electrons 
is produced by the commutation
relations that appear among the density
fluctuations as a result of the projection into the lowest Landau level.
The phenomenology of the transport properties is shown in the
Fig. \ref{qhe_transport_data.eps}. As we have discussed the
transversal resistance $R_{xy}$ (also known as the Hall resistance) shows 
a series of plateaus at values $\sigma_{H}=\nu e^2/h$ where 
$\nu$ is either an integer or a rational number, while the longitudinal 
resistance $R_{xx}$ shows an alternation of vanishing regions and 
$\delta$-like peaks. This effect
takes the name of Quantum Hall effect due to its connection
with the Classical Hall effect. 

To put our work in context, we have reviewed some of the theoretical approaches
developed  for the calculation
of the response function of these systems. In this thesis we have chosen
a hydrodynamical approach that allows us to treat
$\nu$ as a free parameter without restricting ourselves to the case
of $\nu$ being an integer or a rational number. 
This allows us to describe the dynamics
of the quasi-particles as charge density fluctuations localized 
at the edges of the system.
We have then developed the algebra to calculate the response
functions by using these density fluctuations and its relations with the quasi-particle
creation and annihilation operators. Our approach gives us the possibility 
to introduce, in a clear way, the interaction between the quasi-particles.
We have shown in detail how our model can recover the linear relation
between the filling factor $\nu$ and the transversal conductance. 

This research started as a collaboration with the experimental group
of F. Beltram, V. Pellegrini, and S. Roddaro
in Pisa. The focus of the experimental research was the development
of devices to detect and use the quasi-particle fractional
charges. The experimental device used is reasonably well modelled by
our idealized scheme (see Fig. \ref{fig_probe_theo.eps}). In studying 
this problem we had to face with
two distinct problems. The first has been the possibility of tunneling between
the edges and the other has regarded the possibility that the presence of 
the constriction can affect the transport properties.

We have thus decided to start by including the effect of the constriction
directly in the Hamiltonian by considering the inter-edge 
interaction potential in a piece-wise
form. The model in such a case can be easily solved. We have shown that 
the presence of a finite size constriction does not affect the low energy
transport properties thus recovering the ideal relation between the conductance and 
the filling fraction. This result can be physically understood by observing
that in the low energy limit, which corresponds to the long wavelength limit,
the constriction becomes fully transparent and the transmission coefficient
goes to $1$.

We have then discussed the possibility of tunneling between the edges inside
the constriction. The presence of the constriction introduces a time scale,
$t_0=d/c$. $t_0$ is the time a charge density wave needs to travel through the constriction.
This time introduces then a separation in the behaviors for long and  short
(compared with $t_0$) times. We have discussed the fact that for short times
the effect of the constriction becomes negligible.
The low energy behavior is mainly determined
by the long times behavior and this in turn is modified by the presence of 
the constriction. We have compared our work
with the previously developed theory of Wen, who considers
the case of tunneling between the edges in a translationally invariant 
system \cite{Wen1990, Wen1991a, Wen1991b}. Many differences have been
pointed out. In the regime of low bias voltage both theories predict the presence of 
a large central maximum. Near this maximum two deep minima develop. 
The structure of these minima differs in the two theories: in particular 
we predict that these minima must be much deeper than in the Wen's theory.  
At larger bias voltages we predict the presence of small
oscillations whose period $\delta V=h/e^* t_0$ allows a direct measure of
the traveling time and of the quasi-particle charge.  

We have then discussed the comparison of our result with some
recent measurements of the tunneling conductance. We have shown
that our model can explain some characteristics of the experimental 
results. Finally we have described possible future research 
lines.  
     
%\input{ac}
%\listoffigures
\appendix 
\chapter{Proof of the commutation relations}\label{commutationrel}

In this appendix we discuss the commutation relations for
the $\dra$ operators and show how to obtain it directly from the 
Lowest Landau Level projection and the hydrodynamical approximation.
We consider separately the two cases of sharp and smooth edge. 

\section{Sharp edge case}
Let us consider  a single left edge. 
We imagine that the region $y<0$ is empty and there is a sharp
wall at $y=0$. Then we have the condition on the wave number
$k<0$.  We suppose that the length scale of variation
of the confining potential in the $y$ direction is smaller than
$l$, hence we can neglect the states with $k$ around $0^+$ and
approximate the number density with a $\theta$ function
\be
n(k)=\nu \theta(-k).
\label{a1.6} 
\ee 
Notice that the negative value of $k$ are
given by the condition $y+kl^2=0$ which identifies the maximum of
the gaussian packet in the LLL and then the position of the
particle. There is a subtle point in this definition. The function
$n(k)$ is discontinuous in $k=0$. We take $n(0)=\nu/2$. This
assumption can be justified using the definition of the theta
function such that $\theta(0)=1/2$. We define the transverse
fluctuation 
\be 
\delta\hat\rho(x)=\int\limits_{-\infty}^{\infty}dy
\delta\hat\rho(x,y) 
\ee 
so we have the commutation relation between
this new operators
\be
\begin{split}
[\delta\hat\rho(x),\delta\hat\rho(x')]=&\int\limits_{-\infty}^\infty dy~dy'~
\sum_{k\not=h,m\not=l}\left(\hat c^\dagger_k \hat c_l
\delta_{h,m}-\hat c^\dagger_h \hat c_m\delta_{k,l}\right)\\
&\times\varphi_k^*({\bf r})\varphi_h({\bf r})\varphi_m^*({\bf r}')
\varphi_l({\bf r}')\\
=&\sum_{k\not= h}n(k)\int\limits_{-\infty}^\infty dy~dy'~
\left[\varphi_k^*({\bf r})\varphi_h({\bf r})\varphi_h^*({\bf r}')\varphi_k({\bf r}')\right.\\
&-\left.\varphi_h^* ({\bf r})\varphi_k({\bf r})\varphi_k^* ({\bf r}')\varphi_h({\bf r}')\right].
\label{difference}
\end{split}
\ee
If we use the expression for the function $\varphi_k({\bf r})$ in the
LLL, in the given gauge,
 \be 
\varphi_k(x,y)=\frac{1}{(\pi
l^2)^\frac14}e^{-ikx}\exp\left[-\frac{(y+kl^2)^2}{2l^2}\right] \ee
we obtain \be \int\limits_{-\infty}^{\infty}dy~ \varphi_k^*
(x,y)\varphi_h(x,y)=e^{-i(k-h)x}\exp\left[-\frac{(k-h)^2l^2}{4}\right]
\ee 
and this allows us to calculate the commutator
\be
\begin{split}
[\delta\hat\rho(x),\delta\hat\rho(x')]=&
\sum_{k\not=h}n(k)\left(e^{-i(k-h)(x-x')}
-e^{i(k-h)(x-x')}\right)\\
&\times\exp\left[-\frac{(k-h)^2 l^2}{2}\right]\\
=&\sum_{q}e^{-iq(x-x')}\exp\left[-\frac{q^2l^2}{2}\right]\sum_{h}
\left(n(q+h)-n(h-q)\right)\\
=&-\sum_{q} \frac{\nu}{2\pi} q e^{-iq(x-x')}\exp\left[-\frac{q^2l^2}{2}\right]\\
\stackrel{l\to 0}{=}&-\frac{i\nu}{2\pi}\frac{d}{dx'}\delta(x-x').
\end{split}
\ee
The last expression has been obtained by observing that 
in the limit $l\to 0$ the gaussian may be approximated by a delta
function. Notice also that $h$ is evaluated at $-q$ and $q$ which
corresponds to $k=0$. This leads to a factor $1/2$ according to
the definition of $n(0)$.

In the presence of two edges the derivation is similar but in this
case the values allowed for the wave vectors in the expressions
for the density fluctuations are different. Is such a case the
Kr\"onecker delta given by the commutator of the fermion operators
cannot be fulfilled if the densities belong to different edges.
Then the commutator is zero. Hence we obtain, if
$\alpha,\beta\in\{L,R\}$ are the indexes of the edge, 
\be
\left[\delta\rho_\alpha(x),\delta\rho_\beta(x')\right]=-\frac{i
\nu}{2\pi} \frac{d}{dx}\delta(x-x')\sigma_{\alpha,\beta}^z.
\label{commutation-2} 
\ee 
The appearance of the Pauli matrix
$\sigma^z$ is justified by the different forms of the
$n_\alpha(k)$ functions. In fact for the number densities we have,
if $D$ is the distance between the edges,
\begin{eqnarray}
&&n_L(k)=\left\{
\begin{array}{l}
\nu \theta(-k)\\
\nu/2~\mbox{if}~k=0
\end{array}
\right.\\
&&n_R(k)=\left\{
\begin{array}{l}
\nu \theta(k+D/l^2)\\
\nu/2~\mbox{if}~k=-D/l^2
\end{array}
\right.
\end{eqnarray}
hence when we integrate over $h$ they assume a different sign
\begin{eqnarray}
&&\int\limits_{-\infty}^{\infty}\frac{dh}{2\pi}\left[n_L(h+q)-n_L(h-q)\right]=-\frac{q}{2\pi},\\
&&\int\limits_{-\infty}^{\infty}\frac{dh}{2\pi}\left[n_R(h+q)-n_R(h-q)\right]=\frac{q}{2\pi}.
\end{eqnarray}
This completes the proof for the commutation relations in the case
of a sharp edge. 

\section{Smooth edge case}
Next we will consider the case of the smooth edge. In this case we
rewrite the equation (\ref{commutation-1}) in the form 
\be
[\delta\rho({\bf r}),\delta\rho({\bf r}')]=\sum_{k,h}(n(k)-n(h))\varphi_k^\dagger({\bf r})
\varphi_k({\bf r}')\varphi_h^\dagger({\bf r}')\varphi_h({\bf r})
\ee 
and in the case of slowly varying density function we have 
\be
n(k)\simeq n(h)+(k-h)\partial_k n(k). 
\ee 
The commutator of the density fluctuations reads 
\begin{eqnarray}
[\delta\rho({\bf r}),\delta\rho({\bf r}')]&=&\sum_{k,h} (k-h)\partial_k n(k)
\varphi_k^\dagger({\bf r})\varphi_k({\bf r}')\varphi_l^\dagger({\bf r}')\varphi_l({\bf r})\nonumber\\
&\stackrel{l\to 0}{=}& -\frac{i}{2\pi l^2}\partial_x \partial_k
n(k)\delta(y-y')\delta(x-x').
\end{eqnarray}
Now by remembering that $n(k)=2\pi l^2 \rho_0(y)$ and
$kl^2=y~\Rightarrow dk=dy/l^2$ we have 
\be
[\delta\rho({\bf r}),\delta\rho({\bf r}')]=-i l^2 \partial_x (\partial_y
\rho_0(y))\delta(x-x')\delta(y-y') 
\ee 
and integrating over the edge we finally obtain
\begin{eqnarray}
[\delta\rho_\alpha(x),\delta\rho_\beta(x')]&=&\int_\alpha dy\int_\beta dy'
[\delta\rho({\bf r}),\delta\rho({\bf r}')]\nonumber\\
&=&-i l^2\partial_y\sigma^z_{\alpha,\beta}\rho \delta(x-x')
\end{eqnarray}
where $\rho_0(y)$ is the equilibrium density and $\rho$ is the
bulk density. The apparition of the $\sigma^z$ matrix is justified
by the order of the limit of integration in the edge: We have
\begin{eqnarray}
&&\int_L dy~\partial_y \rho(y)=\rho(d)-\rho(0)=\rho,\\
&&\int_R dy~\partial_y \rho(y)=\rho(R+d)-\rho(R)=-\rho
\end{eqnarray}
if $d$ is the length over which the density goes from zero to the
bulk value (supposed for simplicity equal for the two edges). The
final observation
that $\nu=2\pi l^2\rho_0$ completes the proof.

\chapter{Properties of the eigenvalues problem}
\label{eigenvalueproperties}

In this appendix we want to study some analytical properties of the
equation (\ref{varphiequationofmotion}).  First of all let us define the operators
\begin{eqnarray}
M_{\alpha,\beta}&=&i\sigmaz\partial_{x},\\
H_{\alpha,\beta}&=&\frac{\nu}{2\pi}\int\limits_{-\infty}^{\infty}dx~
\partial_{x}V_{\alpha,\beta}(x,x')\partial_{x'}.
\end{eqnarray}
With this definition we rewrite the equation of motion
(\ref{varphiequationofmotion}) in the compact form 
\be 
\omega M \varphi=H\varphi.
\ee
It is easy to see that $H$ and $M$ are hermitian operators and we
request that $H$ is positive definite (this will assure the stability
of the physical system).

Let us define the auxiliary function 
\be 
\Psi=H^\frac12\varphi 
\ee
which is a solution of the equation 
\be
\frac{1}{\omega}\Psi=\left(H^{-\frac12}MH^{-\frac12}\right)\Psi=\tilde{M}\Psi
\label{new_prob} 
\ee 
if $\varphi$ is a solution of (\ref{varphiequationofmotion}). Because
$\tilde{M}$ is an hermitian operator we have the results:
\begin{enumerate}
\item the set $\{\Psi\}$ of solutions forms a complete base of the
Hilbert space,
\item the orthonormality condition is 
\be
\sum_{\alpha}\int dx
\Psi_{n,\alpha}^*(x)\Psi_{m,\alpha}(x)=\delta_{n,m}, 
\ee
\item the completeness relation 
\be
\sum_{n}\Psi_{n,\alpha}(x)\Psi_{m,\beta}^*(x')=\delta_{\alpha,\beta}
\delta(x-x'). 
\ee
\end{enumerate}

Because there is a one-to-one relation between $\varphi$ and $\Psi$ we
have the following properties of the solutions of equation
(\ref{varphiequationofmotion}):
\begin{enumerate}
\item the solutions $\varphi$ form a complete base of the Hilbert
space,
\item they are orthogonal with respect to the scalar product 
\be
(\varphi_n,\varphi_m)=\sum_{\alpha}\int dx~ \omega_n
\varphi_{n,\alpha}^*(x)M_{\alpha,\beta}\varphi_{m,\beta}(x), 
\ee
\item they satisfy the completeness relation 
\be -i \sum_n \omega_n
\varphi_{n,\alpha}(x)\varphi_{n,\beta}^*(x')s_\beta\partial_{x}=
\delta_{\alpha,\beta}\delta(x-x'). 
\ee
\end{enumerate}
We obtain the relations reported in the text if we normalize the
functions $\varphi_n$ as $\varphi_n/\sqrt{|\omega_n|}$.

Now we want discuss the degeneracy of the eigenvalues of the equation
(\ref{varphiequationofmotion}):
\begin{itemize}
\item If $\varphi_{n,\alpha}(x)$ is a solution with given eigenvalue
$\omega_n$ then the function $\varphi_{n,\alpha}^*(x)$ is also a
solution with eigenvalue $\omega_{-n}=-\omega_n$.
\item If $\varphi_{m,\alpha}(x)$ is a solution with given eigenvalue
$\omega_m$ then the function $\sigma_{\alpha,\beta}^x
\varphi_{m,\beta}(x)$ is also a solution with eigenvalue $-\omega_m$.
\end{itemize}
Then we have that if $\varphi_{m,\alpha}(x)$ is a solution then
$\sigma_{\alpha,\beta}^x\varphi_{m,\beta}^*(x)$ is still a solution with
the same eigenvalue thus the solutions of problem (\ref{varphiequationofmotion})
are doubly degenerate.

\chapter{Landauer-B\"uttiker transport formalism}
\label{Landauer-Buttiker}
In this appendix we want to discuss briefly the Landauer-B\"uttiker
formalism for the transport in one-dimensional system and its
application to the Quantum Hall edge transport \cite{Buttiker1988}.

We will discuss before the case of the Integer Quantum Hall phase where many
edge states are present. Every such state is a possible channel for
the current transport.
In the clean case, when there is not tunneling between different edges, in a six 
probes geometry similar to the one showed in the Fig.~\ref{fig_probe.eps}, we can 
write the balance equations 
\be
%\begin{split}
\left(
\begin{array}{c}
I_1\\
I_4\\
I_2\\
I_3\\
I_5\\
I_6
\end{array}
\right)
=
\frac{e^2}{h}(N+1)
\left(
\begin{array}{cccccc}
1 & 0 & 0 & 0 & 0 &-1 \\
0 & 1 & 0 & -1 & 0 & 0\\
-1 & 0 & 1 & 0 & 0 & 0\\
0 & 0 & -1 & 1 & 0 & 0\\
0 & -1 & 0 & 0 & 1 & 0\\
0 & 0 & 0 & 0 & -1 & 1
\end{array}
\right)
\left(
\begin{array}{c}
V_1\\
V_4\\
V_2\\
V_3\\
V_5\\
V_6
\end{array}
\right)
\ee
where we have assumed that the current in the $i$-th terminal is determined
only by its potential and the potential of the contact directly ``down-stream" with
respect to  the versus of propagation of the edge which leaves the terminal. Notice
also that we have ordered differently the reservoir to have a direct comparison
with the matrix conductance given by the Eq.~(\ref{gij}). We 
have also considered the possibility that $N+1$ propagation channels are present.
To solve this set of equations one fixes the current of the terminal $2,~3,~5$ and
$6$ to zero then the total current is given by $I=I_1=-I_4$. We fix
also the potential of the terminal $4$ to zero. We obtain, by defining $\sigma_{i,j}=I/(V_i-V_j)$ 
\be
\begin{split}
&\sigma_{2,3}=\sigma_{5,6}=0,\\
&\sigma_{2,5}=\sigma_{2,6}=\sigma_{3,5}=\sigma_{3,6}=\frac{e^2}{h}(N+1).
\end{split}
\ee
It is possible to generalize such approach to the case of the FQHE by 
substituting the number of possible channels $N+1$ with the filling fraction
$\nu$.

It is also possible to consider the effect of the tunneling between 
two edges. We model this tunneling by defining a transmission
$T$ and a reflection $R$ probability. The reflection coefficient is
given by the probability that a particle moving
from $2$ to $3$ or from $5$ to $6$ tunnels to the other edge. 
We have
\be
\left(
\begin{array}{c}
I_1\\
I_4\\
I_2\\
I_3\\
I_5\\
I_6
\end{array}
\right)
=
\frac{e^2}{h}\nu
\left(
\begin{array}{cccccc}
1 & 0 & 0 & 0 & 0 &-1 \\
0 & 1 & 0 & -1 & 0 & 0\\
-1 & 0 & 1 & 0 & 0 & 0\\
0 & 0 & -T & 1 & -R & 0\\
0 & -1 & 0 & 0 & 1 & 0\\
0 & 0 & -R & 0 & -T & 1
\end{array}
\right)
\left(
\begin{array}{c}
V_1\\
V_4\\
V_2\\
V_3\\
V_5\\
V_6
\end{array}
\right).
\ee
Notice that the gauge invariance and the current conservation require that
$R+T=1$. We can solve this system with the same requirements for
the setup of the currents and voltages to obtain the conductance matrix 
in this case. In the limit $R\to 0$ we recover the previous result.

A more complex matrix can be obtained if one considers the reservoirs 
coupled as we have discussed in Eq. (\ref{gij}). 
If we consider the non interacting case $u=1,~v=0$ from the Eq. (\ref{gij})
we have
\be
\label{gij-notin} 
G_{ij}=\frac{e^2}{h}\nu\left(
\begin{array}{cccccc}
1  &  0 & 0 		     & 0 		& 0 		      & -1               \\ 
0  &  1 & 0 		     & -1 		& 0 		      & 0	       \\ 
-1 &  0 & 1 		     & 0     		& 0      	      & 0  	    \\ 
0  &  0 &-1+\delta g_{21}& 1 & \delta g_{23}      & 0\\
0  & -1 & 0    & 0     & 1  & 0\\
0  &  0 & \delta g_{41}    & 0     & -1+\delta g_{43} & 1
\end{array}
\right).
\ee 
The gauge invariance and the conservation of the current fix 
the allowed values of the conductances $\delta g_{21}$, $\delta g_{23}$, 
$\delta g_{41}$, $\delta g_{43}$: we have then
\be
R=\delta g_{21}=\delta g_{43}=-\delta g_{41}=-\delta g_{23}=i\lim_{\omega \to 0}
\frac{\tilde M_T(\omega)}{\omega}e^{-2\theta}.
\ee
The presence of the constriction couples all the edges and we must solve the
linear problem given by the Eq.~(\ref{gij}). In such case it is not possible to define
a single reflection coefficient and express the solution only in terms of it.

\chapter{Evaluation of integrals}\label{calculationofg}
In this appendix we provide a few details concerning the evaluation of
the integral occurring in the calculation of the Fourier transform of
the response function $G_-(t)$.  In the zero temperature case, we must
evaluate an integral of the form 
\be
\label{integral1}
\int\limits_{t_0}^\infty dt (\delta\pm it)^{-\alpha} e^{i\omega t} 
\ee
where $\alpha$ is a positive real number. To do this we go in the
complex plane of the variable $t$ and consider, for positive frequency
$\omega$, an integration path as the one shown in
Fig. \ref{integrationpath.eps}. 
\begin{figure}[!h]
\begin{center}
\includegraphics[clip,width=7cm]{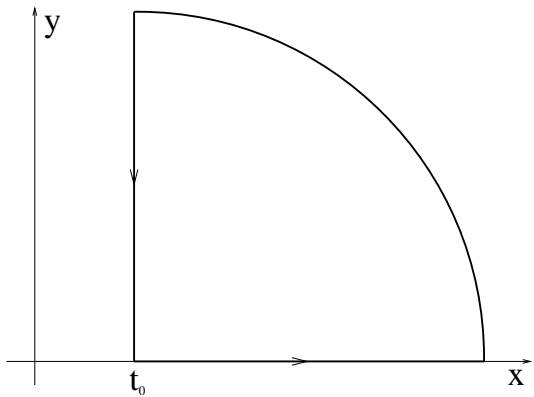}
\caption{The integration path used to evaluate the integral
(\ref{integral1}) when the frequency is positive. A similar path,
closed in the lower plane, is used in the case $\omega<0$.}
\label{integrationpath.eps}
\end{center}
\end{figure}
A specular path in the lower half
plane must be used for negative frequency.  We observe that the
integrand function has no poles in the complex half-plane of $t$ with
a positive real part. Hence the integral on the whole path is
zero. The integral on the arc vanishes by letting the radius go to
infinity.

As a result we get 
\be \int\limits_{t_0}^\infty dt (\delta\pm
it)^{-\alpha} e^{i\omega t}\stackrel{\delta \to 0}{=} i (\mp
1)^{-\alpha} \omega^{\alpha-1} \Gamma(1-\alpha,-i\omega t_0).  
\ee 
The
case $t_0=0$ can be carried out by calculating the convolution product
between the Fourier transform of the $\Theta(t)$ function and the
integral 
\be
\begin{split}
\label{completeintegral}
\int\limits_{-\infty}^\infty dt (\delta\pm it)^{-\alpha} e^{i\omega
t}=&2 \sin(\pi\alpha) e^{i\frac{\pi}{2}\alpha}|\omega|^{-1-\alpha}\\
&\times \Gamma(1+\alpha)(\mp 1)^{-1-\alpha},
\end{split}
\ee 
obtained by cutting the complex plane along the imaginary axis,
starting from $t=\pm i\delta$ as is shown in the Fig.~\ref{integrationpath2.eps} 
when the singularity is located in the upper plane 
and for positive frequency. Similar paths were used in the other cases.    
\begin{figure}[!h]
\begin{center}
\includegraphics[width=10cm,clip]{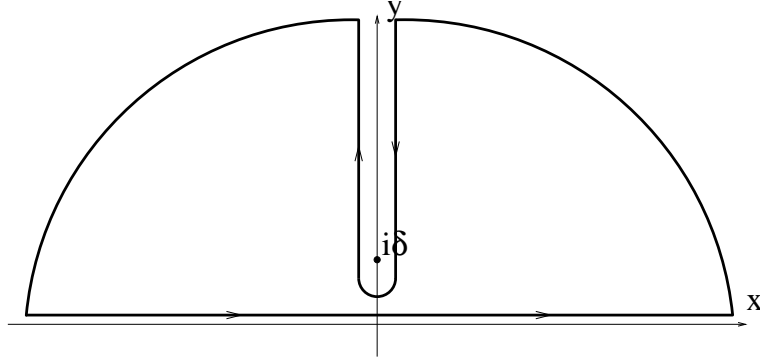}
\caption{The integration path used for the evaluation of the integral~
(\ref{completeintegral}) when the frequency is positive. A specular path
with respect to the real axis is used when the singularity is located in the lower plane. Similar paths were used in the other cases.}
\label{integrationpath2.eps}
\end{center}
\end{figure}

In the finite temperature case, we need to calculate the integral 
\be
\int\limits_{t_0}^\infty dt~e^{i\omega t}\frac{{(\pi
T)}^\alpha}{\sinh(\pi T t)^\alpha}.  
\ee One can easily obtain the
result reported in the text by means of the substitution $s=e^{-2\pi
Tt}$ which reduces the above integral to the definition of the
hypergeometric $F$ function of four
arguments~\cite{Abramowitz1964, Gradshteyn1965}. The integral with $t_0=
0$ can be easily obtained in the limit $t_0\to 0$ by using the
corresponding limiting expression of the hypergeometric function $F$
in terms of the Euler beta function.

\addcontentsline{toc}{chapter}{References}
\bibliographystyle{apsrev} 
\bibliography{../qhe-biblio}

\end{document}